\documentclass[fleqn,usenatbib]{mnras}

\DeclareRobustCommand{\VAN}[3]{#2}
\let\VANthebibliography\thebibliography
\def\thebibliography{\DeclareRobustCommand{\VAN}[3]{##3}\VANthebibliography}

\DeclareSymbolFont{CMletters}{OML}{cmm}{m}{it}
\DeclareMathSymbol{v}{\mathord}{CMletters}{`v}

\usepackage{graphicx}	
\usepackage{amsmath}	
\usepackage{amssymb}
\usepackage{enumitem}
\usepackage[dvipsnames]{xcolor}
\usepackage{ulem}

\usepackage{newtxtext}
\usepackage[varvw]{newtxmath}
\usepackage[T1]{fontenc}
\usepackage{bm}

\defcitealias{paper1}{Paper~I}

\title[Structure Formation in SFDM with Repulsive SI]
{Cosmological structure formation in scalar field dark matter with repulsive self-interaction: \textit{The Incredible Shrinking Jeans Mass}}

\author[P. R. Shapiro, T. Dawoodbhoy, \& T. Rindler-Daller]{
Paul R. Shapiro,$^{1}$\thanks{E-mail: shapiro@astro.as.utexas.edu}
Taha Dawoodbhoy,$^{1}$\thanks{E-mail: tahad@astro.as.utexas.edu}
and Tanja Rindler-Daller$^{2}$\thanks{E-mail: tanja.rindler-daller@univie.ac.at}
\\
$^{1}$Department of Astronomy and Texas Cosmology Center, The University of Texas at Austin, Austin, TX 78712-1083, USA\\
$^{2}$Institut f\"ur Astrophysik, Universit\"atssternwarte Wien, University of Vienna, T\"urkenschanzstr.17, A-1180 Vienna, Austria
}

\date{Accepted XXX. Received YYY; in original form ZZZ}
\pubyear{2021}

\begin{document}
\label{firstpage}
\pagerange{\pageref{firstpage}--\pageref{lastpage}}
\maketitle

\begin{abstract}

Scalar Field Dark Matter (SFDM) comprised of ultralight ($\gtrsim 10^{-22}$ eV) 
bosons is an alternative to standard, collisionless Cold Dark Matter (CDM)
that is CDM-like on large scales but inhibits small-scale structure formation.  
As a Bose-Einstein condensate, its free-field (``fuzzy’’) limit (FDM) 
suppresses structure below the de~Broglie wavelength, $\lambda_\text{deB}$, 
creating virialized haloes with central cores of 
radius $\sim\lambda_\text{deB}$, surrounded by CDM-like envelopes, 
and a halo mass function (HMF) with a sharp cut-off on small scales. 
With a strong enough repulsive self-interaction (SI), 
structure is inhibited, instead, below the Thomas-Fermi (TF) radius, $R_\text{TF}$  (the size 
of an SI-pressure-supported ($n=1$)-polytrope), 
when $R_\text{TF} > \lambda_\text{deB}$. 
Previously, we developed tools to describe SFDM dynamics on scales 
above $\lambda_\text{deB}$ and showed 
that SFDM-TF haloes formed by Jeans-unstable collapse from non-cosmological initial conditions 
have $R_\text{TF}$-sized cores, surrounded by CDM-like envelopes.  
Revisiting SFDM-TF in the cosmological context, we simulate halo formation by cosmological 
infall and collapse, and derive its transfer function from linear perturbation theory
to produce cosmological initial conditions and predict statistical measures of structure formation, such as the HMF. 
Since FDM and SFDM-TF transfer functions both have small-scale cut-offs, we can
align them to let observational constraints on FDM proxy for SFDM-TF, 
finding FDM with particle masses $1 \lesssim m/(10^{-22} \text{ eV}/c^2) \lesssim 30$ corresponds 
to SFDM-TF  with $10 \gtrsim R_\text{TF}/(1  \text{ pc}) \gtrsim 1$,
favoring sub-galactic (sub-kpc) core-size. 
The SFDM-TF HMF cuts off gradually, 
however, leaving more small-mass haloes:   its Jeans mass shrinks so fast, scales filtered early can still 
recover and grow!

\end{abstract}




\section{Introduction}
\label{sec:Introduction}

The nature and origin of cosmic dark matter is still unknown.  The standard Cold Dark Matter (``CDM'') model is actually a place-holder with certain assumed properties -- e.g. it behaves
during the structure formation era as a
nonrelativistic, collisionless gas of particles that couple only weakly to each other and to other
components of the Universe, via gravity alone,
contributing to the expansion rate of the Universe as a pressure-free component whose mass density, conserved over time, is measured astronomically.  A further assumption is made 
that the CDM particles are born so ``cold'' -- i.e. with such small peculiar motion -- that their free-streaming could only have washed out primordial density fluctuations on the tiniest
of scales, many orders of magnitude below the scale that affects the formation of galaxies or even their internal structure. With these assumptions, it is possible to work out, within the
context of the standard big bang model and inflationary paradigm, how structure formation proceeds from linear-perturbation initial conditions through nonlinear formation of galaxies, clusters and the cosmic web.  This model has successfully accounted for many observational properties of the Universe on large scales, from galaxy formation scales on up.
It has, however, been challenged by over-predicting
the level of small-scale structure inside galaxies
compared to that observed in a variety of systems (see, for instance, \citealp{BBK17} and references, therein).  This may reflect
our incomplete knowledge of the impact on galaxy
formation of the baryonic component, coupled via
gravity to CDM.  However, it has also led
to a consideration of microscopic properties
of CDM that go beyond the standard assumptions
described above, in ways that can affect
small-scale structure without spoiling its success on larger scales.  

The best-studied origin story
of CDM involves 
Weakly Interacting Massive Particles (``WIMPs'') with masses in the range $\sim$ GeV - TeV, predicted
by theories of supersymmetry, which form as big bang thermal relics whose abundance is set
when their annihilation at a weak-interaction-like rate becomes slower than the Hubble expansion rate.
This so-called ``standard CDM'' -- more
properly referred to as ``WIMP-CDM'' -- 
is more specific than the CDM whose
generic properties we identified above, with additional observable consequences, and, as such,
is more tightly constrained.  
In particular, it predicts that the same annihilations might lead to their indirect detection from astronomical sources 
(e.g. by gamma-rays), while their 
scattering interactions with nucleons 
might permit their direct detection.  So far,
however, despite considerable effort,  
there are no confirmed detections, and the allowed
range of particle models and parameters that remain
viable, consistent with that nondetection, has
shrunk \citep[see e.g.][]{Schumann19}.  This has added some urgency to the consideration of alternative models for
the origin and microscopic nature of CDM, even
apart from the possible conflicts it has of
over-predicting small-scale structure on
galactic scales.  We emphasize this last point
at the outset, because it is often missed when
evaluating alternatives to WIMP-CDM, by
dismissing the alternatives as soon as it
appears they are merely ``CDM-like'' at
all scales of interest, i.e. failing to
provide the ``Deus ex machina'' of
reducing the apparent over-prediction of small-scale structure by CDM, while retaining
all of its successes on larger scales. 
That is, as long as the origin of CDM is
unknown, any microscopic model
that is observationally indistinguishable
from it is just as successful as any other, until it is disfavored
either by further theoretical considerations, or by predictions of additional observables
that can distinguish it from the standard model.  

One such alternative model of great interest is Scalar Field Dark Matter (``SFDM'') comprised of ultralight ($m \gtrsim 10^{-22}$ eV$/c^2$) 
bosons, originally suggested because of its 
novel structure-formation dynamics as a Bose-Einstein condensate and quantum superfluid, described by coupled Schr\"odinger-Poisson equations.\footnote{The field of SFDM study has an extensive literature, well-beyond our ability to cite it here.  A few early references not otherwise cited in this paper include \citet*{Sin94,LK96,MGU00,GU04,MVM09,SC17,DKR18}. Some reviews with additional references therein include \citet*{SRM14,RDS14,Marsh16}.} 
In the free-field (``fuzzy'') limit of SFDM, or fuzzy dark matter (``FDM''), the small-scale 
structure which we described above for 
``generic'' CDM is suppressed below the 
de~Broglie wavelength, $\lambda_\text{deB}$, but resembles CDM on larger scales.  
Virialized FDM haloes have been shown to have
solitonic cores of radius $\sim\lambda_\text{deB}$,
surrounded by CDM-like envelopes (\citealp*{SCB14, SNE16}; \citealp{Mocz17}). When a strong enough repulsive self-interaction (SI)
is also present, 
structure can be inhibited below a second length scale, $\lambda_\text{SI}$, 
with $\lambda_\text{SI}>\lambda_\text{deB}$ -- referred to as the Thomas-Fermi (``TF'') regime.
This TF regime of SFDM (henceforth, ``SFDM-TF''),
assuming a quartic potential for the repulsive SI, has been referred to variously as ``BEC-CDM'' \citep{RDS12,RDS14},
SFDM \citep*{Li14,Li17}, ``Repulsive Dark Matter'' \cite{Goodman2000}, or more generally ``Fluid Dark Matter'' in \cite{Peebles2000}.\footnote{
Note that SFDM-TF should not be confused with another CDM alternative known as self-interacting dark matter (SIDM). 
The latter behaves as a collection of point-particles undergoing elastic scatterings, which
amounts to adding 2-body elastic collisions to the standard, collisionless gravitational dynamics of CDM, while the former is a BEC, best thought of as a classical field evolving under a continuous self-interaction potential (along with the gravitational potential).
The two models are phenomenologically distinct, as is evident by the r\^ole SI plays in their respective non-relativistic fluid equations. In the SFDM-TF model (as discussed in \S\ref{subsection:fluidapprox}), SI enters as a potential or pressure term in the momentum equation (equation~\ref{eq:1momLagrangian}), whereas in the SIDM model, SI enters as a heat conduction term in the energy equation \citep*{BSI02,AS05}.}

Structure formation in the TF regime differs significantly from FDM.  
In \citet*{paper1} \citepalias[henceforth][]{paper1}, we discussed the nonlinear dynamics and internal structure 
of haloes that form from gravitational instability in the TF regime.  Here we revisit this in the 
context of halo and large-scale structure formation from cosmological perturbations, 
including observational constraints. As this is a companion paper (``Paper II''), we 
will sometimes refer the reader to \citetalias{paper1} for details, 
for the sake of brevity and to avoid
redundancy.  

In \citetalias{paper1}, we developed equations
and tools to calculate the dynamical evolution of inhomogeneity
in response to gravitational instability in this model, 
starting from the nonlinear Schr\"{o}dinger equation (``NLSE'').
A key development was to take
full account of both quantum pressure (associated with inhomogeneity on the tiny
scale of the de~Broglie wavelength) and SI, based upon a
reformulation of the NLSE, smoothed on scales larger than 
the de~Broglie scale. In that case,
the NLSE reduced to the collisionless Boltzmann equation
that describes CDM, except with
the SI potential added to
the usual gravitational potential. 
From this, a fluid approximation 
for SFDM-TF was derived, which
added a new pressure force for
SI to equations we 
previously derived for CDM.
This enabled us to describe the scalar field by the familiar 
hydrodynamical conservation
equations for a $\gamma=5/3$ ideal gas, with an ideal gas pressure,
$P_{\!\sigma}$, 
associated with particle velocity dispersion (which accounts
for the large-scale effects of
quantum pressure), and with a
second pressure, $P_\text{SI}$, which accounts for the
repulsive SI force. This
fluid approximation allowed us to
evolve SFDM-TF computationally
in spherical-1D with the very 
high resolution and dynamic
range which are required
(but not currently possible)
for full 3D simulations of SFDM-TF.
With this tool, we solved the Newtonian problem of virialized halo formation by the nonlinear collapse of a Jeans-unstable perturbation,
from non-cosmological (i.e. static)
initial conditions, for the
first time. These simulations in 
\citetalias{paper1} showed that
haloes formed in this way have 
core-envelope density profiles, with
flattened cores of size $R_\text{TF}$ -- 
the radius of the ($n=1$)-polytrope that forms for a self-gravitating fluid in hydrostatic equilibrium when supported by SI pressure alone, referred to as the Thomas-Fermi radius --
surrounded by an envelope just like the one that
formed when CDM dynamics 
was simulated, instead, 
from the same initial
conditions. This explained why, in the case of FDM, when haloes form in 3D simulations (which are possible
for FDM, where the de~Broglie wavelength is large enough to be 
fully resolvable, unlike SFDM-TF,
for which de~Broglie wavelengths are unresolvably small),
density profiles are also CDM-like
outside their solitonic cores (of size close to the de~Broglie wavelength).

In \citetalias{paper1}, we showed
how SFDM-TF haloes with the right value of $R_\text{TF}$ could resolve
the cusp-core and ``too-big-to-fail'' (TBTF) problems
of CDM, without suffering
from the problem identified for FDM
by \cite*{VBB19} in which the
solitonic cores of larger-mass FDM haloes were too dense to be consistent with the rotation curves
of larger-mass galaxies, if the FDM
particle mass were tuned to 
$m\simeq10^{-22}$~eV$/c^2$ to match observations of smaller-mass galaxies.  We found
that $R_\text{TF}\gtrsim1$ kpc is
required.  It was left to Paper II,
here, to determine if such a value
of $R_\text{TF}$ was consistent
with the cosmological formation of
those haloes.

In what follows, we study the cosmological context for structure formation in the SFDM-TF model, 
both by simulating individual halo formation cosmologically,
and by deriving the initial conditions for Gaussian-random
primordial fluctuations from
linear perturbation theory. 
In \S\ref{sec:basicequations}, we
present the model and basic equations we use to study SFDM-TF
cosmologically.  In \S\ref{sec:infall}, we
apply these equations to simulate
individual halo formation by spherical infall and collapse in a cosmologically expanding SFDM-TF universe.  The overall scenario for cosmic structure in the SFDM-TF model is developed in \S\ref{sec:LinPertsTransferFn}, including
linear perturbation theory to
derive the transfer function and
power spectrum, as well as their
application to statistical
measures of structure formation like
the halo mass function.  We apply
the same approach to the FDM
model, as well, for comparison.
In \S\ref{sec:results}, we present
the results of this linear perturbation theory and use 
observational constraints on the
FDM model, fully parameterized by
its particle mass $m$,
as a proxy to derive
constraints on the SFDM-TF model,
parameterized by $R_\text{TF}$.
Our summary and conclusions are presented in \S\ref{sec:Conclusion}.

\section{The model and basic equations}
\label{sec:basicequations}

\subsection{The fully relativistic description}
\label{sec:fullyrelativistic}

We model SFDM as a classical, massive, complex
scalar field $\phi$ with a quartic, repulsive self-interaction (SI), which obeys the Klein-Gordon equation
\begin{equation}
    \Box \phi + \frac{m^2 c^2}{\hbar^2}\phi + \frac{2 g}{\hbar^2} |\phi|^2 \phi = 0
    \label{eq:KGE}
\end{equation}
where $\Box \equiv \nabla_\mu \nabla^\mu$ is the d'Alembertian for a general 4D spacetime metric, $m$ is the boson mass, 
and $g$ is the SI coupling strength. This equation is coupled to the Einstein equations, with a metric tensor given 
by the Friedmann-Lemaitre-Robertson-Walker metric for a homogeneous and isotropic background universe, 
perturbed by primordial fluctuations.  The scalar field contributes terms to the stress-energy tensor in the Einstein equations, which contain the field and its derivatives.  

As we described in 
\citet{Li14,Li17}, the evolution of the homogeneous background scalar field corresponds to that of a perfect fluid. However, the scalar field oscillates 
with an angular frequency which eventually
exceeds the Hubble expansion rate, when 
\begin{equation}
    \omega \simeq \frac{mc^2}{\hbar} \sqrt{1 + \frac{2g}{m^2 c^2}|\phi|^2 } \gg H
\end{equation}
\citep[see equation 21 in][]{Li14},
where the square of the modulus of the field, $|\phi|^2$, is, at late times, related to the SFDM mass density by  $\rho_b=\langle|\phi|^2\rangle=\rho_b(a=0)a^{-3}$ (the subscript ``$b$'' refers to background), and the Hubble parameter $H(t) = (da/dt)/a$ with scale factor $a=a(t)$ depending on cosmic time $t$.  
Prior to this rapid oscillation epoch, the 
field evolution was shown to start from a kinetic-energy-dominated, stiff, 
relativistic phase during the slow-oscillation period when $\omega \ll H$, but eventually to transition 
over time to a radiation-like, still relativistic phase during the rapid oscillation period, 
when the SI term in equation~(\ref{eq:KGE}) dominates the quadratic mass term.
Finally, at later times, when the mass term dominates, there is a transition to a nonrelativistic, matter-like phase.  More precisely, as a perfect fluid, the field can be characterized by an equation of state parameter $w$ defined as the ratio of pressure to energy density, which varies over time to reflect these phases of evolution.  Initially, $w=1$ (stiff phase), which must end prior to the epoch of big bang nucleosynthesis (BBN), in order not to spoil the agreement between predictions of the standard model involving primordial nucleosynthesis in a radiation-dominated (RD) universe with 
the observed relic abundances of the light elements. Eventually, when the field oscillates much more rapidly than the Hubble expansion rate,
we can average the pressure $P$ and energy-density $\varepsilon$ of the field over the oscillation period and take the ratio of these average values to give the average value of the equation
of state parameter $w$, according to the following \citep[which is equation 26 in][]{Li14}:
\begin{equation}
    \langle w \rangle \equiv \frac{\langle P \rangle}{\langle \varepsilon \rangle} = \frac{1}{3} \left( \frac{1}{1 + \frac{2m^2c^2}{3g \langle |\phi|^2 \rangle}} \right)
    \label{eq:averagew}
\end{equation}
When the SI term dominates during this phase,
$\langle w \rangle \simeq 1/3$ (radiation-like), but at late times, before the
universe made its transition from radiation-
to matter-dominated (MD), the mass term
must come to dominate and $\langle w \rangle \simeq 0$ thereafter, rendering SFDM matter-like, as it should be in order to account for the MD era.\footnote{This description above
from \citet{Li14,Li17}, of the evolution of the EOS
of the homogeneous scalar field, from stiff,
relativistic ($w=1$) at early times, before
the rapid-oscillation phase, to radiation-like
\big($\langle w \rangle \simeq 1/3$\big) when rapid oscillation begins and SI dominates, and finally
to nonrelativistic, matter-like 
\big($\langle w \rangle \simeq 0$\big),
was derived for a complex scalar
field.  However, we are only interested here
in the rapid-oscillation phase, during which
the EOS is radiation-like when the SI term 
dominates over the mass term in the Lagrangian,
then transitions to matter-like when the
latter term dominates.  As such, 
what happens before the rapid-oscillation 
phase begins does not affect our calculations here.  
In that case, we do not expect our calculations
to depend upon whether the field is complex or
real, since the description above of the EOS
evolution during the rapid-oscillation phase
applies equally well to the real and complex cases.}

\subsection{Matter-like regime and the Newtonian approximation}
\label{sec:matterlike}

In this latter \big($\langle w \rangle \simeq 0$\big) regime, in the limit of weak-field (i.e. Newtonian) gravity and even in the presence of inhomogeneities, as long as they are
on scales well-within the horizon (i.e. $\ll c/H$), we can adopt the Newtonian approximation in which gravity obeys the Poisson equation. The equation of motion of the scalar field in this nonrelativistic regime then simplifies to the nonlinear Schr\"odinger equation (NLSE) in an expanding background, by factorizing the field into a slowly-varying ``mean'' field $\psi$ and an oscillatory component, as follows:\footnote{
If $\phi$ is a real field, then one should add to the right side of equation~(\ref{eq:meanfield}) the complex conjugate, i.e. $\phi = \psi e^{-imc^2t/\hbar} + \psi^* e^{imc^2t/\hbar}$, where $\psi$ is still a complex field.}
\begin{equation}
    \phi = \psi e^{-imc^2t/\hbar} \quad \quad \quad |\partial_t \psi| \ll \left| \frac{mc^2}{\hbar} \psi \right|
    \label{eq:meanfield}
\end{equation}
In terms of the comoving coordinate $\bm{x}$, defined with respect to the proper coordinate $\bm{r}$ such that
\begin{align}
    &\bm{x} = \bm{r}/a \\
    &\nabla \equiv \nabla_{\bm{x}} = a\nabla_{\bm{r}}
\end{align}
and
\begin{equation}
    \partial_t \equiv (\partial_t)_{\bm{x}} = (\partial_t)_{\bm{r}} + H \bm{r} \cdot \nabla_{\bm{r}}
\end{equation}
where $(\partial_t)_{\bm{x}~(\bm{r})}$ represents the partial time derivative at fixed comoving (proper) coordinate $\bm{x}$ ~($\bm{r}$),
the NLSE and Poisson equations are then
\begin{equation}
    i\hbar \left(\partial_t + \frac{3}{2}H \right)\psi = \left( -\frac{\hbar^2}{2ma^2} \nabla^2 + \frac{g}{m}|\psi|^2 + m\Phi \right) \psi
\end{equation}
and
\begin{equation}
    \nabla^2 \Phi = 4\pi G a^2 (\rho - \rho_b)
\end{equation}
with
\begin{equation}
    \rho = |\psi|^2
\end{equation}
where $\Phi$ is the peculiar Newtonian gravitational potential, $\rho$ is the mass density of the scalar field, and $\rho_b=\rho_b(a=0)a^{-3}$ is the homogeneous background value.\footnote{
We shall generally assume here, unless otherwise stated, that the nonrelativistic matter density is dominated by the scalar field dark matter density, so we do not need to keep separate track of other nonrelativistic mass components like the baryons.}

These equations were presented in \citetalias{paper1}
in the Newtonian limit in the
absence of cosmological expansion.  The difference
here is that we represent derivatives and spatial coordinates in the comoving frame,
which introduces the homogeneous expansion term involving Hubble parameter $H(t)$,
and the peculiar gravitational potential $\Phi$.
If the scale factor were constant, $a(t) = 1$,
we would, of course, recover the equations of \citetalias{paper1}.

As we discussed in detail in \citetalias{paper1}, 
there is a fully equivalent formulation of this
nonrelativistic NLSE as quantum hydrodynamics
equations (``QHD'').  In particular, the NLSE
can be transformed into a pair of hydrodynamical
mass and momentum conservation equations via the Madelung transformation,
in which the field is expressed in polar form as
\begin{equation}
    \psi = |\psi|e^{iS}
    \label{eq:Madelung}
\end{equation}
and the conservation of current density
allows us to identify the peculiar bulk fluid velocity in the current flux density as
\begin{equation}
    \bm{v} = \frac{\hbar}{ma} \nabla S
\end{equation}
This leads to
\begin{equation}
    \partial_t \rho + 3H\rho + \frac{1}{a} \nabla \cdot (\rho \bm{v}) = 0
\end{equation}
and
\begin{equation}
    \partial_t \bm{v} + H\bm{v} + \frac{1}{a} (\bm{v} \cdot \nabla) \bm{v} + \frac{1}{a} \nabla (Q + V_\text{SI} + \Phi) = 0
\end{equation}
where the ``quantum potential'' $Q$ and
self-interaction potential $V_\text{SI}$ are given
by 
\begin{equation}
    Q = -\frac{\hbar^2}{2m^2 a^2} \frac{\nabla^2 \sqrt{\rho}}{\sqrt{\rho}}
\end{equation}
and
\begin{equation}
    V_\text{SI} = \frac{g \rho}{m^2}
\end{equation}
These may also be written as pressure terms:
\begin{equation}
    \nabla Q = \frac{1}{\rho} \nabla \cdot \bm{\Pi}
\end{equation}
where $\bm{\Pi}$ is the momentum flux density tensor, also known as the ``quantum pressure'' tensor, given by
\begin{equation}
    \Pi_{ij} = \Big(\frac{\hbar}{2m a}\Big)^2 \Big( \frac{1}{\rho}\frac{\partial \rho}{\partial x_i}\frac{\partial \rho}{\partial x_j} - \frac{\partial^2 \rho}{\partial x_i \partial x_j} \Big)
    \label{eq:QPT}
\end{equation}
and
\begin{equation}
    \nabla V_\text{SI} = \frac{1}{\rho} \nabla P_\text{SI}
    \label{eq:SIpressure}
\end{equation}
where $P_\text{SI}$ is the self-interaction (SI) pressure, given by
\begin{equation}\label{eq:SIpressure2}
    P_\text{SI} = \frac{g\rho^2}{2m^2}
\end{equation}
In the matter-like phase when $\langle w \rangle \simeq 0$, the parameter $g/m^2$ is related to the radius $R_\text{TF}$ of the ($n=1$)-polytrope which results when gravity balances SI pressure in hydrostatic equilibrium, sometimes referred to as the Thomas-Fermi (TF) radius,
according to
\begin{equation}
    R_\text{TF} = \pi\sqrt{\frac{g}{4 \pi G m^2}} \simeq 1.1 \text{ kpc} \left( \frac{g/(mc^2)^2}{2\times 10^{-18} \text{ eV}^{-1}\text{cm}^3} \right)^{1/2}
    \label{eq:RTF}
\end{equation}

\subsection{The TF regime and the limit of small de~Broglie wavelength}
\label{sec:smalldeB}

As discussed in \citetalias{paper1}, the same
QHD equations can also by derived by
a reformulation of the NLSE in phase space,
involving momentum moments of a quantum phase space distribution function.  In one version
of that approach, involving a smoothed phase space
distribution, we showed that when the
smoothing scale is much larger than the de~Broglie
wavelength $\lambda_\text{deB}$, the NLSE is
equivalent to the Collisionless Boltzmann 
Equation (``CBE''), except with the gravitational
potential term replaced by the sum of the
gravitational and SI potentials.  This 
reduction of the NLSE to the CBE in the
small-$\lambda_\text{deB}$ limit, for the free-field case \textit{without} the modification to account for SI introduced in \citetalias{paper1},
was previously discussed by \citet{Mocz18}, in regards to the FDM model,
following earlier work by \citet{WK93} and \cite*{SKV89}.  Since the
description of standard CDM as a gas of point-particles which are nonrelativistic, 
noninteracting (except gravitationally), and
collisionless (i.e. the 2-body gravitational relaxation time is much longer than the
lifetime of the system) is consistent with
the CBE, this explains why FDM behaves just like CDM, dynamically,
on scales much
larger than $\lambda_\text{deB}$,
i.e. when viewed in this coarse-grained way that does not
resolve substructure on scales smaller than
$\lambda_\text{deB}$.  
For example, numerical simulations of 
the NLSE in the FDM limit (no SI) show that,
when haloes form gravitationally and virialize, 
they have a core-envelope structure in which
a solitonic core of size $\sim\lambda_\text{deB}$
(with a centrally-flattened density profile 
like the ground-state,
static equilibrium solution of the 1D, spherical
NLSE) is surrounded by an envelope
with a profile like that which would have 
occurred in standard CDM.  With strong enough SI
to be in the TF regime, however, we showed
in \citetalias{paper1} that 
the solitonic core would be replaced
by a core,  much larger than 
$\lambda_\text{deB}$ (the latter is arbitrarily small
in this TF regime), 
supported instead by SI pressure, while
outside this SI-supported core, 
a CDM-like envelope would still be expected. 
However, to solve the latter problem in this TF regime, 
by the same kind of 3D simulations as were done for FDM, 
is much more difficult, since it would 
require fully resolving length scales well-below
$\lambda_\text{deB}$ in order to produce the correct CDM-like behavior
outside the core. In the TF regime, $\lambda_\text{deB}$ is
much smaller than the size of the SI-smoothed halo
cores, which are, themselves, much smaller than the 
haloes, and even smaller relative to 
the large-scale structure from which they formed, 
so the required resolution and dynamic range becomes prohibitively expensive.
We solved
this problem by finding an approximate fluid
description for SFDM-TF, as described below.

\subsection{The fluid approximation for SFDM-TF}
\label{subsection:fluidapprox}

In \citetalias{paper1}, we were tasked with calculating the dynamical formation of virialized haloes in SFDM-TF by evolving the scalar field
from a nearly-uniform, linearly-perturbed, initial condition through many orders of magnitude of gravitational collapse, all the while
keeping track of the dominant dynamical effects involving quantum pressure,
which result from inhomogeneities on the scale of the unresolvably-small $\lambda_\text{deB}$.
The fact discussed there (noted above
in \S\ref{sec:smalldeB}), that, in
the small-$\lambda_\text{deB}$ limit, the NLSE
for SFDM-TF can be shown to reduce to
the same CBE as for CDM, except 
modified by adding the SI potential to
that for gravity, suggests that,
in principle, an SI-modified N-body method 
might be developed to solve this problem,
if we were willing to forego knowledge
of the detailed substructure on the scale of
$\lambda_\text{deB}$, while still 
accounting for the effects
of quantum pressure on larger scales in
a coarse-grained way.
However, such an N-body method would still require
an extremely large dynamic range 
to simulate this in 3D, in order to
resolve the small-scale suppression effects
of SI pressure on scales below $R_\text{TF}$, even in this limit in which 
the (much smaller) de~Broglie scale is 
left unresolved -- a high computational barrier.
Instead, we avoided the complicated challenge
of such a grand computational solution
by closing the infinite hierarchy of the 
momentum moments of the CBE by following the approach developed for standard CDM dynamics by \citet{AS05}.

\citet{AS05} reduced the CBE equation for
CDM to a set of equations 
identical to the hydrodynamical 
equations of conservation of mass, momentum, and
energy for an ideal gas with ratio of specific
heats $\gamma = 5/3$, in 1D, spherical symmetry, by assuming the velocity distribution is
skewless and isotropic in the frame of bulk motion. As discussed there, these are assumptions that 
approximate the results of CDM N-body simulations 
reasonably well, for the virialized regions
identified as haloes.\footnote{Outside of
virialized regions, infall is highly supersonic; 
the velocity of bulk peculiar motions 
greatly exceeds that associated with 
the velocity dispersion due to random motions 
in that frame. Hence, these assumptions make
no appreciable difference in the dynamics
outside virialized regions.} 
In the case of SFDM-TF, however,
as we showed in \citetalias{paper1}, we must 
replace the gravitational potential
in the CDM case of \citet{AS05}
by the sum of the gravitational and SI potentials.
As a result, a new kind of 
pressure force appears in the momentum equation,
in addition to the ideal gas pressure, namely 
the SI pressure force, involving the gradient of
the SI pressure given above in equations~(\ref{eq:SIpressure}) and (\ref{eq:SIpressure2}).  

The resulting spherically-symmetric fluid equations for SFDM-TF are presented below in proper coordinates, in Eulerian form:
\begin{align}
        \text{[continuity]}& \quad \quad \frac{\partial \rho}{\partial t} + \frac{1}{r^2}\frac{\partial (r^2 \rho v)}{\partial r} = 0 \label{eq:0momSpherical}\\
        \text{[momentum]}& \quad \quad \frac{\partial v}{\partial t} + v \frac{\partial v}{\partial r} + \frac{1}{\rho}\frac{\partial}{\partial r}(P_{\!\sigma}+P_\text{SI}) + \frac{\partial \Phi}{\partial r} = 0 \label{eq:1momSpherical} \\
        \text{[energy]}& \quad \quad \frac{\partial}{\partial t}\Big(\frac{3P_{\!\sigma}}{2\rho}\Big) + v \frac{\partial}{\partial r}\Big(\frac{3P_{\!\sigma}}{2\rho}\Big) + \frac{P_{\!\sigma}}{\rho r^2}\frac{\partial (r^2 v)}{\partial r} = 0
        \label{eq:2momSpherical}
\end{align}
where $P_{\!\sigma} = \rho \sigma^2$ is an isotropic pressure associated with the velocity dispersion $\sigma$ obtained from velocity moments of the phase space distribution function. This
pressure, which behaves like the pressure
of an ideal gas with ratio of specific heats $\gamma = 5/3$ and, hence, contributes 
a pressure-gradient force to the 
momentum equation, while changing
its specific internal energy $u \equiv (P_{\!\sigma}/\rho)/(\gamma -1)$ to account for
$P_{\!\sigma} dV$-work, thereby accounts for the large-scale effects of the quantum pressure tensor ($\bm{\Pi}$), as 
discussed in \citetalias{paper1}.
If we define the mass interior to radius $r$ as $M_r$, according to
\begin{equation}
    M_r \equiv M(\leq r) = \int_0^r 4\pi (r')^2 \rho~dr'
    \label{eq:interiormass}
\end{equation}
we can rewrite equations~(\ref{eq:0momSpherical}) - (\ref{eq:2momSpherical}) in Lagrangian form, 
though still in proper coordinates, as follows:
\begin{align}
        \text{[continuity]}& \quad \quad \frac{D \rho}{D t} = -\frac{\rho}{r^2}\frac{\partial (r^2 v)}{\partial r}  \label{eq:0momLagrangian}\\
        \text{[momentum]}& \quad \quad \frac{D v}{D t} = -4\pi r^2 \frac{\partial}{\partial M_r}(P_{\!\sigma}+P_\text{SI}) - \frac{G M_r}{r^2} \label{eq:1momLagrangian} \\
        \text{[energy]}& \quad \quad \frac{D}{D t}\Big(\frac{3P_{\!\sigma}}{2\rho}\Big) = -P_{\!\sigma}\frac{D (1/\rho)}{D t} \label{eq:2momLagrangian}
\end{align}
where the Lagrangian, or total, time derivative is given by
\begin{equation}
    \frac{D}{Dt} \equiv \frac{\partial}{\partial t} + v \frac{\partial}{\partial r}
\end{equation}
In \citetalias{paper1}, we also stated the
adiabatic shock jump conditions that 
accompany these differential conservation
equations when there are discontinuities
present.  The reader is referred to \citetalias{paper1} for those.  

\section{Halo formation by cosmological infall: spherical collapse and virialization}
\label{sec:infall}

\subsection{The non-cosmological case: nonlinear Jeans instability from a static initial condition}
\label{sec:staticICPaperI}

In \citetalias{paper1}, we solved the equations above to follow the spherical collapse of a Jeans-unstable linear perturbation in a static background. We simulated the nonlinear
evolution of a region containing a mass much larger than the Jeans mass, initially at rest, 
from a uniform background density, perturbed
by a spherically-symmetric perturbation that is slightly concentrated toward the center.
As it was gravitationally unstable,
the mass collapsed, first at the center, 
until the rising SI pressure inside the central region of radius $\simeq R_\text{TF}$ halted the gravitational collapse of the innermost
mass shells, and a strong accretion shock 
formed there to decelerate other 
incoming shells.  As more
mass collapsed and would have, if entirely pressure-free, 
reached the origin in a finite time, with
shells originating at successively larger radii
predicted to reach the
origin at successively later times, each
shell, instead, encountered the shock at
a finite radius and was halted by it.
Outside this accretion shock, 
the infall was highly supersonic and the
ideal gas ``temperature'' (i.e. the velocity dispersion in momentum space, as 
measured in the frame of bulk radial motion) was
very small, so the ideal gas pressure associated with it was too small to halt the collapse. The
SI pressure was also low there, so it, too, was
unable to decelerate the infall.  When
a mass shell crossed the accretion shock, however,
its ``temperature'' and ideal gas pressure jumped 
by orders of magnitude, 
while its density jumped only by a factor
of four (i.e. the value for a strong, adiabatic
shock jump in an ideal gas with a ratio of specific heats $\gamma = 5/3$).  As a result, 
the post-shock gas was dominated by ideal gas pressure, high enough to bring the mass shell to
an approximate rest in hydrostatic 
equilibrium (``HSE'').
In this post-shock HSE, ideal gas pressure balanced gravity everywhere inside the shock, except inside $\sim R_\text{TF}$, 
where it was SI pressure that provided the support, instead,
and the ideal gas pressure was unimportant. 
Over time, more mass shells reached the
accretion shock, which gradually moved outward,
encompassing an ever-increasing mass.
As new mass shells passed through the shock and
steadily added their mass to that of the previously-shocked shells, they extended the
radius of the post-shock region in hydrostatic equilibrium as the shock
moved outward. 
The
structure interior to the post-shock region
corresponds to the ``virialized halo'' in
cosmological simulations.   

In \citetalias{paper1},
we compared the outcomes of
two simulations from the same initial
conditions, one that included the SI
pressure for SFDM-TF and one that did not, 
so the latter was entirely ``CDM-like'', 
therefore. Since the initial conditions for
both of these simulations was that for a
Jeans-unstable density perturbation on a
static homogeneous background, the
collapse that ensued left a post-shock
density profile in the CDM-like case
(i.e. with no SI) with a radial dependence
$\rho\propto r^{-12/7}$.  As we explained there,
this slope results when cold, infalling
gas with a density profile with this
slope -- the same as reported in earlier work on infall from instability in a spherical, Jeans-unstable gas cloud
\citep [e.g.][]{Penston69} -- encounters the strong accretion shock. The shock jump converts the kinetic energy of this cold, supersonic infall into
ideal-gas internal energy, with enough post-shock gas pressure to establish hydrostatic equilibrium.
According to the strong, adiabatic shock jump conditions, the density must also increase,
jumping by a constant factor of 4 
(for $\gamma = 5/3$).  As a result, the static
post-shock density profile inside the shock 
must have the same power-law profile as the pre-shock infall, outside the shock, namely, $\rho\propto r^{-12/7}$.
The SFDM-TF solution was 
shown to be identical to
this CDM solution everywhere outside of the
SI-pressure-supported core of radius equal 
to $\sim R_\text{TF}$.  Inside this radius, however,
the density profile was flattened, following
that of the $(n=1$)-polytropic profile that would result
if SI pressure alone supported the scalar
field against its own self-gravity. Its 
central density was set by the condition that
it conserved the mass that would have been
there inside that radius if the CDM-like
profile had continued onward to the origin,
uninterrupted by SI pressure.  

In short, the
generic structure of the haloes that formed
as the nonlinear outcome of Jeans instability
in SFDM-TF, therefore, could be described
simply as the core-envelope structure in which
a flattened, SI-polytropic core of 
radius $\sim R_\text{TF}$
was surrounded by the same CDM-like envelope
it would have had if there were no SI pressure,
and the mass inside the polytropic core radius was
the same as it would have been in that CDM-like
profile in the absence of SI.  As such, this
also served to compare the SFDM-TF profile with
the corresponding FDM (i.e. ``free-field'' limit) profile that would have
resulted from the same initial conditions if $\lambda_\text{deB}$ is small, since the
latter was shown in \citetalias{paper1} to be
identical to the CDM profile, when smoothed 
on scales large compared with $\lambda_\text{deB}$.

We note that, although the description above
applies to the entire post-shock region at
a given time, we can make our description
apply, as well, to cosmological haloes,
which are customarily identified as the
``virialized regions'' bounded by 
a ``virial radius'' inside of which
the average mass density is equal 
to some fixed constant multiple of the 
unperturbed background density
of the universe at that cosmic time.  In the Jeans-unstable collapse
calculations in \citetalias{paper1} from static
initial conditions, there is not a cosmological
background density to relate to the overdensity
inside some radius within
the post-shock region, but we are still free
to adopt an overdensity like the conventional
definition, relative to the unperturbed
background density, in order to determine
the radius (within the post-shock sphere) which we can 
call the ``virial radius''.  

\subsection{The cosmological case: spherical infall in the expanding universe}
\label{sec:cosmologicalcase}

The description above is a guide to
what we can expect below, when we re-consider the
nonlinear Jeans instability in SFDM-TF here,
to place it in the proper cosmological context of
the expanding universe in the MD era, i.e. after matter-radiation equality, $a > a_\text{eq}$.  To do this, we will 
solve the same equations, as written
in proper coordinates, but now with cosmological initial and boundary conditions,\footnote{
Although our dynamical equations (\ref{eq:0momSpherical}) - (\ref{eq:2momSpherical}) 
do not appear to be cosmological, for a spherically-symmetric distribution of 
matter in the nonrelativistic and weak gravitational field limits, 
Birkhoff's Theorem implies that Newtonian gravity is sufficient for 
describing cosmological dynamics. Therefore, our cosmology is taken 
care of by the Newtonian potential and the inclusion of 
Hubble flow (and its perturbation)
in the initial velocity.}
to follow the spherical collapse of a linear perturbation in the expanding universe.
This is fully cosmological, for structure
formation on scales that are well within the
horizon when we evolve them.  
As we shall see, the novel effect of cosmological
expansion is to allow mass shells that were
initially inside the characteristic radius
of the effects of SI pressure to expand outward
for a time, carrying them beyond the reach of this repulsive force, and then to fall back again, 
to re-encounter it after turn-around and collapse.
An imprint of this initial encounter remains, 
however, and has important consequences for their
final resting place.  

\subsubsection{CDM-like initial conditions: the infall rate that makes an NFW profile} \label{sec:NFW}

Our goal in this paper (``Paper II'')
is to generalize the non-cosmological results in \citetalias{paper1} to investigate the cosmological formation of SFDM-TF haloes.
Our first approach in this direction follows
the tradition of previous study of structure formation in the free-field limit of the
SFDM model (FDM),
in which the first cosmological
simulations assumed scalar field dynamics
but initial conditions like those for the CDM model. In particular, we will simulate the formation of individual haloes from the same initial conditions for both CDM and SFDM-TF, in order to learn what difference results 
from scalar field dynamics with repulsive SI, 
as opposed to the collisionless dynamics of CDM. 

It is useful to begin by summarizing what is
known about the CDM case and why it is that
we have confidence in the 1D, spherical 
collapse calculations we propose to use to
study the contrast between CDM and SFDM-TF
halo dynamics.
In \cite*{AAS03}, \citet{Shapiro04}, \citet{AS05} and \citet{Shapiro06}, we showed that the 1D, spherical infall model involving the same fluid 
approximation as used here and in \citetalias{paper1}, derived by \citet{AS05}  
from the 
collisionless Boltzmann equation, 
can explain
the universal structure of haloes in N-body simulations of nonlinear structure formation from 
Gaussian-random-noise density fluctuations in
the CDM universe. As a first step, 
\citet{AS05} applied
this fluid approximation for CDM to derive
an analytical similarity solution for halo formation from spherical infall. As described below, this
produced a density profile inside the accretion shock in this solution which was a
surprisingly good match to various 
empirical formulae developed to fit density profiles of CDM haloes in N-body simulations, e.g. the NFW profile \cite*{NFW97}. 
However, while the agreement
was impressive at a given snapshot of time,
the N-body results showed that, 
if captured at different times in the mass assembly history (``MAH'') of the same halo,
the shape of its profile would evolve,
while that of a self-similar profile does not.
As \citet{AS05} pointed out, this reflected
the fact that infall rates
of individual haloes derived by fitting the Lagrangian MAH of haloes in CDM N-body simulations 
\textit{vary} over the lifetime of each halo --
rapid early, then slowing down late -- 
while the rate in their self-similar
solution ($d \ln M/d \ln a = 6$) is fixed, corresponding to the early, rapid-infall phase of the N-body results.  
To better match the \textit{non-self-similar} shape evolution of the N-body haloes, in \citet{AAS03} we replaced the self-similar initial perturbation
in \citet{AS05} by a \textit{non-self-similar} spherically-symmetric perturbation profile calculated to make a time-varying
infall rate that matched the MAH of CDM
N-body haloes.  Solving the 1D fluid
approximation equations for CDM in this case
required a numerical hydrodynamics
code.  The haloes that resulted were shown
to be in excellent agreement with 
the universal average behavior of the haloes in N-body simulations.  Since this earlier work,
also summarized in \citet{Shapiro04} and
updated in \citet{Shapiro06}, is central to
our adoption of CDM-like initial conditions here
to demonstrate the difference between SFDM and
CDM dynamics on cosmological halo formation,
we describe it in more detail below.

\subsubsection{CDM haloes from self-similar spherical infall: $\epsilon=1/6$}
\label{sec:selfsimilar}

In an Einstein-de Sitter universe, a scale-free,
spherically-symmetric linear perturbation 
$\delta M/M \propto M^{-\epsilon}$ leads to
self-similar halo formation around the density
peak at the center. According to the 
theory of halo formation
from peaks in the density field which result from Gaussian-random-noise initial density fluctuations,
the average density profile around such peaks can be related to the shape of the fluctuation power spectrum \citep{HS85,Bardeen86}.  These density peaks in the linear regime are the progenitors of nonlinear structures like haloes.  It is useful to
describe such peaks in terms of the RMS mass fluctuation, $\sigma_\textsc{m}$, of the overdensity field smoothed on some mass scale $M$. The average linear overdensity profiles around peaks of 
overdensity $\nu\sigma_\textsc{m}$ are triaxial, in general, but approach spherical symmetry as $\nu$ increases  (e.g. for $\nu \geq 3$), with cumulative overdensity $\Delta_0$ inside radius $r$, given by $\Delta_0 \propto r^{-(n+3)}$, where $n$ is
the effective index of the (transferred) 
power spectrum $P(k)$ if approximated as a power-law $P(k)\propto k^n$ at the wavenumber $k$ that corresponds to that halo mass.
In terms of fractional mass perturbation, 
$\Delta_0 = \delta M/M \propto  M^{-(n+3)/3}$, 
so a power-law $P(k)\propto k^n$ leads 
naturally to a scale-free initial condition with $\epsilon = (n+3)/3$.  While the actual transferred
$P(k)$ for CDM is not a power-law, it is
possible to use the actual shape to derive an effective power-law index which varies with halo mass $M$ as a good approximation.  \citet{AS05} calculated this $n_\text{eff}$ for the current $\Lambda$CDM universe, for the full range of halo masses from $10^3$ to $10^{15} \text{ M}_\odot$, showing that galactic haloes are well-approximated by $n = -2.5$, since $n\cong-2.5\pm0.2$ for $M$ in the range from
$10^3 \text{ M}_\odot$ to $10^{11} \text{ M}_\odot$.
For haloes of galactic mass, formation by
spherical infall from a self-similar initial perturbation with $\epsilon=1/6$ is therefore a reasonable
first approximation (for the
MD era, for which the universe
is well-approximated as Einstein-de~Sitter). 

The resulting self-similar density profile agreed with that for the NFW profile to within
$10\%$, at radii $r \lesssim 0.6 R_{200}$, for a best-fitting concentration parameter of 3, including a density cusp at small radii with a logarithmic
slope $\approx{-1.27}$ \big(for $4 \times 10^{-3} < r/R_{200} < 1.4 \times 10^{-2}$\big), where $R_{200}$
is the radius within which the average density is
200 times the cosmic mean density.   Furthermore,
as we showed in \citet{Shapiro06}, the analytical similarity solution yields a profile for the quantity $\rho/\sigma_V^3$, sometimes referred to as the ``phase-space density'', which is 
best-fit by a power-law $\rho/\sigma_V^3\propto r^{-1.91}$, in close agreement at all times with
the universal profile reported in the literature
for CDM N-body simulation results cited there. 
However, the
radius of the accretion shock that bounds the virialized region in this solution is somewhat smaller than $R_{200}$, located instead at the
radius which encloses an average overdensity of
564 times this mean density, at $R_{564} \simeq 0.6 R_{200}$.  This suggests that the rapid early phase of MAH in the N-body results, which deposits the
mass of the density profile at inner radii
that remains there at later times, is well-represented by this self-similar infall solution, but the departure from self-similarity,
especially as related to the late
tapering-off of the mass accretion is not.  

\subsubsection{CDM haloes from non-self-similar spherical infall}
\label{sec:nonselfsimilar}

If the infall rate was given, instead, by 
the spherically-averaged, Lagrangian MAH 
of individual 
haloes in CDM N-body simulations, however, then
the resulting haloes shared most of the
empirical CDM N-body halo structural
properties \textit{and} their time-evolution.
In particular, in \citet{AAS03}, 
we took as our initial
condition, a spherical density perturbation
whose radial dependence ensured that
the rate at which infalling mass reached the virial radius of the emerging halo followed
the empirical MAH derived by \cite{Wechsler02}
from N-body simulations of CDM.  With this
simple assumption, we found that the virialized
haloes which result have density profiles that
are well-fit by the NFW density profile
(at all radii $r/R_\text{vir}\geq0.01$), and 
the radius of the accretion shock is safely
outside $R_{200}$. These
halo density profiles also \textit{evolve} like
CDM N-body haloes, with concentration
parameter that grows with time just as it
does for CDM N-body haloes.  In addition, 
as shown in \citet{Shapiro06},
the resulting halo phase-space density profile, $\rho/\sigma_V^3$, best-fit by a power-law
$\propto r^{-1.93}$, was found to be in
remarkable agreement at all times with 
the universal profile reported for CDM
N-body haloes.  We concluded, therefore,
that the time-varying mass accretion rate,
or equivalently, the initial perturbation
profile shape around the density peak that
led to the formation of the halo, is the
dominant influence on the internal structure
of CDM haloes, which can be well-understood
from this simple spherical infall model involving
smooth accretion and 
the fluid approximation derived by \citet{AS05}
from the CBE.

\subsubsection{SFDM vs. CDM haloes from non-self-similar spherical infall}
\label{sec:SFDMnonselfsimilar}

In \citetalias{paper1}, we showed that
the fluid approximation used above to model CDM halo
dynamics also models SFDM dynamics
in the free-field limit of FDM, as long as we
do not need to resolve substructure
on scales as small as the de~Broglie 
wavelength but still wish to account faithfully
for the influence of quantum pressure on
larger scales.
This, then, is a useful tool for the study
of halo formation in the FDM regime, 
as long as we do not need to resolve the
solitonic core in the central region.  
The results described above, which
explain how the NFW profile
and its related universal properties
in the CDM model can arise as a simple consequence
of spherical infall of smoothly-accreted matter, 
also explain why 3D simulations
of FDM haloes report NFW-like density profiles 
\textit{outside} their solitonic cores. 

It remains
for us to determine here, what happens to this
result when a repulsive SI is added to 
SFDM.  In \citetalias{paper1},
we \textit{generalized} the
fluid approximation  of \citet{AS05}
to include the effects of SI pressure in the
TF regime, which corresponds to
the regime in which the de~Broglie wavelength
can be assumed to be arbitrarily small,
just right for application of the fluid
approximation derived from the CBE.
Our first approach to putting halo formation
in the SFDM-TF model in a cosmological context
will be to adopt the spherical infall
model described above for the CDM model and
solve the 1D, fluid conservation equations for 
SFDM-TF of \citetalias{paper1} with cosmological initial conditions to determine
what the dynamical effect of adding SI to the scalar
field is.   

We consider a spherically-symmetric, adiabatic
density perturbation as our initial condition.  We 
take our initial time to be when the 
scale factor $a_i = a_\text{eq}$, and further
assume that the universe remains matter-dominated, thereafter, at the critical density, corresponding
to an Einstein-de Sitter (EdS) cosmology. 
The initial linear density perturbation profile
can be described in terms of the mass perturbation $\Delta_\textsc{l}(M,a_i)$, where $M$ is
the mass inside a sphere of initial (perturbed) radius $r(M,a_i)$, related to the mass $M_b(M,a_i)$ that would have been inside a sphere of that radius in the unperturbed background universe, according to
\begin{equation}
    \Delta_\textsc{l}(M,a_i) \equiv \frac{M-M_b(M, a_i)}{M_b(M,a_i)}
    \label{eq:DeltaLMass}
\end{equation}
where $M_b(M,a_i) = (4\pi/3) r^3(M,a_i) \rho_b(a_i)$, and $\rho_b(a_i)$ is the background matter
density at the initial redshift, given by 
$\rho_b(a_i) = \rho_\text{crit}(a_i) \equiv 3H^2(a_i)/8\pi G$, where the Hubble parameter
evolves according to $H^2(a)=H^2_0a^{-3}$.
We adopt the value $H_0 = 67$ km/s/Mpc for the present-day Hubble parameter.
The initial radius of the Lagrangian
mass shell which encloses that mass $M$ is then given by
\begin{equation}
    r(M,a_i) = \Bigg(\frac{M}{\frac{4\pi}{3} \big[1+\Delta_\textsc{l}(M,a_i)\big] \rho_b(a_i)} \Bigg)^{1/3}
    \label{eq:initialradius}
\end{equation}
while its initial velocity corresponds to the 
perturbed Hubble flow for the growing mode of the 
perturbation with that linear overdensity, given by \citep*[see, e.g.,][\S5.1.1]{MVW10}
\begin{equation}
    v(M,a_i) = H(a_i) r(M,a_i)\big(1 -\Delta_\textsc{l}(M,a_i)/3\big)
\end{equation}

The quantity $\Delta_\textsc{l}(M,a_i)$ is also the
average fractional overdensity inside the sphere which encloses the mass $M$ at $a_i$.  This average overdensity is chosen to be a function of mass $M$
such that, for an initial linear perturbation with this profile in a CDM universe, 
the ensuing growth would lead to infall and
nonlinear collapse that reproduces the average MAH of 
the virialized haloes that form in CDM N-body simulations, as described above.
The density profiles of these N-body simulation haloes are well-fit by NFW profiles\footnote{Note that in \citetalias{paper1}, we denoted $A_\text{NFW}$ as $\delta_\text{NFW}$. } \citep{NFW97}:
\begin{align}
    &\rho = \frac{\rho_\text{crit} A_\text{NFW}}{(c_\textsc{nfw}r/R_{200})(1+c_\textsc{nfw}r/R_{200})^2} \\
    &A_\text{NFW} = \frac{200}{3}\frac{c_\textsc{nfw}^3}{\ln{(1+c_\textsc{nfw})} - c_\textsc{nfw}/(1+c_\textsc{nfw})}
\end{align}
where $R_{200}$ is the radius within which the average density is 200$\rho_\text{crit}$, and $c_\textsc{nfw}$ is the concentration parameter, defined by $c_\textsc{nfw}\equiv R_{200}/R_s$, where $R_s$ is the ``scale radius'' at which the logarithmic slope of the density profile equals $-2$. \cite{Wechsler02} found that the MAH of CDM haloes in N-body simulations followed a simple rule -- the logarithmic slope of the mass accretion is inversely proportional to the scale factor:
\begin{align}
    \frac{d\ln{M_{200}}}{d\ln{a}} &\propto \frac{1}{a} \\
    \Rightarrow M_{200}(a) &= M_\infty\exp{(-s a_f/a)} \label{eq:MAH}
\end{align}
where $M_{200}$ is the mass within $R_{200}$, $M_\infty$ is the asymptotic halo mass as $a\rightarrow\infty$, and $a_f$ is the ``formation scale factor'' at which the logarithmic slope equals $s$. By convention, $s=2$. In \citet{AAS03} and \cite{Shapiro04}, we found an initial perturbation profile for an EdS universe that is consistent with this MAH,
given by
\begin{equation}
    \Delta_\textsc{l}(M,a_i) = \Delta_{\textsc{l},i} \ln{(M_\infty/M)}
    \label{eq:NFWPert}
\end{equation}
where $\Delta_{\textsc{l},i} \equiv \Delta_\textsc{l}\big( M_{200}(s a_f \!) , a_i \big)$ is the initial average linear overdensity within the shell whose ballistic motion brings it to a radius which corresponds to
$R_{200}$ at $a=s a_f$. 
Since $a_f$ corresponds roughly to
the ``formation time'' of the halo, we can linearly extrapolate this initial overdensity to $a_f$, when the (linearly-extrapolated) average overdensity should be roughly unity:
\begin{equation}
    \Delta_{\textsc{l},f} = \Delta_{\textsc{l},i} \frac{a_f}{a_i} \approx 1 \Longrightarrow \Delta_{\textsc{l},i} \approx \frac{a_i}{a_f}
    \label{eq:deli}
\end{equation}
The exact value we use in our code is $\Delta_{\textsc{l},i} = 0.8 a_i/a_f$, which we obtained by calculating the linearly-extrapolated overdensity of a spherical top-hat in an EdS universe that has collapsed to a nonlinear overdensity of 200,
at $a=s a_f$. We show our calculation in Appendix~\ref{sec:deli}.

As described above in \S\ref{sec:nonselfsimilar}, 
the result of adopting this
linear perturbation profile and simulating its
growth and collapse in a CDM universe, using the
fluid approximation of \citet{AS05}, is to
produce a virialized region inside the accretion
shock whose structure and evolution closely 
resembles that reported for CDM N-body haloes. 
For example, not only are the halo profiles that result from these initial conditions well-fit by an NFW density profile at
all times following the formation of that
accretion shock, but the concentration parameter 
of that profile increases over time as the halo grows, according to  
\begin{equation}
    c_\textsc{nfw}=4.25 a/a_f \quad \quad \text{for  } a\gtrsim a_f
    \label{eq:c200}
\end{equation}
\citep{Shapiro04,AAS03}. 
This is close to the relationship reported by \citet{Wechsler02} in their analysis of 3D N-body simulations: $c_\textsc{nfw}=4.1a/a_f$.

In practice, if we want to form an individual
halo with a given mass and concentration parameter
at a given scale factor, we combine  equations~(\ref{eq:MAH}) and (\ref{eq:c200}) and
solve them for the free parameters of this NFW-producing initial condition, ($M_\infty$, $a_f$), that will yield the desired halo.
For example, consider a $10^9 \text{ M}_\odot$ halo with a concentration parameter of 18 at the
present \citep[the median value for this mass as per the mass-concentration relation found by][]{Klypin16}. Equation~(\ref{eq:c200}) then gives $a_f \simeq 0.24$ for this halo, in which case equation~(\ref{eq:MAH}) gives $M_\infty \simeq 1.6 \times 10^9 \text{ M}_\odot$.
The results of a simulation run using these values in the initial condition will be discussed below,
in \S\ref{sec:CDMNFWsims}.

\subsubsection{Numerical Method}
\label{sec:numericalmethod}

All simulations of cosmological halo formation from spherical infall in this paper will use the 
same 1D, Lagrangian hydrodynamics
method and set-up as described in \citetalias{paper1}.   We adopt the same
non-self-similar initial conditions
described above in \S\ref{sec:SFDMnonselfsimilar}
and Appendix \ref{sec:deli}, for both CDM
and SFDM-TF, starting from $a = a_\text{eq}$.  
This is designed 
to allow us to compare the dynamical
consequences of SFDM-TF and CDM directly,
for evolution from the same initial perturbation.
Unless otherwise noted, each simulation uses
2000 Lagrangian shells, logarithmically-spaced
in mass.  

\subsection{Simulation results for cosmological halo formation from spherical infall and collapse}
\label{sec:simulationresults}

\subsubsection{CDM halo formation from NFW-producing infall rate}
\label{sec:CDMNFWsims}

\begin{figure}
    \centering
    \includegraphics[width=\columnwidth]{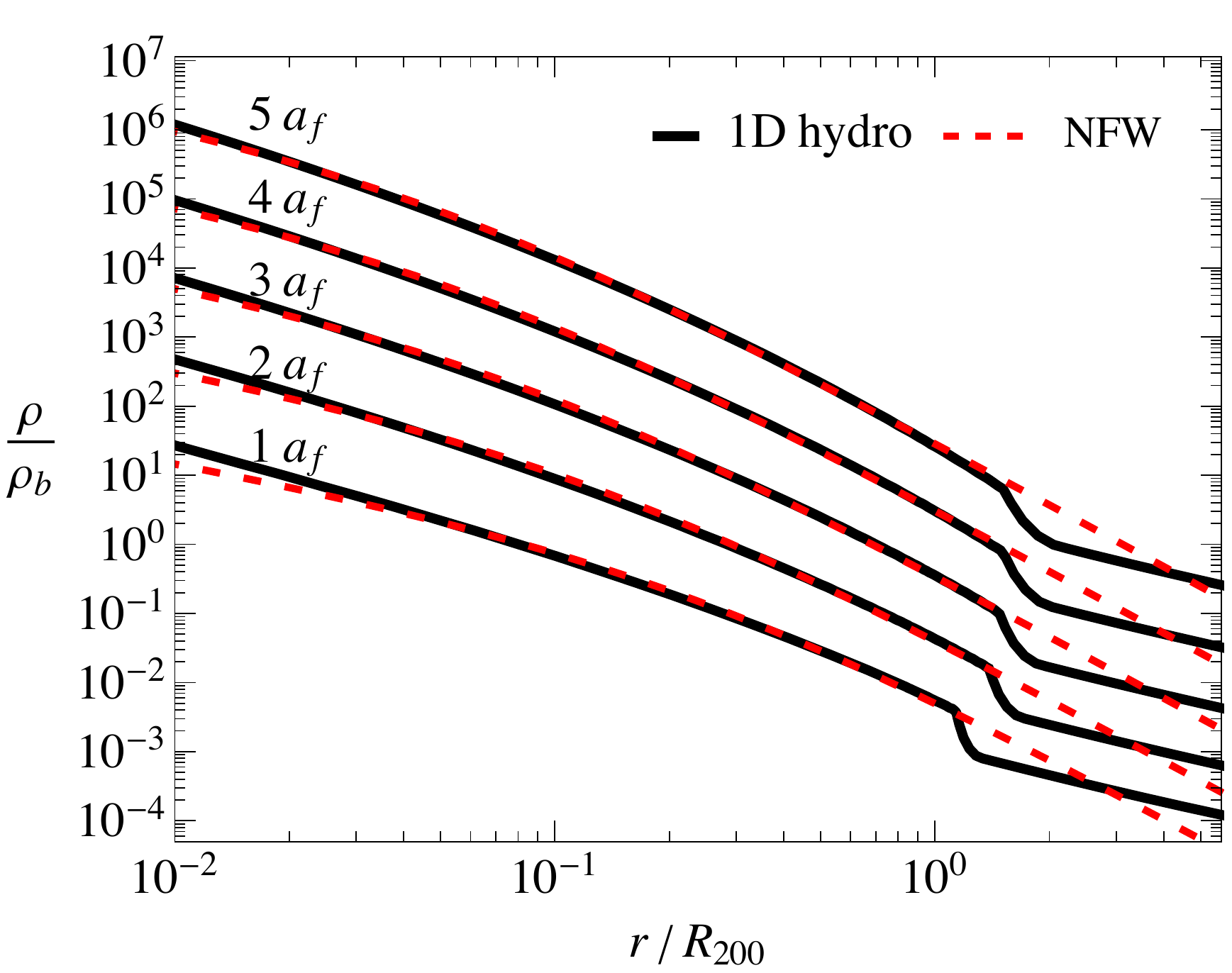}
    \caption{\textbf{The fluid approximation results for halo formation from spherical infall and collapse in CDM vs. the NFW fit to N-body simulations}. Resulting density profile from our 1D hydrodynamics code without repulsive self-interaction ($g=0$) starting from NFW-producing initial conditions. The red dashed line shows the corresponding NFW profile with the same value of $R_{200}$ as the numerical profile, but with $c_\textsc{nfw}$ chosen to fit the numerical profile in the domain $10^{-2} < r/R_{200} < 1$. The profiles are plotted at 5 scale factors, $a/a_f = \{1, 2, 3, 4, 5\}$, for which the best-fit NFW concentration parameters are $c_\textsc{nfw} = \{4.6, 8.5, 12.7, 17.0, 21.2\}$, respectively. The curves for $a/a_f = 1, 2, 3$, and 4 have been shifted downward by 4, 3, 2, and 1 order(s) of magnitude, respectively, for visual clarity. The radius $R_{200}$ and background density $\rho_b$, used to normalize the radius on the x-axis and density on the y-axis, are evaluated at each time-slice, separately.
    As expected, these results reproduce those presented previously in \citet{AAS03}, \citet{Shapiro04}, and \citet{Shapiro06}.  }
    \label{fig:NFW}
\end{figure}

In Fig.~\ref{fig:NFW}, we show results of our 1D
numerical hydrodynamics simulation of the formation of a CDM halo, based upon the fluid approximation of \citet{AS05} for CDM described above in \S\ref{subsection:fluidapprox} and \citetalias{paper1}.  We adopted the non-self-similar cosmological perturbation described in \S\ref{sec:SFDMnonselfsimilar} and Appendix~\ref{sec:deli} as our initial condition,
starting at $a = a_\text{eq}$.  Our purpose here is
to illustrate the fact that this fluid approximation
for CDM (\textit{without} the SI pressure term
which must be added to represent SFDM-TF) 
faithfully reproduces the universal halo properties
and their evolution over time of the NFW-profile
haloes reported for CDM N-body simulations 
when the initial perturbation is tuned to
produce a smooth accretion rate that matches
the MAH from \citet{Wechsler02}. 
The results in Fig.~\ref{fig:NFW} confirm what we
previously reported in \citet{Shapiro04}, \citet{AAS03}, and \citet{Shapiro06}, as they
should, since we are following the same 
procedure.  Our purpose in the \textit{next} 
section will be to show what \textit{difference} 
is made if these same CDM-like initial conditions
are used to simulate the formation of an \textit{SFDM-TF} halo, instead, when 
the important dynamical effects
of both quantum pressure and SI-pressure
are properly taken into account.

\subsubsection{SFDM-TF halo formation from  NFW-producing infall rate: (SFDM-TF dynamics) + (CDM-like initial conditions)}
\label{sec:TFNFWsims}

\begin{figure*}
    \centering
    \includegraphics[width=\columnwidth]{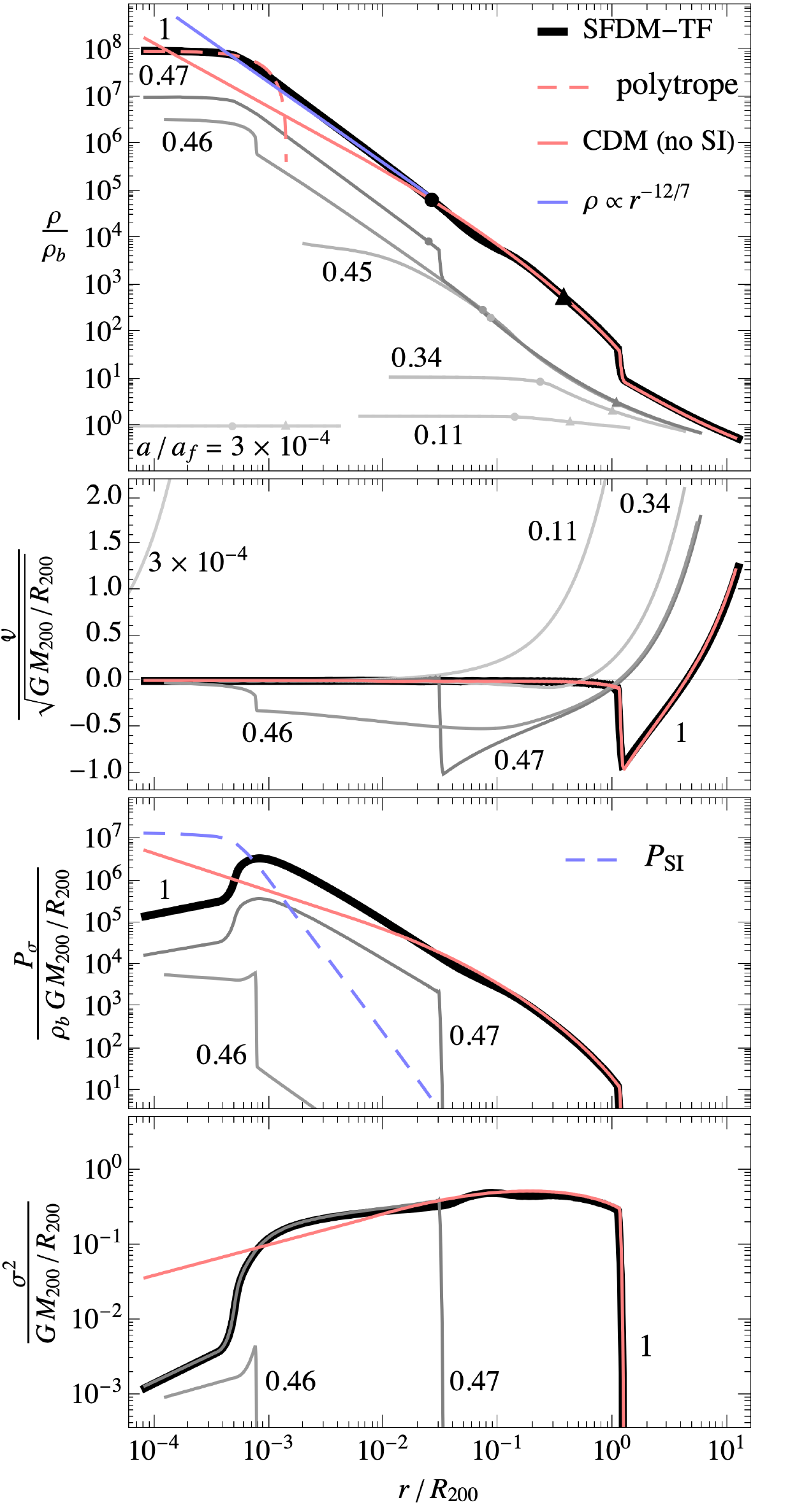}
    \includegraphics[width=\columnwidth]{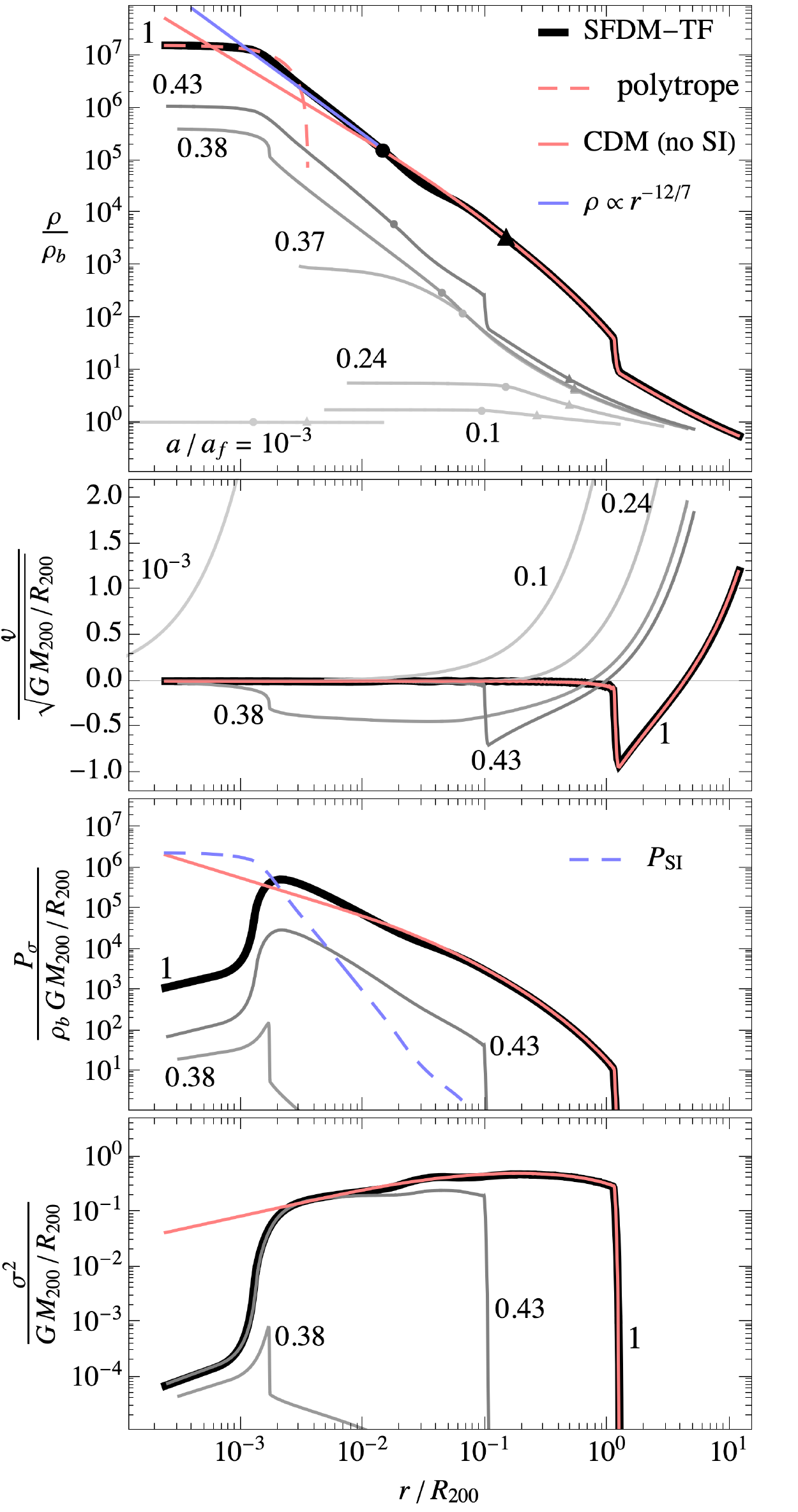}
    \caption{\textbf{The fluid approximation simulation results for SFDM-TF halo formation from spherical infall and collapse of a cosmological perturbation.} 
    Each panel shows profiles of hydrodynamical variables at select times: 
    from top to bottom, (non-dimensionalized) 
    mass density, bulk velocity, ideal gas pressure (i.e. velocity-dispersion ``pressure''), 
    and ``temperature'' (i.e. velocity dispersion) 
    vs. radius, for two example SFDM-TF haloes, from the same NFW-producing initial conditions at $a = a_\text{eq}$ as for the CDM simulations in Fig. \ref{fig:NFW}, with
    $\{M_\infty, a_f~(z_f), R_\text{TF}\} = \{5.7\times10^{14} \text{ M}_\odot, 0.87~(0.15), 1 \text{ kpc}\}$ (left) and $\{1.6\times10^{9} \text{ M}_\odot, 0.24~(3.2), 10 \text{ pc}\}$ (right). 
    The thick black lines show the profiles at the ``formation'' scale factor $a = a_f$. The thin grey lines show the profiles at earlier times in the simulation, labeled by $a/a_f$.
    Here, $\rho_b$ is the time-dependent background density, 
    while $\{M_{200}, R_{200}\} \simeq \{7\times10^{13} \text{ M}_\odot, 740 \text{ kpc}\}$ (left) and $\{2\times10^{8} \text{ M}_\odot, 3 \text{ kpc}\}$ (right) are the mass and radius of the halo at $a=a_f$. 
    The discontinuity near $R_{200}$ is the accretion shock, which sets the virial radius of the halo. 
    The thin red lines are profiles for a corresponding CDM run using the same initial conditions. The SFDM-TF density profile closely tracks the NFW-like CDM profile at large radii, but more closely resembles a power-law, $\rho \propto r^{-12/7}$, 
    at smaller radii (blue lines), until, at still
    smaller radii, $r \lesssim R_\text{TF}$, flattens to the shape of an SI-supported polytropic core (red long-dashed lines).
    For SFDM-TF at early times, the centrally-peaked initial density profile was flattened, out to $\sim R_\text{TF}$, by an outward-propagating, SI-pressure-driven sound wave. Circles mark the shell that was located at $r=R_\text{TF}$ at the time the sound wave reached that radius, while triangles mark the shell that was located at $r=R_\text{TF}$ at the initial time $a_i$ -- these are indicators of where the profile deviates from CDM.
    The long-dashed blue line in the $P_{\!\sigma}$ panel shows the profile for $P_\text{SI} = g\rho^2/2m^2$ at $a=a_f$ with the same normalization.}
    \label{fig:NFWSI}
\end{figure*}

\begin{figure}
    \centering
    \includegraphics[width=\columnwidth]{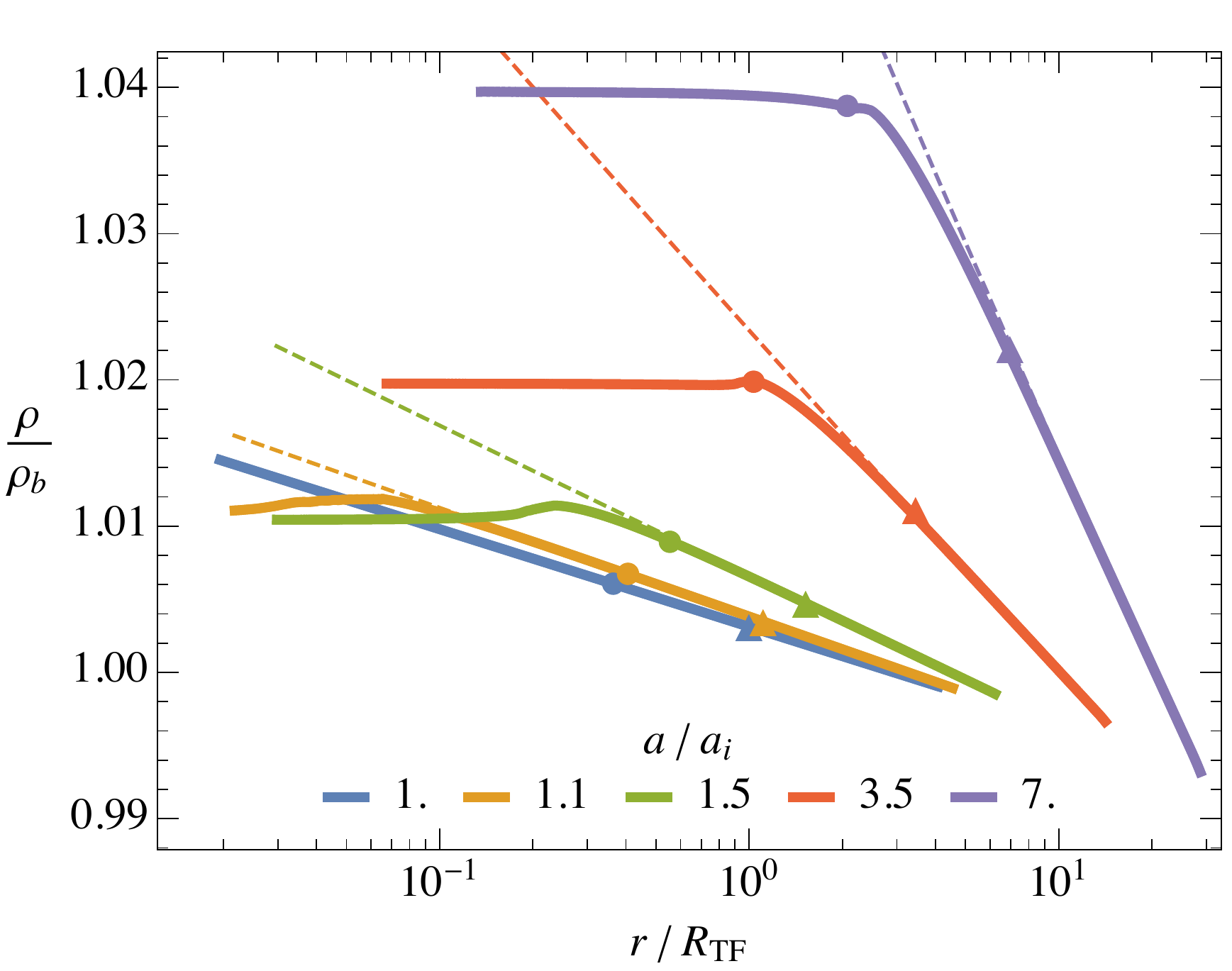}
    \includegraphics[width=\columnwidth]{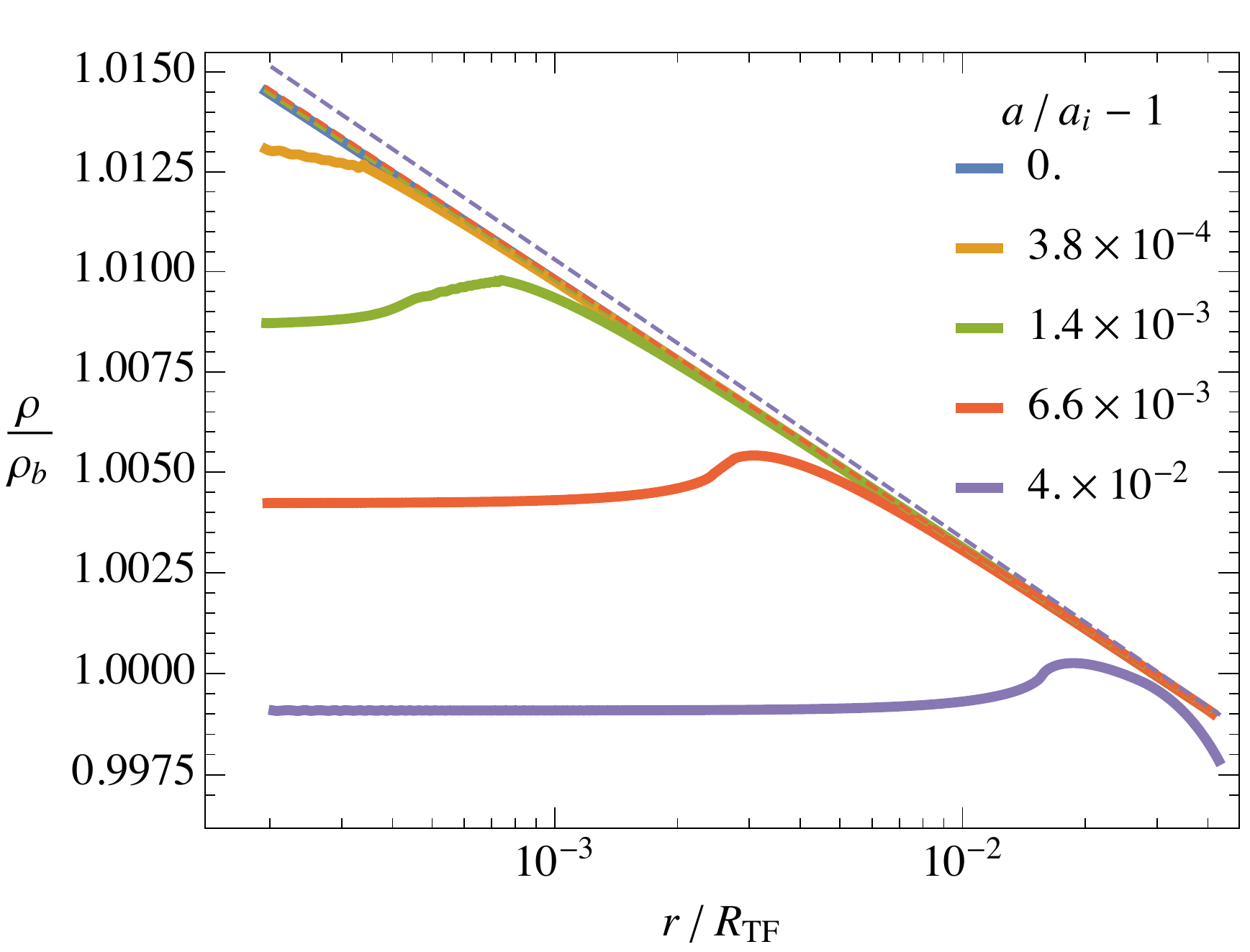}
    \caption{\textbf{The beginning of spherical collapse from NFW-producing initial conditions: early Jeans-filtering by SI-pressure}. Simulation results at several early time-slices, soon after the initial scale factor $a_i$, for two cases with the same
    initial perturbation, with $M_\infty=1.6\times10^{9} \text{ M}_\odot$ and $a_f = 0.24$, but with
    different SI strengths for SFDM-TF: \textbf{(a) (top panel)} $R_\text{TF} = 10$ pc,  and \textbf{(b) (bottom panel)} $R_\text{TF} = 1$ kpc. Corresponding CDM runs are shown for comparison (dashed lines). The density profiles are normalized by the time-dependent background density. The figures illustrate the initial SI sound wave that propagates outward and flattens the density profile out to $R_\text{TF}$. In the case of $R_\text{TF} = 1$ kpc, this sound wave passes over the entire profile (which encompasses the entire computational domain, extending to $M_\infty$), and the perturbation is prevented from growing (i.e. it is Jeans-filtered) within the age of the universe. Circles mark the shell that was located at $r=R_\text{TF}$ at the time the sound wave reached that radius, while triangles mark the shell that was located at $r=R_\text{TF}$ at the initial time $a_i$.}
    \label{fig:SIsmoothing}
\end{figure}

In Figs.~\ref{fig:NFWSI} and \ref{fig:SIsmoothing}, we show illustrative
results of our 1D numerical hydrodynamics simulations of SFDM-TF halo formation from 
the same initial conditions as used in the
previous section for the CDM halo formation
shown in Fig.~\ref{fig:NFW}.
In this case, however, our fluid equations are now
the modified ones for SFDM-TF, which
include the dynamical
effects of both SI pressure $P_\text{SI}$ and
quantum pressure, where the latter
is accounted for by the ideal gas pressure $P_{\!\sigma}$ 
(i.e. CDM-like ``velocity-dispersion'' 
pressure), which represents the dynamical 
effect on large scales that results from inhomogeneity on the tiny, unresolved scale of the 
de~Broglie wavelength. Two cases are shown
in Fig.~\ref{fig:NFWSI}, involving different
SI strengths, parameterized by their 
adopted $R_\text{TF}$-values, with initial perturbations chosen to make haloes of different mass at $a=a_f$ in each
case.  Profiles of hydrodynamical variables vs. proper
radius are plotted at selected times, including 
mass density, bulk velocity, ideal gas pressure 
(i.e. velocity-dispersion ``pressure''), 
and ``temperature'' (i.e. velocity-dispersion), 
for two example SFDM-TF haloes, formed
from the same NFW-producing initial conditions 
at $a = a_\text{eq}$ as for the 
CDM simulations in Fig. \ref{fig:NFW}, with
$\{M_\infty, a_f~(z_f), R_\text{TF}\} = \{5.7\times10^{14} \text{ M}_\odot, 0.87~(0.15), 1 \text{ kpc}\}$ (left) 
and $\{1.6\times10^{9} \text{ M}_\odot, 0.24~(3.2), 10 \text{ pc}\}$ (right).  For comparison, we also plot
the end results for CDM, from the same initial conditions. 

The evolution is similar in both cases, with some elements common to the non-cosmological infall and
collapse we simulated in \citetalias{paper1}, 
from static, Jeans-unstable initial conditions, while
others are uniquely cosmological, as follows:

\begin{enumerate}[labelwidth=2em, leftmargin=2em, listparindent=\parindent, label=(\arabic*)]
    \item Growth of the initial perturbation slows its
    expansion relative to the unperturbed Hubble flow
    far from the center, so the perturbed density 
    in units of the time-dependent mean background density increases with time, even as the latter drops.

    \item Close to the center, a sound wave driven by
    the SI pressure gradient quickly 
    propagates outward until it reaches the physical
    radius $R_\text{TF}$, one sound-crossing-time after
    the initial time at $a_i$, travelling 
    at the SI sound speed given by 
    \begin{equation}
        c^2_{s,\text{SI}} = \frac{\partial P_\text{SI}}{\partial \rho} = \frac{g\rho}{m^2}
        \label{eq:SI_soundspeed}
    \end {equation}
    evaluated at a density roughly equal to the initial
    background density. This sound wave
    smooths out the density as it goes,
    leaving it nearly uniform within this radius,
    and remaining uniform as it expands, i.e. it
    expands homologously. We will say more about this sound-wave-smoothed central region below.
    
    \item Thereafter, all mass shells continue expanding until
    -- at later times for shells that started at
    larger radii -- they reach maximum expansion, halt 
    and collapse inward, supersonically.
    
    \item As the density of these infalling shells
    rises, adiabatic compression
    increases their ideal gas pressure 
    more slowly \big(since $P_{\!\sigma} \propto \rho^{5/3}$\big) than it does their SI pressure \big(for which $P_\text{SI} \propto\rho^2$\big). The 
    latter eventually becomes large enough in the center
    to halt the infall of the innermost
    shells and provide their hydrostatic support
    within a radius $\simeq R_\text{TF}$, with a flattened
    profile like that of an ($n=1$)-polytrope. This
    was also seen in the non-cosmological results of
    \citetalias{paper1}.
    
    \item At this point, by $a/a_f \gtrsim 0.4$, 
    the incoming mass shell just outside this polytropic core
    -- just about to slam into it -- is halted by the 
    formation of a strong accretion shock that converts
    its kinetic energy of infall into ``thermal energy'' (i.e. energy due to dispersion).
    This boosts its ideal gas pressure to be
    comparable to the SI pressure within the core
    and extends the hydrostatic inner profile of the
    latter to larger radii, bounded by the shock, 
    supported outside the core by post-shock
    $P_{\!\sigma}$, instead of $P_\text{SI}$. 
    This, too, is similar to the non-cosmological case
    in \citetalias{paper1}.
    
    \item By $a/a_f = 1$, the accretion shock has moved 
    outward in mass, just beyond the radius $R_{200}$, 
    to encompass a virialized halo, within which the
    ``temperature'' profile associated with velocity dispersion
    is approximately isothermal, except inside the polytrope,
    where it is cold, because it remained interior to the 
    mass shell at which the shock formed.  However, matter interior to that shock is everywhere hydrostatic, supported
    inside the polytropic core by SI pressure and outside
    it by post-shock ideal gas pressure.
    
    \item The post-shock density profile, thereafter,
    has three distinct regions in 
    radius:
    \begin{enumerate}[label=(\alph*)]
        \item a flattened, polytropic core inside $R_\text{TF}$,
        \item an intermediate zone with power-law density profile, $\rho \propto r^{-12/7}$, and 
        \item an outer envelope which
        follows the same NFW-profile as for CDM
        from the same initial conditions, inside
        the advancing accretion shock, which remains
        just outside of the time-varying $R_{200}$.
    \end{enumerate}
    Regions 7(a) and (b) were also present in the non-cosmological results in
    \citetalias{paper1}, while 7(c) is unique to the
    cosmological case. 
    
    \item Outside the shock, the density follows
    the pre-shock, ballistic infall profile of a cold, pressure-free gas, just like the CDM case.
\end{enumerate}

As described in \S\ref{sec:staticICPaperI},  \citetalias{paper1} found a generic core-envelope
structure for the virialized objects formed
by the nonlinear evolution of the Jeans 
instability in SFDM-TF from non-cosmological 
(i.e. static) initial conditions.  As in
the cosmological case described above, an accretion
shock bounded the virialized region. However,
unlike the cosmological case above, the entire envelope
between the polytropic core and the shock
had the same power-law density profile $\rho \propto r^{-12/7}$ as reported above for the intermediate 
zone 7(b), and there was no NFW-like zone 7(c).
This reflected the fact that, in this \textit{non}-cosmological
case, the Jeans-instability problem involved a nearly uniform initial density, which gravitational instability
turned into an infall density profile 
(outside the accretion shock) with
that power-law.  The condition of hydrostatic
equilibrium inside the post-shock region then combined with
the constant (shock-velocity-independent) factor
of density jump at the shock to make the
profile of \textit{post-shock} density the same
as that of the \textit{pre-shock} infall.  
Indeed, for this reason, our simulation from
the same non-cosmological initial conditions but
with CDM dynamics, instead, 
produced the exact same power-law
profile inside the shock, except with no SI-flattened 
polytropic core in the center.  In fact, the mass
inside the core in the SFDM-TF case, with polytropic
radius $R_\text{TF}$, was the same as that 
inside that radius in the CDM-like power-law profile.
So, apparently, outside of $R_\text{TF}$, where 
SFDM-TF and CDM dynamics was the same, it was the non-cosmological \textit{initial} conditions
of the SFDM-TF and CDM simulations that
resulted in making their infall 
profiles $\rho \propto r^{-12/7}$, instead of
the cosmological infall profile derived from
the halo mass accretion rate in CDM N-body 
simulations, which makes the post-shock profile NFW-like.  

From that description, we might reasonably have expected
halo formation from
\textit{cosmological} initial conditions to be
similar, with the same polytropic
core, surrounded again by a CDM-like envelope.
In particular, we might
have expected this CDM-like envelope
to be \textit{NFW-like}, as it is in our CDM simulation 
(without SI) for this same cosmological case.
Then why did we find, as reported 
in item (7), that while this is true for 
regions (a) and (c), there is an intermediate zone (b) 
in which the density profile follows 
the same power-law profile \big($\rho \propto r^{-12/7}$\big)
as in the \textit{non-cosmological} simulations of \citetalias{paper1}, instead?   
The answer to this question encapsulates
the fundamental distinction between static and
cosmological Jeans instability in SFDM-TF and
is at the heart of our cosmological story-line here
in Paper II. 

What distinguishes a \textit{cosmological} from a \textit{non-cosmological} density perturbation is their different initial velocity perturbations.  
Within the Newtonian approximation, our 
spherically-symmetric Jeans instability problem
is transformed from a non-cosmological one into a cosmological one entirely by the presence of perturbed Hubble expansion in the latter that replaces the static initial condition of the former.  This small change has
profound consequences, however, because the SI Jeans 
length is fixed in physical coordinates by the constant
value of $R_\text{TF}$.  For CDM in an Einstein-de Sitter
universe, a spherical perturbation that is overdense
will expand more slowly than unperturbed Hubble expansion, until it eventually
reaches a maximum expansion radius and recollapses.
This will be true in SFDM-TF, as well, if it is overdense
across a region much larger than $R_\text{TF}$.  
But in that case,
matter close to the center of the perturbation,
which is initially expanding, will be affected by the
presence of SI pressure, too, which responds to density
inhomogeneity by driving a sound wave outward from
the center, as far as $R_\text{TF}$, which smooths out the density perturbation as it goes.  
This is what we found in item (2) above.
The effect of this early SI-smoothing imprints the
evolution to come for the mass shells inside the
sphere smoothed-out by the sound wave, even though they
continue to expand homologously, thereafter, 
which carries them far outside the physical radius 
originally overtaken by the sound wave.

In Fig. \ref{fig:SIsmoothing}, we illustrate this
early SI-sound-wave smoothing by
two examples from the same NFW-producing
CDM-like initial perturbation, with $M_\infty=1.6\times10^{9} \text{ M}_\odot$ and $a_f = 0.24$,
but with different SI strengths, by plotting density
profiles for a sequence of early time-slices of 
our SFDM-TF simulation, compared with
their CDM counterparts. Let us focus first
on the top panel, Fig. \ref{fig:SIsmoothing}(a),
the case for $R_\text{TF} = 10$ pc. 
The central density profile is quickly 
made nearly uniform out to $R_\text{TF}$, by a sound
wave from the origin that reaches $R_\text{TF}$ 
long before that shell reaches maximum expansion.  Thereafter, this inner
sphere evolves just like a top-hat perturbation.
As such, it continues to expand until it reaches maximum expansion, turns around, and recollapses.
At maximum expansion, the spherical region is 
momentarily at rest, so it is at that
moment essentially the same as the initial perturbation
of the \textit{non-cosmological} problem in \citetalias{paper1}!
Its evolution, thereafter, must, therefore, be similar to
what we found there.  It should yield 
the same core-envelope structure inside the
advancing accretion shock as in \citetalias{paper1}, with
polytropic core of proper radius $R_\text{TF}$
and the same power-law density profile, $\rho \propto r^{-12/7}$, outside the core, but only out through
the mass that was initially smoothed by the SI-pressure-driven sound wave.  As we can see in Fig. \ref{fig:SIsmoothing}, the sound wave smooths out the density as it travels to $\sim R_\text{TF}$. Each time-slice in Fig. \ref{fig:SIsmoothing}(a)
is marked (by a filled circle) to identify the particular mass shell that was located at $R_\text{TF}$ when the 
sound wave reached that radius.  The mass within that 
radius was basically the SI-Jeans mass of the background universe at that time, since the density 
perturbation was only linear then. 
The time-varying radius so marked,
which contains this fixed mass as it
evolves, provides a rough estimate (i.e. lower-limit) 
for the outer radius of the intermediate zone 7(b)
within which we expect the final 
density profile to follow the power-law 
$\rho \propto r^{-12/7}$, 
once this mass is inside the shock.  We also
mark this radius on the density
profiles in the top panels of Fig. \ref{fig:NFWSI},
for all the time-slices, to confirm this.  

The transition from power-law to NFW-like post-shock
density profiles inside the shock extends beyond the 
filled circles, encompassing a somewhat larger mass.
(The filled circles in 
Fig. \ref{fig:NFWSI} enclose masses of $5\times10^{11} \text{ M}_\odot$ (left) and $5\times10^5 \text{ M}_\odot$ (right), for $R_\text{TF} = 1$ kpc and $10$ pc, respectively, as compared with their Jeans masses at the initial time of $2\times10^{13} \text{ M}_\odot$ and $2\times10^7 \text{ M}_\odot$.)
We can understand why and bound this transition from
above by a second estimate, as follows, which we 
mark with filled triangles on the density profiles
in both Figs. \ref{fig:NFWSI} and \ref{fig:SIsmoothing}(a).
As we shall see, in \S\ref{sec:LinPertsTransferFn},
the cosmological SI Jeans mass at any scale factor 
is the mass inside a sphere of fixed physical radius 
$R_\text{TF}$, containing the mean density at that time, 
a mass which drops with scale factor as $a^{-3}$
(see equation \ref{eq:Jeansmass}).  As
a result, even the small, finite time for
the sound wave to reach $R_\text{TF}$ is enough
time for the Jeans mass to drop significantly from its value at the initial scale factor $a_i$.
Therefore, the effects of SI Jeans-filtering extend to
encompass a mass which is between this 
(larger) \textit{initial} Jeans mass 
(marked by filled triangles)
and the (smaller) Jeans mass when the sound
wave reached $R_\text{TF}$ (marked by filled
circles). 

Now we are ready to focus on the second example of
SI-smoothing, plotted in the bottom panel
Fig. \ref{fig:SIsmoothing}(b). This case differs from
that in the top panel Fig. \ref{fig:SIsmoothing}(a)
only by having a much stronger SI, corresponding to
$R_\text{TF} = 1$ kpc, instead of $10$ pc.  As with
the latter case, the initial SI sound wave propagates outward and flattens the density profile out to $R_\text{TF}$. In the case of $R_\text{TF} = 1$ kpc, 
however, this sound wave passes over the entire computational domain, which extends to $M_\infty$ 
(i.e. the radial position of the outermost shell containing $M_\infty$ is $\ll R_\text{TF}$), 
and the perturbation is, thereby, prevented from growing 
to form the collapsed virialized object it would
have formed in the CDM case.

This reveals the essential factor that distinguishes
the outcomes of non-cosmological vs. cosmological initial conditions for the formation of an SFDM-TF halo with 
a given mass at a given epoch. The former yields a
virialized halo in the post-shock region, with a
polytropic core of radius $R_\text{TF}$ surrounded by
a CDM-like density profile that extends to the
accretion shock radius at that epoch.  The mass encompassed
by that shock grows with time just by adding new
mass shells to the outside, at the
gravitational free-fall times defined by the mean
mass densities initially interior to each mass shell.
The mass in the core is fixed by
extrapolating the CDM-like profile inward from $R_\text{TF}$
and integrating out to that radius.
SI-smoothing in this case is confined to the region within
a fixed physical radius $R_\text{TF}$, 
and to the mass which ends up inside this
radius at the final epoch, while the dynamics outside this
radius -- and the final halo structure there -- are CDM-like.
For the \textit{cosmological} case, however, 
the mass that is SI-smoothed is the much larger 
amount that was inside that same sphere of physical radius $R_\text{TF}$ at a time
close to the \textit{initial} time, at the 
mean matter density of the universe at that time,
before expansion carried it well beyond
$R_\text{TF}$, enroute to its turn-around radius.   
Since SI-smoothing quickly flattens the initial perturbation
out to $R_\text{TF}$,
the turn-around and recollapse of any shell interior
to the one which bounds the smoothed mass
will be delayed relative to their timing without
SI-smoothing. That bounding sphere and every shell inside it
will expand, turn-around and recollapse like
a uniform, top-hat perturbation with the same mean density.
Since SI-smoothing out to $R_\text{TF}$
must conserve mass, the average overdensity inside
the smoothed sphere must be about
the same as in the unsmoothed perturbation, 
so the bounding sphere and its contents
turn-around and recollapse just like the
bounding sphere in the case without SI-smoothing.
As long as the latter is destined to collapse,
that means SI-smoothing does not \textit{prevent}
the smoothed mass from collapsing, 
it only \textit{delays} the collapse of interior
shells to the time at which the bounding sphere, itself,
would have, in the absence of SI-smoothing.

This assumes, of course, that the average overdensity
inside the bounding sphere in the initial
perturbation was positive,
as it would be for the self-similar perturbations described above in \S\ref{sec:selfsimilar}. 
The CDM-like perturbation that leads to an NFW-like
post-shock profile, however, is a \textit{non}-self-similar one,
in which the overdensity is ``compensated'' by an underdensity at large radii, so that
the average overdensity
inside the sphere which encompasses mass $M_{\infty}$
is zero.   Spherical shells outside that mass will,
therefore, remain unperturbed and cannot collapse.
In that case, if the mass affected by 
SI-smoothing exceeds $M_{\infty}$, as it does
in Fig. \ref{fig:SIsmoothing}(b), then the 
bounding sphere described above is actually 
\textit{prevented} from collapsing, altogether.  
For those cases that \textit{do} collapse, however,
our simulations above tell us that
the virialized object that forms inside the accretion shock
will have an inner portion that 
resembles that in \citetalias{paper1} from  \textit{non-cosmological} (static) initial conditions,
encompassing the initial SI-smoothed mass, 
with the core-envelope structure reported there
\big(polytropic core of radius $R_\text{TF}$, 
surrounded by a power-law density profile, $\rho \propto r^{-12/7}$\big), but beyond the radius 
containing this initial SI-smoothed mass, 
it will transition to the NFW-like profile. 

This description above
assumes we started from CDM-like initial conditions.
We made this assumption as the first step in
understanding the fully nonlinear
dynamics of SFDM-TF halo formation.  
That was a reasonable first step for a range
of perturbation mass scales, in view of our
discussion in \citetalias{paper1} of the fact
that, at scales beyond $R_\text{TF}$, the dynamics
of SFDM-TF is CDM-like.  As such, we were able to
confirm that, beyond the inner mass smoothed by SI,
SFDM-TF halo profiles simulated with nonlinear
dynamics will follow the same NFW-like profile
expected for the CDM model.  However, we
also learned that this SI-smoothed inner
mass can be as large as the mass inside
the fixed radius $R_\text{TF}$ at early
times, when the universe was much denser,
and that there is an intermediate zone of
density profile outside the polytropic core
of radius $R_\text{TF}$ in the final object, which
differs from the NFW profile.  Such a feature
might be a signature of SFDM-TF that distinguishes
it observationally from CDM.  However, it remains to
be seen from 3D-cosmological SFDM-TF
simulations, 
from initial conditions 
that are self-consistently based upon
linear perturbation theory for SFDM-TF,
if this intermediate zone will survive, or else
relax to the NFW-like profile over time.

{ In addition, such 3D simulations are required to determine what MAH results for halo growth from SFDM initial conditions and how it differs from that for CDM.
The results we
present here by assuming the smoothed accretion rate derived from N-body simulations of the CDM model are likely to be robust on the smallest and largest scales with respect to replacing the initial conditions with the proper MAH of SFDM. The shells at large radii are likely to follow CDM-like trajectories even with correct initial conditions, while the shells at small radii are quickly smoothed by SI Jeans filtering and so should also follow the correct trajectories.}

{ As we shall see below, from linear perturbation analysis, SFDM initial conditions differ from those of CDM by suppressing power on small scales, due to Jeans filtering.   Halos can form, however, on all mass scales, even those below that filter scale, although with lower abundance for those smaller masses. Filtering may smooth the initial density perturbation responsible for building a halo by spherical infall, out to a larger radius than in the case with CDM-like initial conditions, resulting in a more nearly uniform collapse.  In that case, we might expect the nonlinear collapse to produce a post-shock density profile outside the polytropic core, which follows the power-law $\rho \propto r^{-12/7}$, over a larger range of radii than in our results above.  } 

In the meantime,  we need linear 
perturbation theory to determine the statistical 
measures of structure formation, e.g. how many haloes of a given mass are likely
to form by a given epoch, in order to constrain
the model parameter $R_\text{TF}$ against observations.  
We consider that in the remainder of this
paper.

\section{Linear perturbation theory and the transfer function for SFDM-TF}
\label{sec:LinPertsTransferFn}

\subsection{Qualitative overview}
\label{sec:LinPertOverview}

The growth of density perturbations in a universe with complex SFDM in the 
presence of a repulsive SI differs from both that in the 
CDM model and that in the Fuzzy Dark Matter (FDM) limit of SFDM.    
In all models, the assumed, primordial initial condition involves 
small-amplitude, 3D, Gaussian-random density fluctuations with zero mean, 
characterized by a scale-invariant\footnote{More generally, 
$P(k,t_i) \propto k^{n_s}$ where $n_s$ is the primordial spectral index of scalar perturbations, and the value we adopt here is
$n_s = 1$, the Harrison-Peebles-Zeldovich spectrum, which is consistent with current observational constraints.} power spectrum  
\begin{equation}
    P(k,t_i) = |\delta_k(t_i)|^2 \propto k
\end{equation}
where the linear overdensity $\delta_k (t)$ of a Fourier mode of 
wavenumber $k$ at a cosmic time $t$ is related to its 
amplitude $\delta_k (t_i)$ at initial time $t_i$, according to
\begin{equation}\label{eq:dP}
    \delta_k(t) = \mathcal{T}_k(t,t_i)\delta_k(t_i)
\end{equation}
in terms of the transfer function $\mathcal{T}_k(t,t_i)$.
Differences in the structure formation histories of these models arise while 
they are still in their linear amplitude phase, as characterized by different 
transfer functions for each model.
They have one very important thing in common, however.  
When fluctuation modes of different proper wavelength $\lambda$ (or, equivalently, comoving $k=2\pi a / \lambda$) 
first re-enter the horizon \big(i.e. near the time
$t=t_\textsc{h}$ at which their proper wavelength equals the Hubble diameter, 
i.e. $\lambda(t) = 2c/H(t)$, in which case
comoving $k = k_\text{H} \equiv \frac{\pi}{c}a_\textsc{h}H(a_\textsc{h})$\big), they do so with the same, wavelength-independent, fractional overdensity, $\delta_k(t_\textsc{h}) \ll 1$, for all models. 
Although modes of different $k$ enter the horizon at different times 
$t_\textsc{h}$ (later for smaller $k$), and it is even possible that 
some modes of the same $k$ in the different models enter the horizon at different times, 
the value of $\delta_k(t_\textsc{h})$ in all cases is the same for all values of $k$ and 
all values of $t_\textsc{h}$. 

The value of $t_\textsc{h}$ for a given wavenumber depends only on the time-dependence 
of the Hubble parameter, so $t_\textsc{h}(k)$ is the same in all models only if $H(t)$ 
is the same at all times in different models.  We expect this to be the case for 
CDM and SFDM models, in general, in our observed Universe, for modes that re-enter the 
horizon during the radiation-dominated (RD) and matter-dominated (MD) eras of interest to us here, since all models are required 
by observational constraints to share a common expansion history from the time the wavenumbers
associated with galaxy formation began to enter the horizon during the RD era through the present.
This condition is not met for all of the models at all times. For modes that enter before 
the RD era -- e.g. for complex SFDM, with or without SI, there is an early stiff phase ($w=1$), 
prior to the RD era, in which the scalar field dominates.  
Hence, even though the Hubble parameter during the RD and MD eras, for example, is basically the same at a given scale factor for all models, it differs during the SFDM \textit{stiff} phase at earlier times, in that case, when much 
smaller scales entered the horizon. In fact,
even \textit{during} the RD epoch, but well before the epoch of matter-radiation equality, 
$a_\text{eq}$ (for both complex and real field
cases), 
the Hubble parameter can be slightly enhanced relative to its value in CDM and FDM, 
by a small boost to the total energy density contributed by SFDM in the presence of SI,
when its EOS is radiation-like \big($\langle w \rangle=1/3$\big), between its stiff- and matter-like phases of evolution, 
but this small enhancement is gone well before $a_\text{eq}$.\footnote{When we
describe the EOS parameter $w$ of the 
scalar field during its rapid-oscillation phase, 
in which the value averaged over oscillation
periods, $\langle w \rangle$, evolves
from $\langle w \rangle = 1/3$ to 
$\langle w \rangle = 0$, we will, henceforth,
generally just write
$w$ to refer to this averaged quantity and
omit the brackets.}
                           
What happens to that amplitude next, after a mode is already inside the horizon, depends upon the microscopic nature of the dark matter and other components that contribute to the total mass-energy of the Universe. It also depends on whether the background Universe is RD or MD during the phase where the galaxy- and galaxy-cluster-scale fluctuations are still linear.

In the standard CDM model, dark matter is a nonrelativistic gas of cold, collisionless particles, so the ratio $w$ of its pressure to its energy-density is well-approximated by $w = 0$.  In that case, its fluctuation growth rate after horizon entry depends upon whether the Universe is in the RD or MD era.  For the scales of interest for structure formation, spanning the range of galaxy- and cluster-mass objects, horizon entry is during the RD era. At that time, 
radiation dominates the gravitational potential fluctuations that drive perturbation growth in CDM, but fluctuations
in the radiation component do not amplify during this phase, because a relativistic fluid with $w = 1/3$ has a speed of sound close to $c/\sqrt{3}$,  and density inhomogeneities are thus smoothed out by pressure disturbances that can travel a distance comparable to the Hubble radius within the age of the universe at that time. As a result, gravitational fluctuation growth in the CDM component is driven only by its own fluctuation as a subdominant contribution to the total energy-density. This yields only a relatively small amount of growth between horizon entry at scale factor $a_\textsc{h}$, and matter-radiation equality at $a_\text{eq}$, which is a logarithmic function of those scale factors. 

We adopt cosmological parameters from \citet{planck18} as given in Table~\ref{tab:params}.
\begin{table}
    \centering
    \begin{tabular}{c|c}
        \hline\hline
        parameter & value \\
        \hline
        $h$ & 0.67 \\
        $\Omega_\text{dm} h^2$ & 0.12 \\
        $\Omega_m h^2$ & 0.14 \\
        $\Omega_\Lambda h^2$ & 0.31 \\
        $\Omega_\text{rad} h^2$ & $4\times10^{-5}$ \\
        \hline
    \end{tabular}
    \caption{\textbf{Cosmological parameters} from \citet{planck18}, which we adopt for our linear perturbation analysis in \S\ref{sec:LinPertsTransferFn}.}
    \label{tab:params}
\end{table}
The scale factor of matter-radiation equality is given by
\begin{equation}
    a_\text{eq} \simeq 2.9 \times 10^{-4}
\end{equation}
For a comoving wavelength $\lambda_0$ that enters the Hubble volume during the RD epoch,
\begin{equation}
    a_\textsc{h} \simeq 10^{-6} \left( \frac{\lambda_0}{\text{1 Mpc}} \right)
\end{equation}
And the unperturbed mass of nonrelativistic dark matter contained in a sphere of proper radius $\lambda(t)/2$, where $\lambda(t)/a(t) = \lambda_0/a_0$, is given at all times by
\begin{equation}
    M = \frac{4\pi}{3} \rho_{\text{dm},0} \left( \frac{\lambda_0}{2} \right)^3 \simeq 1.45 \times 10^{11} \text{ M}_\odot \big(\Omega_\text{dm} h^2\big) \left( \frac{\lambda_0}{\text{1 Mpc}} \right)^3
\end{equation}
where the mean dark matter density at present is given by
\begin{equation}
    \rho_{\text{dm},0} \equiv \Omega_\text{dm}\frac{3 H_0^2}{8\pi G} \simeq 2.8 \times 10^{11}~\frac{\text{M}_\odot}{\text{Mpc}^{3}} \big(\Omega_\text{dm} h^2\big)
\end{equation}
We can also write the scale factor when a given mass fills the Hubble volume, according to
\begin{equation}
    a_\textsc{h} \simeq 1.9 \times 10^{-6} \big(\Omega_\text{dm} h^2\big)^{-1/3} \left( \frac{M}{10^{12} \text{ M}_\odot} \right)^{1/3}
\end{equation}
The comoving wavelength that entered the horizon at $a_\text{eq}$ is given by
\begin{equation}
    \lambda_\text{0,eq} \simeq 290 \text{ Mpc}
\end{equation}
and the mass\footnote{If we had chosen to define the mass associated with a horizon-filling mode as the mass inside a sphere of comoving radius 
$\lambda_0$, instead of  $\lambda_0/2$ as we did above, then the mass associated with the mode that enters at $a_\text{eq}$ would be 8 times larger than this, or $3 \times 10^{18} \text{ M}_\odot$.  
Alternatively, if we had, instead, defined the mass associated with a mode as just $\rho_{\text{dm},0} \lambda_0^3$,  rather than $4\pi/3  \rho_{\text{dm},0} (\lambda_0/2)^3$, as above, then the mass above, which is evaluated based upon the latter, should be multiplied by $6/\pi \simeq 1.91$, instead, which would make $M_\text{eq} \simeq 8 \times 10^{17} \text{ M}_\odot$.   These are just matters of definition, of course, as the horizon-filling wavelength and the unperturbed mass associated with a given wavelength mode are not quantities of precise meaning, but we should try to compare different mode-associated masses using the same choice of definition.} that filled the Hubble volume at $a_\text{eq}$ is, therefore, given by
\begin{equation}
    M_\text{H,eq} \simeq 3.5 \times 10^{18} \text{ M}_\odot \big(\Omega_\text{dm} h^2 \big)
\end{equation}
which, for $\Omega_\text{dm} h^2=0.12$, yields $M_\text{H,eq} \simeq 4 \times 10^{17} \text{ M}_\odot$.

For CDM, modes of mass scale $M$ that enter the horizon during RD experience only logarithmic growth between $a_\textsc{h}$ and $a_\text{eq}$, given by (see equation~\ref{eq:deltaCDM})
\begin{equation}
    \frac{\delta_\text{eq}}{\delta_\textsc{h}} \approx 15\ln \left(\frac{a_\text{eq}}{a_\textsc{h}}\right)
    \label{logarithmicgrowth}
\end{equation}
where $\delta_\text{eq} \equiv \delta_k (a_\text{eq})$ and $\delta_\textsc{h} \equiv \delta_k (a_\textsc{h})$.
Henceforth, we will omit the subscript $k$ when referring to the Fourier mode overdensity $\delta$.
By $a_\text{eq}$, all modes of interest in forming galaxies and clusters had already entered the horizon, so they all experience the same factor of growth between $a_\text{eq}$ and any later $a$ (i.e. for early times during the MD era, growth is the same as in an Einstein-de Sitter universe, in which case $\delta \propto a$).

It is convenient to describe the growth of density fluctuations in alternative dark matter models like SFDM in terms of their growth relative to that in CDM. 
For a range of wavenumbers, the growth of density fluctuations in SFDM with SI in the 
TF regime, for modes of interest to us in galaxy and large-scale structure formation, is the same 
after horizon crossing as for CDM. This is the case
for modes of small enough wavenumber, as we shall see. 

There are two effects, however, that distinguish the two models.   
CDM behaves as a pressure-free, matter-like gas (with $w=0$) at all times, both during the RD era when it is a subdominant component of the total mass-energy density of the universe and during the MD era when it dominates that energy-density.    
SFDM differs in that regard.   
SFDM with repulsive SI is only nonrelativistic and matter-like (with $w = 0$) at late times, after it transitions from being relativistic and  radiation-like (with $w  = 1/3$).  
Although this transition is not instantaneous, it is fast enough that we can characterize it as a sudden transition at scale factor $a_t$.   
As we discuss below, this transition scale factor $a_t$ is set by the value of the ratio $g/m^2$.  
As pointed out by \citet{Li14}, observations of the CMB anisotropy constrain the allowed values of this ratio by requiring that the EOS of the background universe must transition from RD to MD at close to the same value of  $a_\text{eq}$ in both SFDM and CDM.   
This places an upper bound on the ratio, by requiring that $a_t \leq a_\text{eq}$.
As such, i.e. as long as $a_t \leq a_\text{eq}$, modes that enter the horizon before $a_t$, when SFDM-TF behaves like a radiation-like perfect fluid ($w=1/3$),  
will not experience the logarithmic growth between $a_\textsc{h}$ and $a_t$ that the same modes 
experience in CDM.  In fact, these modes that enter 
the horizon prior to $a_t$ should evolve
just like those of the dominant radiation component, which oscillate
like sound waves of a relativistic gas
with sound speed $c/\sqrt{3}$, without changing their amplitude. If so, then these modes that entered the horizon before $a_t$ and before $a_\text{eq}$
would be deprived of the interval of logarithmic
growth which the same modes would have experienced
after horizon entry in CDM between $a_\textsc{h}$ and
$a_\text{eq}$, according to equation (\ref{logarithmicgrowth}).   
In the absence of CDM-like 
growth during this interval of scale factor, therefore, we might expect a net amount of growth for these modes over the entire RD era, 
including the time between $a_t$
and $a_\text{eq}$, equal to that which would result for CDM modes if we replaced $a_\textsc{h}$ in their logarithmic growth factor in equation (\ref{logarithmicgrowth}) by $a_t$.   

There is a second effect, however, which may also interfere with the growth of SFDM modes, if their proper wavelength is below the Jeans length that results from the back reaction of SI pressure forces that act to cancel their gravitational instability.    
To be more specific, we must estimate this Jeans length and its dependence on scale factor, and consider which of these two effects dominates, for which modes, and during which intervals of scale factor. 
 
Let us start by finding the transition scale factor $a_t$ and its dependence on SI strength $g$ and particle mass $m$. 
Suppose we identify the transition between $w = 1/3$ and $w = 0$ in equation (\ref{eq:averagew}), which goes through these values monotonically with increasing scale factor, by solving that equation for the scale factor at which $w=1/9$,
assuming that, for small $w$, it is sufficient to take the homogeneous background value of 
$\langle|\phi|^2\rangle$ as given by
$\langle|\phi|^2\rangle\simeq\rho_\text{dm}=\rho_\text{dm,0}a^{-3}$,
which then yields  
\begin{equation}
    a_{1/9} = \left(3\rho_\text{dm,0} \frac{g}{m^2c^2}\right)^{1/3}
\end{equation}
We can then write the scale factor for any 
$w$
as follows: 
\begin{equation}
    a_w = a_{1/9}\left (\frac{1}{6w}-\frac{1}{2}\right)^{1/3}
\end{equation}
In terms of $R_\text{TF}$ in equation~(\ref{eq:RTF}), we can write the transition scale factor as follows:
\begin{align}
    a_{1/9} &= \left( \frac{12G \rho_\text{dm,0} R_\text{TF}^2}{\pi c^2} \right)^{1/3} \nonumber \\
    &\simeq 1.8\times 10^{-5} \left( \frac{\Omega_\text{dm} h^2}{0.12} \right)^{2/3} \left( \frac{R_\text{TF}}{\text{1 kpc} } \right)^{2/3}
\end{align}
If we now take $a_t = a_{1/9}$, then $a_t(R_\text{TF})$ follows
\begin{equation}
    a_t(R_\text{TF}) \propto R_\text{TF}^{2/3}
    \label{eq:atransitionRTF}
\end{equation}
In the radiation-dominated era, for which $H(a) \propto a^{-2}$, the mass that fills the Hubble volume is
\begin{equation}
    M_\text{H,RD}(a) = \frac{4\pi}{3}\rho_\text{dm,0}a^{-3} \Big(\frac{c}{H(a)}\Big)^3 \propto a^3
\end{equation}
which, evaluated at $a = a_t$ for a given value of $R_\text{TF}$, is then
\begin{equation}
    M_\text{H,RD}(a_t(R_\text{TF})) \propto a_t^3(R_\text{TF}) \propto R_\text{TF}^2
\end{equation}
Combining this with the relationship above between $a_t$ and $R_\text{TF}$, we can write the horizon-filling mass at the transition scale factor in terms of $R_\text{TF}$ as
\begin{equation}
    M_\text{H,RD}(a_t(R_\text{TF})) = M_\text{H,eq}\left(\frac{R_\text{TF}}{R_\text{TF,eq}}\right)^2
\end{equation}
where $R_\text{TF,eq}$ is the value that makes $a_t(R_\text{TF,eq}) = a_\text{eq}$.

As discussed in \citetalias{paper1}, 
when $w\simeq0$ 
(i.e. $a\gtrsim a_t$), the Jeans length scale below which gravitational instability is opposed by the back-reaction of SI pressure is the proper wavelength
$\lambda_\text{J,0} = 2 R_\text{TF}$
(where the subscript ``0'' refers to $w=0$ and
the subscript ``SI'' is removed for simplicity 
but is implied).
The associated SFDM-TF Jeans mass scale is given by the dark matter mass inside a sphere of radius $R_\text{TF}$, at any scale factor $a$, according to
\begin{align}
    M_\text{J,0}(a) &= \frac{4\pi}{3} \rho_\text{dm,0} a^{-3} R_\text{TF}^3 \nonumber \\
    &\simeq 4.6\times 10^{12} \text{ M}_\odot \left( \frac{\Omega_\text{dm} h^2}{0.12} \right) \left(\frac{a}{10^{-3.5}}\right)^{-3} \left(\frac{R_\text{TF}}{\text{1 kpc}}\right)^3
    \label{eq:Jeansmass}
\end{align}
For future reference, it will be useful to define the scale factor $a_\textsc{j,0}(M)$ at which the mass $M$ equals the Jeans mass, by inverting equation~(\ref{eq:Jeansmass}) to write
\begin{align}
    a_\text{J,0}(M) &= R_\text{TF} \left( \frac{ 4\pi \rho_\text{dm,0} }{ 3M } \right)^{1/3} = \frac{R_\text{TF} k}{\pi} \nonumber \\
    &\simeq 5\times10^{-4} \left( \frac{\Omega_\text{dm} h^2}{0.12} \right)^{1/3}\left( \frac{M}{10^{12} \text{ M}_\odot} \right)^{-1/3} \left(\frac{R_\text{TF}}{\text{1 kpc}}\right)
    \label{eq:aJeans0}
\end{align}

\begin{figure}
    \centering
    \includegraphics[width=\columnwidth]{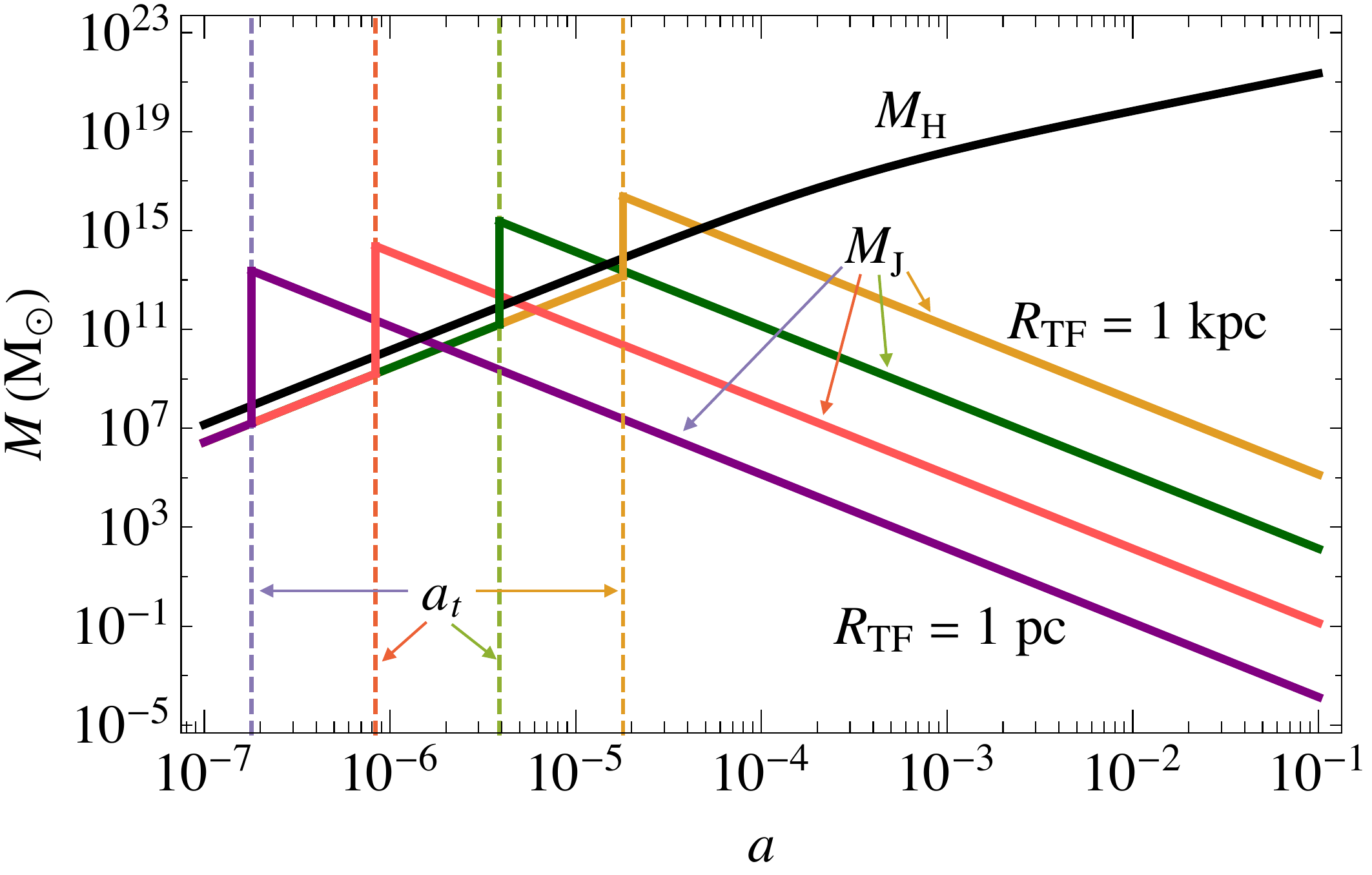}
    \caption{\textbf{Cosmic evolution of the SFDM-TF Jeans-filter scale: \textit{The Incredible Shrinking Jeans Mass}.}  The Jeans mass $M_\text{J}$ \big(i.e. $M_\text{J} = \frac{4\pi}{3} \rho_\text{dm,0} ( \pi/k_\text{J})^3$\big) associated with the critical wavenumber $k_\text{J}$ for linear density fluctuation modes in SFDM-TF that marks the boundary between growth (for $M>M_\text{J}$) and oscillation (for $M<M_\text{J}$), vs. scale factor $a$, for 
    several illustrative values of the Thomas-Fermi radius $R_\text{TF}$.  
    Curve going from lower left to upper right is $M_\text{H}(a)$, the mass that fills the horizon 
    (i.e. the Hubble volume, of radius $c/H(a)$) at each scale factor.  The diagonal curves from upper left to lower right represent the curves of $M_\text{J}(a)$ for different values of the $R_\text{TF}$  (1000, 100, 10, and 1 pc, in order from top to bottom).  The dashed vertical lines and the discontinuities on each of these curves for different values of $R_\text{TF}$ locates the value of the transition scale factor $a_t$ ([1.8, 8.4, 39, 180] $\times 10^{-7}$, in order from left to right)
    for the SFDM EOS, to the
    left of which $w=1/3$, while to the right $w=0$.}
    \label{fig:JeansMass}
\end{figure}

We are now in a position to compare this Jeans mass scale $M_\text{J,0}$ given above as a function of scale factor, for a given value of  $R_\text{TF}$, with the horizon-filling mass at each scale factor, to show the range of fluctuation mass-scales that are free of Jeans filtering from the moment those fluctuations re-enter the horizon during the RD era.   
For those fluctuation modes, CDM-like perturbation growth can be assumed to apply to the SFDM fluctuations, as well, as long as their horizon entry took place after SFDM transitions from a radiation-like to a matter-like EOS.   
We plot these quantities in Fig.~\ref{fig:JeansMass} for 4 different values of $R_\text{TF} = \{1000, 100, 10, 1\}$ pc.    
As the figure indicates, all modes that enter the horizon super-Jeans (i.e. $M > M_\text{J}$ when $M < M_\text{H}$)  do so when the SFDM EOS was matter-like.   
That means that the Jeans mass calculated above
for the case when SFDM is matter-like is applicable, as we have assumed.  
Modes of smaller wavelength, for which horizon entry preceded the transition scale factor of the SFDM EOS, were relativistic and radiation-like upon entry, so their Jeans filtering during that phase should be calculated for a relativistic fluid, instead, because their effective sound speed is close to the speed of light and the Jeans scale is comparable to the horizon size, at which scale the Newtonian approximation breaks down. 

Roughly speaking, when $w = 1/3$ 
(i.e. $a \lesssim a_t$), 
the Jeans length is equal to the distance a 
relativistic sound wave can travel at the
sound speed $c/\sqrt{3}$ in a Hubble time:
\begin{equation}
    R_\text{J,1/3}(a) = \frac{c}{H\sqrt{3}}
\end{equation}
The mass enclosed by this radius
is the same as the mass inside the usual
acoustic horizon during the RD era, which is
just smaller than the mass inside the Hubble
volume, according to
\begin{align}
    M_\text{J,1/3}(a) &= \frac{4\pi}{3} \rho_\text{dm,0} a^{-3} R_\text{J,rad}^3 \nonumber \\
    &\simeq 8.7\times 10^{16} \text{ M}_\odot \left( \frac{\Omega_\text{dm} h^2}{0.12} \right) \left(\frac{a}{10^{-3.5}}\right)^{3}
\end{align} 
\textit{independent of} $R_\text{TF}$,
aside from the dependence of transition scale factor $a_t$ on it (see, for example, equation~\ref{eq:atransitionRTF}).
This radiation-like Jeans mass defines the $a<a_t$ segment of each colored solid line
in Fig.~\ref{fig:JeansMass}.

For any given mode, we can describe its cosmological evolution by drawing a horizontal line across the box in Fig.~\ref{fig:JeansMass}
at that mass scale and following it in time
from small scale factor to large, for a fixed
choice of SI strength, expressed by the
value of $R_\text{TF}$, as follows.  At early
times such that $a<a_t$, the SFDM-TF is 
radiation-like ($w\approx{1/3}$), so its fluctuations behave during this phase
just like those of fluctuations of the
same wavenumber $k$ in the dominant radiation component.  They are indistinguishable from the
latter, in fact 
(i.e. Lagrangian volume changes in the
scalar field and radiation are the same for these
modes, during this phase of their evolution, so
the fluctuation in dark matter \textit{mass} density $\delta_\text{SFDM}$ for these modes
is simply related to 
the energy-density fluctuation in the radiation
$\delta_\text{rad}$ by $\delta_\text{SFDM}= (3/4)\delta_\text{rad}$). The smaller the
value of $R_\text{TF}$ is, the earlier this
radiation-like phase of EOS will end. 
However, there is always 
a range of modes for which the
mass is small enough (i.e. $k$ is large enough) to make horizon entry early enough that $a_\textsc{h}< a_t$ for those modes, 
so they enter while the EOS of the scalar field is radiation-like. According to Fig.~\ref{fig:JeansMass}, these modes are sub-Jeans ($M<M_\text{J,1/3}$) once they
are within the horizon.  Radiation fluctuations that enter the horizon during this epoch
(which is RD), all
start from the same $k$-independent amplitude when they are far outside the horizon, but
oscillate like sound waves with constant amplitude after entry. This also describes
the behavior of SFDM fluctuations of the 
same wavenumber, therefore, until the
end of their radiation-like phase at $a=a_t$.
At this time, the EOS of the
scalar field transitions to matter-like ($w\approx0$) and, thereafter, the character
of the evolution departs from that of the
dominant radiation component, as described 
below.  

There are three kinds of 
modes to track, as we follow their evolution \textit{after}
$a=a_t$, when the EOS turns matter-like. There
are the smaller-mass modes, described above, that entered the horizon prior to this time.  At the other extreme, there are  
modes that are so massive that they are matter-like and
\textit{super-Jeans} when they enter the horizon
(i.e. $M>M_\text{J,0}$ when $M<M_\text{H}$).  
In between these two mass ranges, there are
modes which are matter-like and \textit{sub-Jeans} upon horizon entry (i.e. $M<M_\text{J,0}$ when $M<M_\text{H}$).
In the first category, the
high-$k$ modes that entered prior to this
phase will still be sub-Jeans initially, after
$a_t$, but with the Jeans mass now given by 
$M_\text{J,0}$, rather than $M_\text{J,1/3}$. 
Their fluctuations will transition
from the radiation-like behavior described
above to a new, matter-like behavior, 
thereafter. For the range of modes that entered the horizon after $a=a_t$, however, 
the EOS is matter-like at all times, thereafter.  Their behavior once within the horizon 
depended upon whether they were above or below
the Jeans mass $M_\text{J,0}$ 
at the time of their horizon entry.  
We expect perturbation modes that were super-Jeans upon horizon entry (i.e. modes with $M > M_\text{J}$ when $M < M_\text{H}$) during this phase
to experience CDM-like growth, thereafter.  For the intermediate-mass modes that were sub-Jeans ($M<M_\text{J}$) and
matter-like ($w\approx0$ at $a>a_t$) when they were \textit{superhorizon}
($M>M_\text{H}$), their evolution was also no different from that of CDM.  After their horizon entry, however, 
these sub-Jeans modes oscillated like sound waves 
with an amplitude that grew as $a^{1/4}$, 
as described by \citet{SC15}
(see below).  This sub-Jeans oscillatory
behavior during the matter-like phase ($a>a_t$) 
of the scalar field, with amplitude growing as $a^{1/4}$, was also shared by the higher-$k$ 
modes that entered the horizon \textit{before}
$a=a_t$, although these latter modes were 
already oscillating at \textit{constant} 
amplitude prior to $a_t$, just like fluctuations in the dominant radiation component at that time.

The latter sub-Jeans behavior during the matter-like
phase of SFDM-TF contrasts with the
behavior of fluctuations in another matter component subject to Jeans-filtering by pressure forces, the baryons. For the baryons,
subhorizon baryonic fluctuations below the Jeans scale in the CDM universe also oscillated like sound waves, but with an amplitude that actually
\textit{decayed} during the MD era, prior to their decoupling from photons. 
However, this sub-Jeans behavior of the baryonic component in which the amplitude does not grow but even decays prior to photon decoupling has a much smaller impact on structure formation in the standard CDM model than the corresponding
sub-Jeans behavior of SFDM-TF described above,
even though the latter involves oscillations
that grow rather than decay.  The reason is 
that the baryonic fluctuations are not required
to ``carry the load'' as the driver of structure formation in the standard CDM model.  
In that model, subhorizon fluctuations in the 
\textit{dark matter} were always able to grow.
Hence, in that case, even though sub-Jeans baryonic fluctuations could not, the baryons were 
later able to catch up (after recombination) to those dark-matter fluctuations that had a head-start at growing during the RD era and 
beyond and never damped.  On scales above
the baryonic Jeans-filter mass 
\textit{after} recombination, the CDM fluctuations were never
filtered and so provided potential wells into which the baryons 
were able to fall, even if their own density
fluctuations were initially
negligible on these scales.  
Here, by contrast, SFDM \textit{is} the dark matter (\textit{all} of it, in fact),
and, even though its Jeans-filtering
bears some superficial similarity to the 
Jeans-filtering of baryonic fluctuations in CDM, 
there is no other component on which it
can rely to drive its growth at later times.
Hence, it is not possible for fluctuations in SFDM on these smaller scales to ``re-animate'' later,
as those of the baryons did for the CDM
model, by being driven to catch-up to perturbations in some other component that were able to grow while those in the SFDM did not.    

That said, the Jeans mass in SFDM-TF decreases rapidly with time, so one might wonder whether this would free some fluctuation modes of their Jeans filtering after they transition from sub-Jeans to super-Jeans.  
The Jeans scale in the matter-like phase of
SFDM corresponds to a fixed proper length, so it encompasses a smaller and smaller mass over time, scaling as it does with $a^{-3}$.
As such, whatever a mode’s amplitude is when it reaches this super-Jeans epoch following its sub-Jeans phase, it will be able to grow, thereafter, at the matter-like growth rate.   
It remains to be determined below,
if that will be a significant effect, or,  if not, if the growth of just those modes that were always super-Jeans from their horizon-entry epoch forward was sufficient on its own to form the full mass range of haloes at different redshifts required by astronomical observations.

\subsection{How the small-scale cut-off in the SFDM-TF transfer function depends on the polytrope radius}
\label{sec:cutoff}

The effect of Jeans-filtering described above
is to introduce a small-scale cut-off in the 
transfer function for SFDM-TF, while for larger
scales, the CDM and SFDM-TF transfer functions 
are the same. We will derive this effect in detail
by solving the perturbation equations below.
First, however, we will anticipate the result of those calculations by estimating the 
comoving wavenumber $k_\text{cut}$ and 
associated mass scale $M_\text{cut}$ 
at which the cut-off begins in the resulting
transfer function, such that power is reduced 
relative to that in CDM (for $k \gtrsim k_\text{cut}$, or equivalently, $M \lesssim M_\text{cut}$),
as follows.  
This cut-off will occur roughly 
at the wavenumber of the first mode to enter
the horizon (i.e. the mode with the smallest $k$-value, corresponding to the largest mass 
scale) which is sub-Jeans when it enters.
This is the mode for which 
$a_\textsc{j,0} = a_\textsc{h}$, or equivalently, $R_\text{TF} = c/H(a_\textsc{h})$:  
\begin{equation}
    R_\text{TF} = \frac{c a_\textsc{h}^2}{H_0 \sqrt{\Omega_\text{rad}}} = \frac{ \pi^2 H_0 \sqrt{\Omega_\text{rad}} }{ c k_\text{cut,TF}^2  }
\end{equation}
where $H(a_\textsc{h})$ is evaluated assuming all modes of interest enter the horizon when the Universe is RD, and the last expression follows from our definition of $a_\textsc{h}$: $ck/\pi = a_\textsc{h} H(a_\textsc{h})$. Solving for $k_\text{cut}$ gives
\begin{align}
    k_\text{cut,TF} &= \left(\frac{ \pi^2 H_0 \sqrt{\Omega_\text{rad}} }{c R_\text{TF}}\right)^{1/2} \nonumber \\
    &\simeq 0.2~h~{\rm Mpc}^{-1} \left( \frac{R_\text{TF}}{\rm 1~kpc} \right)^{-1/2} \label{eq:kcut-TF}
\end{align}
\begin{align}
    M_\text{cut,TF} &= \frac{4\pi}{3} \rho_\text{dm,0} \left( \frac{\pi}{k_\text{cut,TF}} \right)^3 \nonumber \\
    &\simeq 1.4\times10^{15}~{\rm M}_\odot \left( \frac{\Omega_\text{dm} h^2}{0.12} \right) \left( \frac{R_\text{TF}}{\rm 1~kpc} \right)^{3/2}
\end{align}

It will be useful to be able to compare the
cut-off scale for SFDM-TF to that which occurs
when SI is not present and the SFDM model is
in the FDM limit, instead.  In the latter case, as
we shall show below, there is a different 
Jeans mass which depends
upon the particle mass $m$, rather than on
the ratio $g/m^2$ involving SI, as it 
does for SFDM-TF.  For FDM, the EOS
is always matter-like during the 
galaxy-formation era (i.e. from horizon entry
forward, for all modes that are relevant), 
and, just as for SFDM-TF and CDM, 
galaxy-formation modes
enter the horizon 
when the Universe is RD.  In that case,
as we shall show in \S\ref{sec:LinPerts}, 
the Jeans mass and its dependence 
on scale factor are then
given as follows:
\begin{equation}
    a_\textsc{j,f} = \frac{\hbar^2 k^4}{16\pi G \rho_\text{dm,0} m^2}
\end{equation}
\begin{equation}
    \lambda_\text{J,F} = \left( \frac{\pi^3 \hbar^2}{ G \rho_\text{dm,0} a^{-3} m^2} \right)^{1/4}
\end{equation}
\begin{align}
    M_\text{J,F}(a) &= \frac{4\pi}{3} \rho_\text{dm,0} a^{-3} \left( \frac{\lambda_\text{J,F}}{2} \right)^3 \nonumber \\
    &\simeq 6.2 \times 10^9 \text{ M}_\odot~\left( \frac{\Omega_\text{dm} h^2}{0.12} \right)^{1/4} \left(\frac{a}{10^{-3.5}}\right)^{-3/4} m_{22}^{-3/2}
\end{align}
where $m_{22} \equiv m/(10^{-22}~{\rm eV}/c^2)$. 
We can then do a similar calculation for FDM
as we did above for SFDM-TF, to estimate
the cut-off scale by setting 
$a_\textsc{j,f} = a_\textsc{h}$, instead:
\begin{equation}
    \frac{\hbar^2 k_\text{cut,F}^4}{16\pi G \rho_\text{dm,0} m^2} = \frac{c k_\text{cut,F} a_\textsc{h}^2}{ \pi H_0 \sqrt{\Omega_\text{rad}} } = \frac{ \pi H_0 \sqrt{\Omega_\text{rad}} }{c k_\text{cut,F}}
\end{equation}
\begin{align}
    k_\text{cut,F} &= \left( \frac{16 \pi^2 G H_0 \rho_\text{dm,0} \sqrt{\Omega_\text{rad}} m^2 }{c\hbar^2} \right)^{1/5} \nonumber \\
    &\simeq 3.9 ~h ~{\rm Mpc}^{-1}~ \left( \frac{\Omega_\text{dm} h^2}{0.12} \right)^{1/5} m_{22}^{2/5} \label{eq:kcut-FDM}
\end{align}
\begin{align}
    M_\text{cut,F} &= \frac{4\pi}{3} \rho_\text{dm,0} \left( \frac{\pi}{k_\text{cut,F}} \right)^3 \nonumber \\
    &\simeq 2.3\times10^{11}~{\rm M}_\odot~\left( \frac{\Omega_\text{dm} h^2}{0.12} \right)^{2/5} m_{22}^{-6/5}
\end{align}

Given this, we can equate equations~(\ref{eq:kcut-TF}) and (\ref{eq:kcut-FDM}) to find the equivalent FDM model that cuts off at the same scale as a given SFDM-TF model, in terms of $m$ and $R_\text{TF}$. That is, the cut-off scale for a FDM model with a particle mass of $m_\text{22,cut}$ is the same as that for a SFDM-TF model with a TF radius of $R_\text{TF,cut}$ given by
\begin{equation}
    R_\text{TF,cut} \simeq 3 \text{ pc }~\left( \frac{\Omega_\text{dm} h^2}{0.12} \right)^{-2/5} m_\text{22,cut}^{-4/5}
    \label{eq:RTFversusmFDM}
\end{equation}

According to equation~(\ref{eq:RTFversusmFDM}),
the FDM model with particle mass $m_{22} = 1$ 
results in suppression of small-scale structure relative to the CDM model below a cut-off scale that is comparable to the
cut-off scale where suppression begins in the SFDM-TF regime if $R_\text{TF}\approx3 \text{ pc}$.  This is interesting, since
this fiducial value chosen for the FDM model is about the value required to flatten the density
profiles of present-day dwarf galaxies relative to the CDM model, by making solitonic cores of
radius equal to the de~Broglie length there,
of order $\sim 1\text{ kpc}$.  According to our results
in \citetalias{paper1} and here in \S\ref{sec:infall}, however, 
when a halo forms in SFDM-TF, it has a 
core-envelope structure in which the envelope
is CDM-like but the center flattens to
the profile of the $(n=1$)-polytropic SI
core, of radius $\simeq R_\text{TF}$.
As such, the value of $R_\text{TF}~\approx~3 \text{ pc}$, which our estimate above tells us will
produce a cut-off of small-scale structure
in the power-spectrum of density fluctuations
relative to that for CDM,
at a similar wavenumber to
that for the FDM model when the latter 
has $m_{22} = 1$, is very much smaller than
that $\simeq 1\text{ kpc}$ solitonic core
size for the corresponding FDM model.

This estimate will guide our expectations from the detailed calculation of linear perturbation growth in the two models below. However, while it would  
be tempting to think we can use this estimate of the correspondence of the two models, with their parameters so chosen to place their small-scale
cutoffs at the same wavenumber, the actual
outcome of structure formation in the two
models depends upon the full spectrum of
perturbation modes and their cosmic evolution, 
down into the nonlinear era. 
A contrast between the two models 
may arise because the TF Jeans mass drops so 
much more rapidly with scale factor than does
the FDM Jeans mass, for example, so 
scales that are Jeans-filtered in SFDM-TF at
some epoch may be able to grow again
at later times, while there is no such
effect in FDM.  Moreover, this effect of
the ``incredible shrinking Jeans mass'' for
SFDM-TF means that, while some structures may be
underabundant relative to CDM as a result of
their suppression in the linear perturbation stage -- an effect which may be similar to
that for the corresponding FDM model --
the internal structures of haloes
that result in the
two models will differ considerably, if 
FDM solitonic cores in that model are much
larger than the polytropic cores in the 
corresponding SFDM-TF haloes.   The true
comparison will require that we first derive 
the transfer function for the full spectrum of modes,
which we do in the next section,
before we can compare and distinguish the actual impact on galaxy and large-scale structure formation of the two models.

\subsection{Linear perturbations in SFDM-TF}
\label{sec:LinPerts}

We are now ready to calculate the growth of linear perturbations for all modes of interest to us in the formation of galaxies and large-scale structure, and from this, the transfer function
for the SFDM-TF model, for comparison with those 
of the CDM and FDM models.   We do this semi-analytically, by adopting several simplifying assumptions.  
The full treatment without approximation would require solving the relativistic 
Klein-Gordon-Einstein equations of motion for the scalar field, in a FLRW background universe, along with all other components of the standard cosmological model (i.e. radiation, baryons, neutrinos). This would be necessary
to be able to consider the
smooth entry of modes into the horizon, in general,
from well-outside to deep within the horizon.  
This full treatment would also need to incorporate the non-standard expansion history which results from the evolution of the SFDM EOS, as described by \citet{Li14,Li17}, which depends upon the contribution of SFDM to the total energy density at each moment and, hence, depends on the value of particle parameters $m$ and $g/m^2$.  In the most general case, the field can be either real or complex.
In both cases, the homogeneous field can be described as a perfect fluid, which makes possible
a fluid description for its evolution.  However,
when treating perturbations in the complex case,
the perfect fluid description can break down in
its stiff phase.   We will, therefore, 
greatly facilitate the semi-analytical solution
by the following approximations:

\begin{enumerate}[labelwidth=2em, leftmargin=2em, listparindent=\parindent, label=(\arabic*)]
    \item We will assume that SFDM model parameters are
    constrained such that the structure formation era,
    when all modes of interest to us enter the horizon
    and evolve, is insignificantly affected by 
    the ``back-reaction'' of the SFDM and its EOS
    evolution on the expansion history of the universe.
    It turns out that this is a very good approximation, since the departures from standard expansion history during this structure formation era are related to the transitions which occur between
    the EOS from matter-like ($w=0$) to 
    radiation-like ($w=1/3$) and, thence,
    either to cosmological-constant-like ($w=-1$) 
    or stiff ($w=1$), as one moves 
    backward in time.  The TF regime under consideration here ensures that the structure formation era is confined to the rapid oscillation phase of the scalar field, and the only transition
    that takes place during this period is that between radiation-like and matter-like.  When that happens, as we said before, it must take place early enough that it does not perturb
    the value of $a_\text{eq}$ inferred from the measured CMB anisotropy.  As pointed out
    in \citet{Li14,Li17},
    this means there is an upper bound to the value of the SI strength
    in terms of $g/m^2$, and, hence, of the value of $R_\text{TF}$.
    In particular, $g/m^2c^4 \leq 4 \times 10^{-17} \text{ eV}^{-1} \text{ cm}^3$ is required
    to satisfy $w(z_\text{eq})\leq0.001$, which
    is equivalent to requiring that $R_\text{TF} \leq 5 \text{ kpc}$. 
    When this bound is placed, there
    is a corresponding limitation placed on the
    contribution of the SFDM radiation-like energy density to the total during the RD era, which
    means that SFDM does not appreciably alter the
    expansion rate of the universe then.
    
    \item We will assume that SFDM behaves as a perfect
    fluid, so a fluid description is possible, with
    pressure and energy density related accordingly, by
    the EOS parameter in equation~(\ref{eq:averagew}). 
    
    \item We will assume that SFDM is radiation-like
    with $w = 1/3$ at $a\leq a_t$ and matter-like with
    $w = 0$ at $a>a_t$, i.e. we assume an instantaneous transition. 
    We will take $a_t=a_{1/9}$.
    
    \item Once modes are well-within the horizon, we will treat them in the Newtonian approximation. 

\end{enumerate}

For each mode, we shall calculate as our
reference solution, that for the growth of dark matter perturbations in the CDM model.
The latter will be treated in the conventional way, as a nonrelativistic, collisionless, pressure-free gas (i.e. ``dust''). 
An analytical solution is available for CDM perturbation modes when they are sub-horizon ($a \gg a_\textsc{h}$), given by \citep[see, e.g.,][\S6.3]{Mukhanov}:
\begin{align}
    &\frac{\delta_\text{CDM}(a)}{\delta_\text{H}} = 6 \left[ \ln \left( \frac{4 y }{ x } \right) + \gamma_\textsc{e} - \frac{7}{2} \right] \left(1 + \frac{3 x}{2} \right) \nonumber\\
    & \quad \quad \quad \quad \quad- 6 \left[ \left(1 + \frac{3 x}{2} \right) \ln \left( \frac{\sqrt{1 + x} + 1}{\sqrt{1 + x} - 1} \right) - 3 \sqrt{1 + x} \right] \label{eq:deltaCDM}\\
    & x \equiv \frac{a}{a_\text{eq}} \\
    & y \equiv \frac{c k a}{H_0 \sqrt{3\Omega_\text{rad}}}
\end{align}
where $\gamma_\textsc{e} \simeq 0.577$ is the Euler-Mascheroni constant, and $\delta_\text{H}$ is the asymptotic superhorizon value for the (mass) density contrast of a given mode.
To match this analytical subhorizon solution in equation~(\ref{eq:deltaCDM}),
valid when a mode is deep within the horizon, to the solution of a more exact, fully-relativistic, numerical solution that follows the perturbation smoothly as it
enters the horizon from far outside to deep within, we
define an effective horizon-entry scale factor for each mode, $a_{\textsc{h},\text{eff}}(k)$,
that makes the two solutions agree
when the modes are deep within
the horizon. This is equivalent
to our adopting a thin-horizon approximation
such that for 
$a < a_{\textsc{h},\text{eff}}$, we treat the perturbations as frozen at their superhorizon values $\delta_\text{H}$, while for $a > a_{\textsc{h},\text{eff}}$, the perturbations grow according to equation~(\ref{eq:deltaCDM}).
To ensure that the density contrast is continuous across horizon-entry, then, we must choose $a_{\textsc{h},\text{eff}}$ such that, by equation~(\ref{eq:deltaCDM}),  
\begin{equation}
    \delta_\text{CDM}(a_{\textsc{h},\text{eff}}) = \delta_\text{H} \label{eq:aHeff}
\end{equation}
for each mode $k$. For SFDM,
as long as the scalar field is matter-like at this value of $a_{\textsc{h},\text{eff}}$ 
(i.e. $a_{\textsc{h},\text{eff}}>a_t$), 
we can also use this condition to set the horizon-entry value for an SFDM perturbation.
That is, the superhorizon and horizon-entry values for a dark matter perturbation should be independent of the dark matter model (i.e. only their subhorizon evolution should differ):
\begin{equation}
    \delta_\text{SFDM}(a_{\textsc{h},\text{eff}}) = \delta_\text{CDM}(a_{\textsc{h},\text{eff}}) = \delta_\text{H}    \label{eq:horentryamp}
\end{equation}
(Henceforth, in \S\ref{sec:LinPerts}, the horizon-entry scale factor will refer to
this $a_{\textsc{h},\text{eff}}$, unless otherwise noted.  Elsewhere, we refer more generally to ``horizon-entry'' in terms of a mode filling the Hubble volume, for which
$k = k_\text{H} \equiv \frac{\pi}{c}a_\textsc{h}H(a_\textsc{h})$.  The two definitions are quantitatively similar; we compare them in Appendix~\ref{sec:aH}, Fig.~\ref{fig:aHeff}.)

For perturbations in SFDM-TF, there are
two possible phases of behavior to consider,
according to whether the scalar field
is radiation-like or matter-like when a given
mode is inside the horizon. 
Modes that enter the horizon when the
scalar field is \textit{radiation-like} (i.e. with $a_{\textsc{h},\text{eff}} \leq a_t$)  
evolve just like those in the radiation component that dominates the total energy density.  
Since we will only consider models for which the field transitions from radiation-like to matter-like at a scale factor $a_t < a_\text{eq}$, the field will behave like radiation in a radiation-dominated universe for $a < a_t$, which has a known analytical solution, given by
\citep[see, e.g.,][\S7.3]{Mukhanov}:
\begin{align}
        \frac{\delta_\text{rad}(a)}{\delta_\text{rad}(0)} = -3 \left[ \left( \frac{2-y^2}{y^2} \right) (\text{sinc }y - \cos y) - \text{sinc } y \right] \label{eq:deltarad}
\end{align}
where $\text{sinc } y \equiv \sin y/y$.
According to this solution, such modes oscillate
like sound waves in a relativistic fluid
with sound speed $c/\sqrt{3}$, with constant
amplitude.  This analytical solution is fully-relativistic and
properly follows the perturbations from far
outside the horizon through horizon entry, so
the time at which oscillations begin as a
function of wavenumber is not arbitrary, but
reflects the synchronization of horizon
entry for all modes\footnote{Our only
approximation in this case is that, in practice,
we assign $a_t$ as the scale factor at which the average EOS parameter $w$ for the scalar field equals $1/9$, and assume that SFDM instantaneously transitions from radiation-like to matter-like at $a_t$.} of a
given $k$.   

For modes that are inside the horizon when
the scalar field is \textit{matter-like}, 
instead \big(i.e. whenever $a > \max[a_{\textsc{h},\text{eff}},a_t]$\big), 
each mode evolves thereafter according to the following differential equation \citep{SM11,Chavanis12}:
\begin{equation}
    \Ddot{\delta} + 2H\Dot{\delta} + \left( \frac{ \hbar^2 k^4}{4 m^2 a^4} + \frac{\bar{c}_{s,\text{SI}}^2 k^2}{a^2} - 4\pi G \rho_\text{dm,0} a^{-3} \right) \delta = 0
    \label{eq:linpertgen}
\end{equation}
where overdots denote differentiation with respect to cosmic time $t$, and $\bar{c}_{s,\text{SI}}$ is the sound speed of the background associated with SI pressure, given by:
\begin{equation}
    \bar{c}^2_{s,\text{SI}} = \frac{g}{m^2} \rho_\text{dm,0} a^{-3}
    \label{eq:SI_soundspeed_bg}
\end{equation}
The first term inside the parentheses in equation (\ref{eq:linpertgen}) is the quantum pressure term, while the second is the effect of SI pressure.  The third term
is responsible for gravitational growth.
The negative sign
in front of that third term means that the first two, both positive-definite, 
act to oppose gravitational growth. The sum of all three terms inside the parentheses can be zero for a certain value of wavenumber, which we
shall call the generalized Jeans wavenumber, 
$k_\text{J}$.  Modes with wavenumber $k<k_\text{J}$ are able to grow, while those with
$k>k_\text{J}$ oscillate 
like sound waves for which
the evolution of their amplitude depends upon
which of those first two terms dominates. 
We note that the scale factor dependences of
those two first terms are different, so,
in principle, if one term dominates at some
time, the other may dominate at other times.

In the TF regime, the
regime of primary interest to us here, we can
take the limit in which ${\hbar k^2 / H m \to 0}$
in equation~(\ref{eq:linpertgen}) above
and drop the first term inside the
parentheses. This means that, in this
linear regime, SI dominates and
quantum pressure is unimportant.  This contrasts
the linear perturbation regime with the nonlinear
regime in which inhomogeneities even on the scale of the de~Broglie wavelength (and their contribution to quantum pressure) cannot be neglected, since they are responsible for transporting momentum and, thereby, affecting larger scales,
even in a spatially-averaged description, as we discussed in \citetalias{paper1}.  
In this linear regime, then, subhorizon
perturbations in SFDM-TF 
during its matter-like phase obey the following
equation:
\begin{equation}
    \Ddot{\delta} + 2H\Dot{\delta} + \left( \frac{\bar{c}_{s,\text{SI}}^2 k^2}{a^2} - 4\pi G \rho_\text{dm,0} a^{-3} \right) \delta = 0
    \label{eq:linpert}
\end{equation}
This equation is the origin of our expression
for the proper TF Jeans wavelength $\lambda_\text{J,0}$ in
\S\ref{sec:LinPertOverview}, which we obtain by solving for the comoving Jeans wavenumber ($k_\text{J,0} = 2\pi a/\lambda_\text{J,0}$)
that makes the terms inside the parentheses in equation~(\ref{eq:linpert}) sum to zero:
\begin{equation}
    k_\text{J,0} = a\sqrt{\frac{4\pi G m^2}{g}} = \frac{\pi a}{R_\text{TF}}
\end{equation}
According to equations~(\ref{eq:linpertgen}) and (\ref{eq:SI_soundspeed_bg}),
if we are in the TF regime at the end
of the evolution period of interest,  
then we were always in the TF regime at all earlier times, as well,
so equation~(\ref{eq:linpert}) remains valid
throughout this phase of matter-like EOS.  

In practice, we solve equation~(\ref{eq:linpert}) for $\delta \equiv \delta_\text{SFDM}(a)/\delta_\text{H}$.
The remaining uncertainty is then about what initial condition to impose on the solution. There are two regimes:
\begin{enumerate}[labelwidth=2em, leftmargin=2em, listparindent=\parindent]
    \item $a_{\textsc{h},\text{eff}} > a_t$ -- modes enter the horizon when the field is matter-like:
    
    In this case, since all dark matter models should behave the same for superhorizon modes, we use the CDM horizon entry amplitude and slope
    \begin{align}
        &\delta_\text{SFDM}(a_{\textsc{h},\text{eff}}) = \delta_\text{CDM}(a_{\textsc{h},\text{eff}}) \label{eq:delICmatter}\\
        &\delta_\text{SFDM}'(a_{\textsc{h},\text{eff}}) = \delta_\text{CDM}'(a_{\textsc{h},\text{eff}}) \label{eq:delprimeICmatter}
    \end{align}
    or, equivalently,
    \begin{align}
        &\delta(a_{\textsc{h},\text{eff}}) = 1 \label{eq:delnormICmatter}\\
        &\delta'(a_{\textsc{h},\text{eff}}) = \delta_\text{CDM}'(a_{\textsc{h},\text{eff}})/\delta_\text{H} \label{eq:delprimenormICmatter}
    \end{align}

    \item $a_{\textsc{h},\text{eff}} \leq a_t$ -- modes enter the horizon when the field is radiation-like:
    
    In this case, since SFDM should behave like radiation prior to $a_t$, we use the (normalized) amplitude and slope of the radiation solution at $a_t$
    \begin{align}
        &\delta(a_t) = \delta_\text{rad}(a_t)/\delta_\text{rad}(0) \\
        &\delta'(a_t) = \delta_\text{rad}'(a_t)/\delta_\text{rad}(0)
    \end{align}
\end{enumerate}

\subsection{The transfer function: SFDM-TF vs. CDM}
\label{Transferfunction}

It is customary to describe the growth of density fluctuations in alternative dark matter models to CDM in terms of a \textit{normalized} transfer function, defined as follows:

\begin{align}
    T_\text{SFDM}^2(k,a) &\equiv 
    \frac{\mathcal{T}^2_{\text{SFDM},k}(a,a_i)}{\mathcal{T}^2_{\text{CDM},k}(a,a_i)} =
    \frac{P_\text{SFDM}(k,a)}{P_\text{CDM}(k,a)} \nonumber\\
    &= \left|\frac{\delta_\text{SFDM}(k,a)}{\delta_\text{CDM}(k,a)}\right|^2
\end{align}
(see also equation \ref{eq:dP}),
where $P(k,a)$ is the linear-fluctuation power spectrum of the dark matter for each model, as labeled.

\subsection{Comparison with FDM}

As a proof of principle, we can also compute the 
transfer function for the FDM model, 
using a procedure similar to that outlined above for SFDM-TF, and compare our results to those of the publicly-available code \textsc{axioncamb}\footnote{https://github.com/dgrin1/axionCAMB}
by \cite{axionCAMB}.
This code is presumably more accurate since it
solves the fully-relativistic perturbation 
equations numerically, from superhorizon through  
subhorizon regimes for each $k$, including their coupling
to fluctuations in other components of energy-density, as
well. To derive the transfer 
function for FDM, 
we drop the SI-term in equation~(\ref{eq:linpertgen}):  
\begin{equation}
    \Ddot{\delta} + 2H\Dot{\delta} + \left( \frac{ \hbar^2 k^4}{4 m^2 a^4} - 4\pi G \rho_\text{dm,0} a^{-3} \right) \delta = 0
    \label{eq:linpertFDM}
\end{equation}
This equation for self-gravitating perturbations 
in the free-field (no-SI) case 
was also discussed elsewhere, e.g. \cite*{KMZ}.
Once again, we obtain the Jeans wavelength and wavenumber by setting the terms in parentheses equal to zero:
\begin{equation}
    k_\text{J,F} = \left( \frac{16\pi G \rho_\text{dm,0} m^2 a}{\hbar^2} \right)^{1/4}
\end{equation}

There are three noteworthy distinctions between the FDM and SFDM-TF models with regards to the solutions of their linear perturbation equations:
\begin{enumerate}[labelwidth=2em, leftmargin=2em, listparindent=\parindent, label=(\arabic*)]
\item when an FDM mode is sub-Jeans, it oscillates with a constant amplitude, whereas a sub-Jeans SFDM-TF mode oscillates with an amplitude that grows like $a^{1/4}$;
\item the FDM Jeans mass shrinks much more slowly over time 
than that for SFDM-TF \big($M_\text{J,F} \propto a^{-3/4}$ vs. $M_\text{J,0} \propto a^{-3}$\big),
so sub-Jeans modes of SFDM-TF can become super-Jeans
sooner and with more time left to grow like CDM; and
\item in the FDM limit, the radiation-like phase is absent,
so the initial conditions are set by equations~(\ref{eq:delICmatter}) - (\ref{eq:delprimenormICmatter}) for all modes.
\end{enumerate}
The first two of these distinctions explain why FDM has a sharper cut-off 
in the transfer function 
than SFDM-TF does.\footnote{The sharp cut-off in the 
transfer function for FDM was also discussed by \cite*{Hu2000},
an important early contribution to this subject. 
They reported an approximate fitting formula for the solution for different wavenumbers 
$k$ at a given scale factor, parameterized in terms of the ratio of that wavenumber to the Jeans wavenumber at that scale factor, $k/k_\text{J}$, which differs somewhat from the solutions presented here, which are numerical solutions of equation~(\ref{eq:linpertFDM}). 
While they do not give the linear perturbation equations they solved numerically, 
they say they treated FDM as a perfect fluid 
with an effective sound speed, which, in its 
rapid-oscillation phase of matter-like EOS, 
is given by
$c_{s,\textsc{f}}^2=\hbar^2 k^2/\big(4 m^2 a^2\big)$. If so, then the dependence of the
Jeans wavenumber on scale factor should agree with ours derived by setting the bracket in our 
equation~(\ref{eq:linpertFDM}) to
zero, namely $k_\text{J}\propto a^{1/4}$, in \textit{both} the RD and MD eras.  
While agreeing with this for the MD era, however, \cite{Hu2000} states 
that $k_\text{J}$ is independent of $a$ during the important RD era, which is not correct.
}

\subsection{The Halo Mass Function}
\label{sec:HMF}

The linear transfer function for the
SFDM-TF model computed above can be used to
calculate the halo mass function (``HMF'') -- defined as the comoving number density of haloes 
per unit logarithm of halo mass $M$, $\frac{dn}{d\ln M}$.  There is an extensive literature on
the various choices for analytical and empirical
prescriptions for computing the HMF from the
transfer function so as to predict the results of
cosmological N-body simulation and their analysis
by different halo-finding algorithms \citep[see, e.g.,][and references therein]{Knebe13,Diemer18}. As our purpose here is to compare and contrast the HMF of
the SFDM-TF model with those of CDM and FDM, and
there are not yet N-body results for the former
with which to perform this comparison (and hardly
any for FDM, either), we will adopt the
most straightforward prescription for calculating
the HMFs semi-analytically 
for all three models, from their respective
linear power spectra $P(k,a)$, 
using the standard Press-Schechter (``PS'') formalism \citep{PS74}. 
The latter is based upon the Gaussian statistics of the initial density perturbations, smoothed on a given mass scale, and the ansatz that a given local overdensity in the smoothed, linearly-perturbed density field will collapse and virialize at the time predicted for a spherical top-hat perturbation of that initial overdensity to reach infinite density.   A more
rigorous underpinning for this remarkably successful ansatz was subsequently provided by the Excursion Set Theory of \citet{Bond91}, including derivation of the PS formula for the HMF for the case of ``sharp $k$-space filtering'', about which we will say more below.

The PS HMF depends, not only on the underlying
power spectrum of the initial, Gaussian-random density fluctuations, 
but also on the shape of the function adopted to
smooth the density field.  The purpose of this
smoothing function is to squelch 
high-frequency modes that correspond to fluctuations of wavelength smaller than that which encompasses a given halo mass scale $M$, on average.  The mass variance $\sigma_\textsc{m}^2(M,a)$ of the linear overdensity field smoothed in coordinate space with a spherical window function 
$W(\bm{x},R)$ -- where $R$ is the comoving radius of a sphere in the unperturbed
background universe which encompasses the mass $M$, $R = \big[M/\big(\frac{4\pi}{3} \rho_\text{dm,0}\big)\big]^{1/3}$ -- is related to
the power spectrum $P(k,a)$ and the Fourier transform of this window function, $\mathcal{W}(k,R)$, as
follows:
\begin{equation}
    \sigma_\textsc{m}^2(M,a) = \frac{1}{2\pi^2} \int_0^\infty P(k,a) \mathcal{W}(k,R) k^2 dk
\end{equation}
The window function is usually taken to be a top-hat filter, expressed in Fourier space as 
\begin{equation}
    \mathcal{W}_\text{top-hat}(k,R) = 3\frac{\sin(kR) - kR \cos(kR)}{(kR)^3}
\end{equation}
However, for models with sharp cut-offs in the power spectrum such as FDM, a sharp-$k$ filter is preferred in order to have the cut-off reflected in the HMF \citep{KO20}:
\begin{equation}
    \mathcal{W}_{\text{sharp-}k}(k,R) =
    \begin{cases}
        1 & k < \alpha/R \\
        0 & k > \alpha/R
    \end{cases}
\end{equation}
where $\alpha$ is a free parameter which we set to 1.9 in order to make the resulting HMF agree with that computed using the top-hat filter at large mass.
The PS HMF is then given by
\begin{equation}
    \frac{dn}{d\ln M} = - \sqrt{\frac{2}{\pi}} \frac{\rho_\text{dm,0}}{M} \frac{\delta_c}{\sigma_\textsc{m}} e^{-\delta_c^2/2\sigma_\textsc{m}^2} \frac{d\ln \sigma_\textsc{m}}{d\ln M}
    \label{eq:PSHMF}
\end{equation}
where $\delta_c = 1.686$ is the critical linear overdensity for halo formation derived from the spherical top-hat collapse model.\footnote{
\citet{KO20} used $\alpha=2.5$ following the work of \citet{Benson13}, who found that value to be a good fit to warm dark matter simulations, and re-scaled $\delta_c$ by a factor of 1.195 in order to produce a sharp-$k$ filtered HMF that was consistent with the top-hat filtered HMF at large mass. In view of the fact that we are here comparing the HMF's of CDM, SFDM-TF and FDM, these choices would not be appropriate for all three models, especially in view of the strong differences we find between SFDM-TF and FDM. However, we also find that adopting these choices would not affect our general conclusions.}

\subsection{Results of linear perturbation theory: SFDM-TF vs. CDM vs. FDM}
\label{sec:results}

\subsubsection{Perturbation growth for a single mode}
\label{sec:pertresults}

\begin{figure}
    \centering
    \includegraphics[width=\columnwidth]{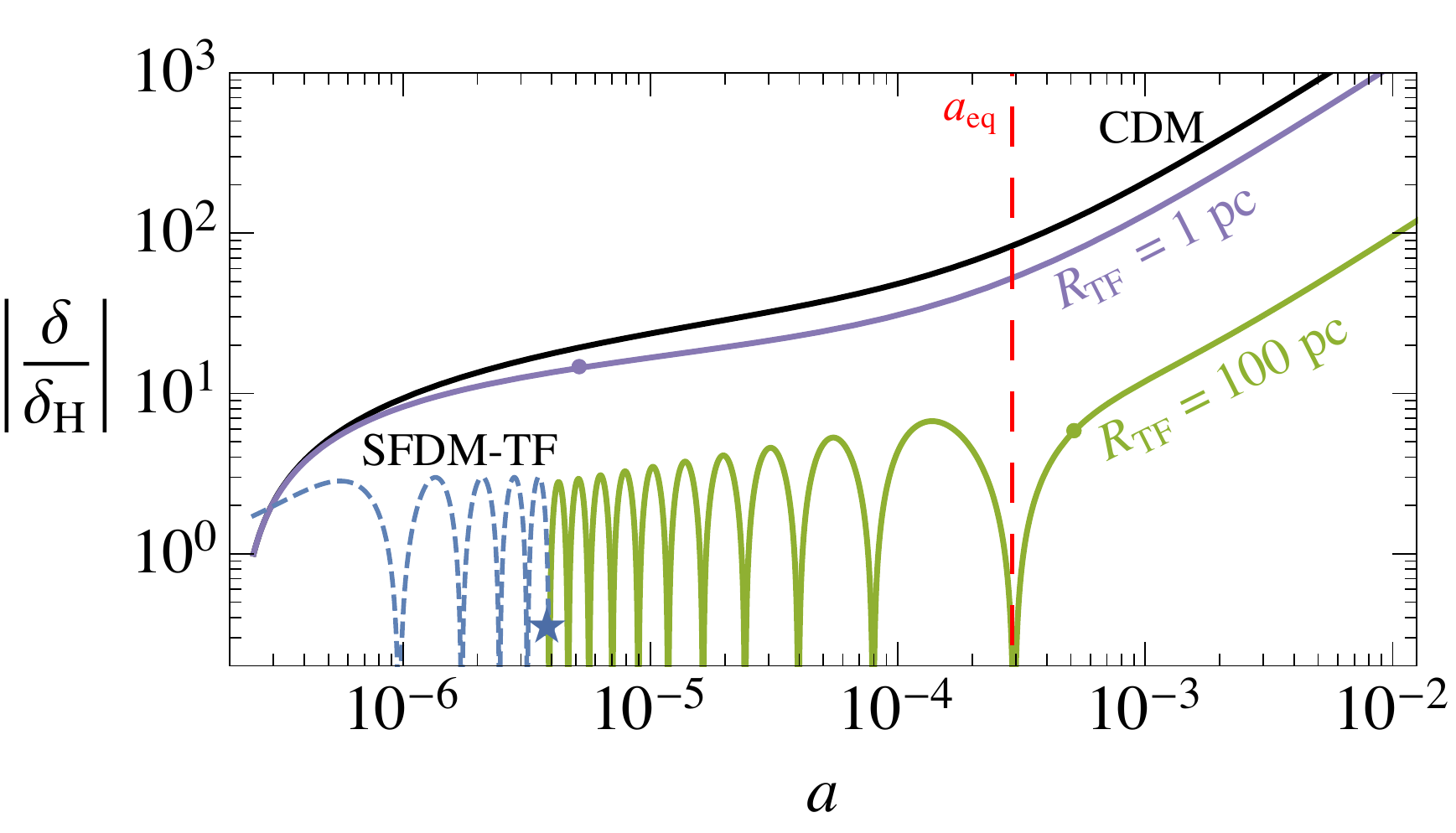}
    \includegraphics[width=\columnwidth]{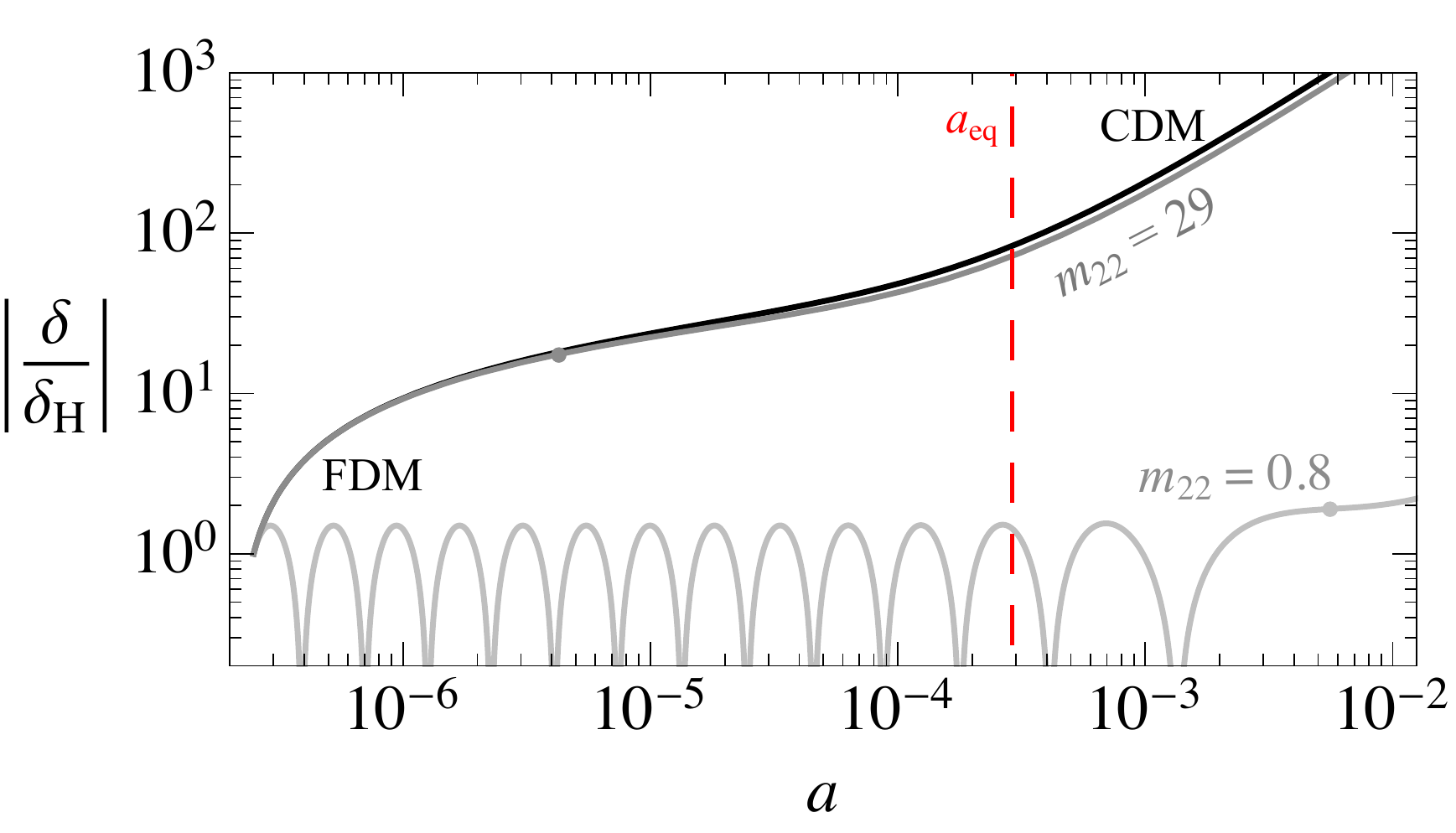}
    \caption{\textbf{Linear perturbation growth after horizon entry, for a single Fourier mode.} Solutions of 
    the differential equations for the evolution of linear-perturbation fractional overdensity $|\delta/\delta_\text{H}|$ vs. scale factor
    $a$ are plotted for a single Fourier mode,
    with wavenumber $k$ associated with a mass
    $M=10^{9} \text{ M}_\odot$ 
    \big(where $M = \frac{4\pi}{3} \rho_\text{dm,0}(\pi/k)^3$ and $k \simeq 16$ Mpc$^{-1}$\big) in the SFDM-TF (top panel) and FDM (bottom panel) models, for
    two different cases for each. 
    The purple and green SFDM-TF lines correspond to $R_\text{TF} = 1$~pc and $R_\text{TF} = 100$~pc, respectively. The light and dark grey FDM lines correspond to $m_{22} = 0.8$ and $m_{22} = 29$, respectively. The black line is the CDM solution. Dots of the same color as the curves on which they reside mark the Jeans scale factors, $a_\textsc{j,0}$ (SFDM-TF) and $a_\textsc{j,\textsc{f}}$ (FDM). The blue dashed line in the top panel shows the SFDM-TF solution for $R_\text{TF} = 100$ pc when the scalar field is radiation-like (i.e. $w=1/3$) (given by equation~\ref{eq:deltarad}), and the blue star 
    marks the transition scale factor $a_t$ for this case, after which SFDM-TF is matter-like (green solid line). The red dashed lines in both panels marks radiation-matter equality.}
    \label{fig:delta_ex}
\end{figure}

In Fig. \ref{fig:delta_ex}, we plot solutions 
of the differential equations described in \S\ref{sec:LinPerts} for the fractional linear overdensity $|\delta/\delta_\text{H}|$ vs. scale factor $a$, 
for the subhorizon evolution of 
a single, representative Fourier mode $k$. 
We compare the analytical
solution for this mode in the CDM model
with our solutions for 
SFDM-TF and FDM, for illustrative cases with different parameters for the two kinds of 
SFDM models. The wavenumber $k$ for our
representative mode is that associated with a mass $M=10^{9} \text{ M}_\odot$ \big(where $M = (4\pi/3) \rho_\text{dm,0}(\pi/k)^3$\big). 
For SFDM-TF, we present two cases, with 
$R_\text{TF} = 1$ and $100$ pc, respectively. For FDM, we present two cases
as well, with values of the particle mass
$m_{22} = 0.8$ and $29$, respectively.

The perturbation evolution for this mode in 
pressure-free CDM plotted in
Fig. \ref{fig:delta_ex} shows the familiar
transition from RD-era logarithmic growth
to the MD-era growth rate ($\delta \propto a$). 
This transition for CDM is gradual, but
is centered around $a = a_\text{eq}$. By contrast,
the perturbation evolution for both the SFDM-TF 
and FDM solutions plotted in Fig. \ref{fig:delta_ex}
shows the effects of pressure and, hence,
Jeans filtering, at early times. 
The evolution changes when this mode passes from 
sub-Jeans ($k>k_\text{J}$) to 
super-Jeans ($k<k_\text{J}$) at 
$a = a_\textsc{j}$, where the values of $a_\textsc{j}$ (marked by a dot on each curve) depend upon the model parameters and differ
for each case.
When a mode of either SFDM-TF or FDM
is sub-Jeans, its fractional overdensity oscillates
like a sound wave.  As such, 
the periods of oscillation in the two models are inversely proportional to their effective sound speeds. For SFDM-TF, this sound speed increases with $R_\text{TF}$ when the EOS is matter-like, but is a constant, independent of $R_\text{TF}$, 
when the EOS is radiation-like.  The scale
factor $a_t$ at which this transition
from radiation-like to matter-like EOS occurs
for SFDM-TF also depends on $R_\text{TF}$,
with $a_t$ increasing with $R_\text{TF}$.   
For FDM, which is always matter-like during
the era we study here, the sound speed decreases
with increasing $m_{22}$. These points allow us
to understand the behavior of the different
curves in Fig. \ref{fig:delta_ex}, as follows.

Consider, first, the curves for
$m_{22} = 0.8$ and $R_\text{TF} = 100$ pc, respectively.  For these values of $R_\text{TF}$ and $m_{22}$, 
the periods of sub-Jeans oscillation are smaller than a Hubble time during that phase of evolution. 
In that case, their fractional overdensities oscillate rapidly enough to 
pass from positive to negative and back again
a number of times, so the
curves of their absolute value displayed in Fig. \ref{fig:delta_ex} appear like trajectories
of a bouncing ball.  For FDM, 
the sub-Jeans oscillations have a constant amplitude, whereas, when the
sub-Jeans SFDM-TF mode oscillates, its amplitude
is initially constant for a time (the blue-dashed portion), during its radiation-like phase,
but then grows like $a^{1/4}$ after it transitions
at $a=a_t$ (as marked on the curve) 
to a matter-like EOS. In both cases, the oscillations cease near $a\approx a_\textsc{j}$, and perturbation
growth rates become CDM-like for $a\gg a_\textsc{j}$.

In the case of the other two 
curves in Fig. \ref{fig:delta_ex},
that for SFDM-TF with $R_\text{TF}= 1$ pc
and FDM with $m_{22} = 29$, the 
oscillatory behavior during their sub-Jeans
phase is apparently absent, but this reflects
the fact that the change of model parameters 
has pushed their sound speeds to be smaller 
and their oscillation ``periods'' to be
larger than a Hubble time.  In that case,
the background universe is expanding too fast for
the density to complete a single oscillation.
These cases therefore follow the CDM trajectory fairly closely, albeit with their growth slightly hampered due to the brief initial sub-Jeans phase.

\begin{figure}
    \centering
    \includegraphics[width=.95\columnwidth]{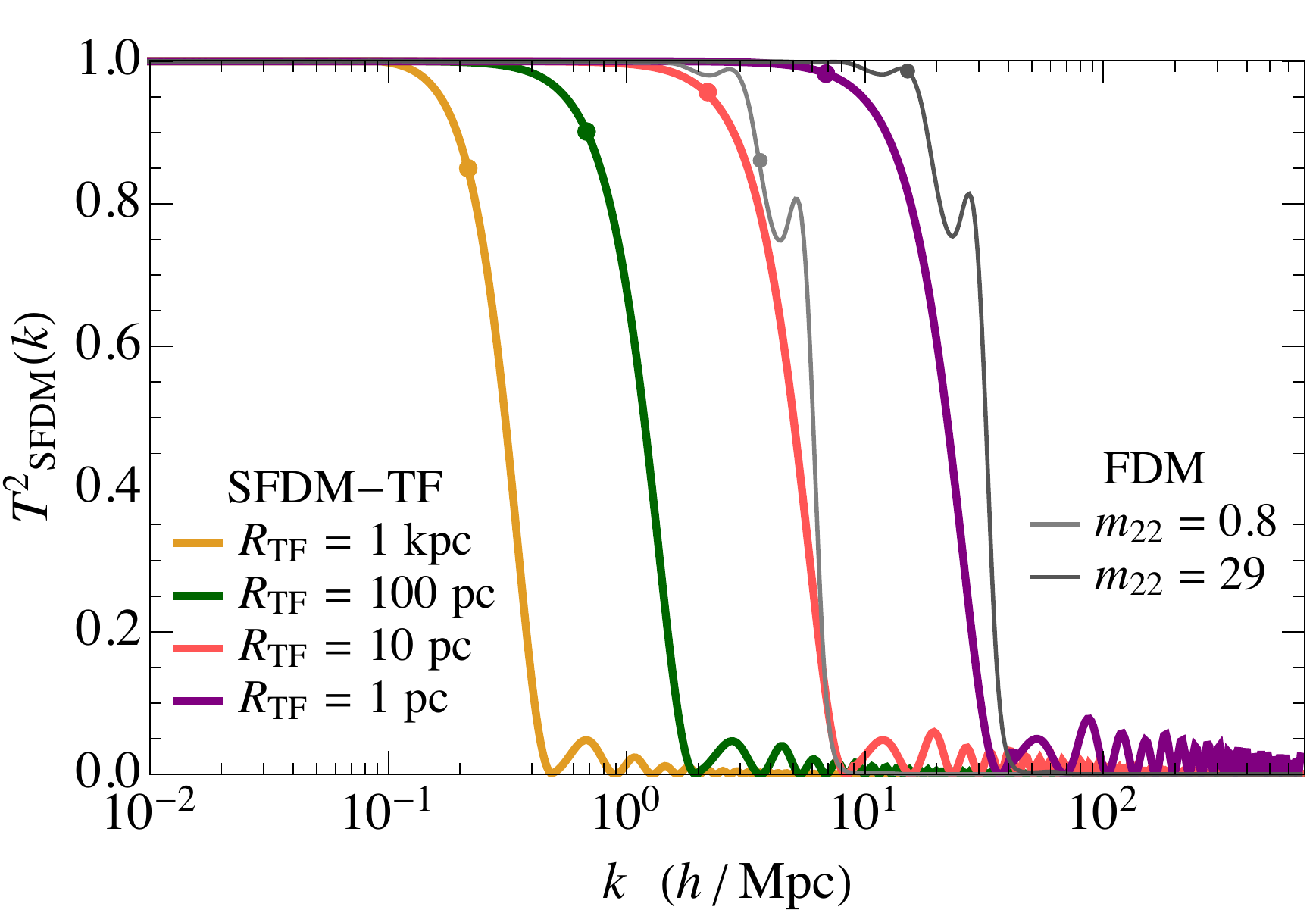}
    \includegraphics[width=.95\columnwidth]{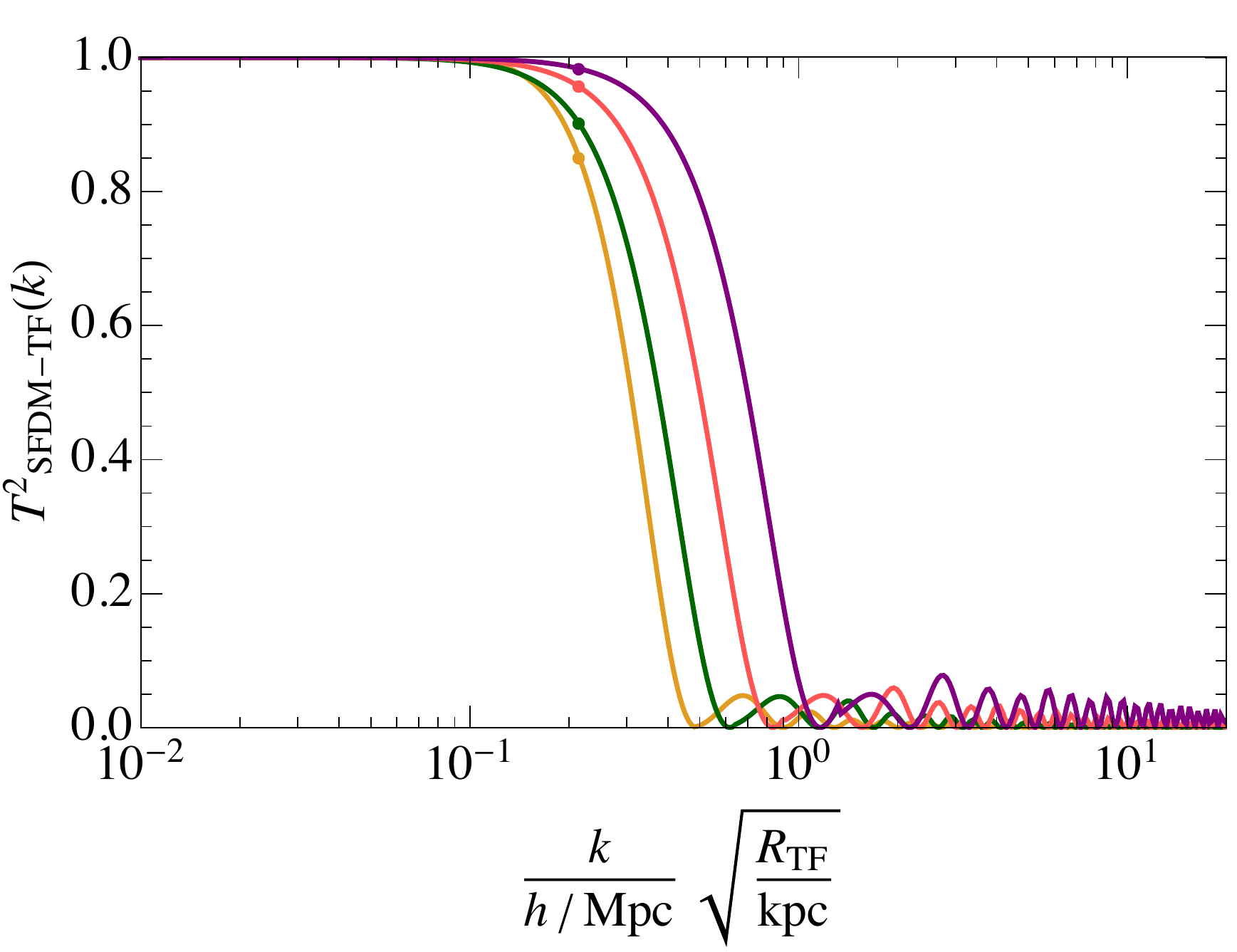}
    \includegraphics[width=.95\columnwidth]{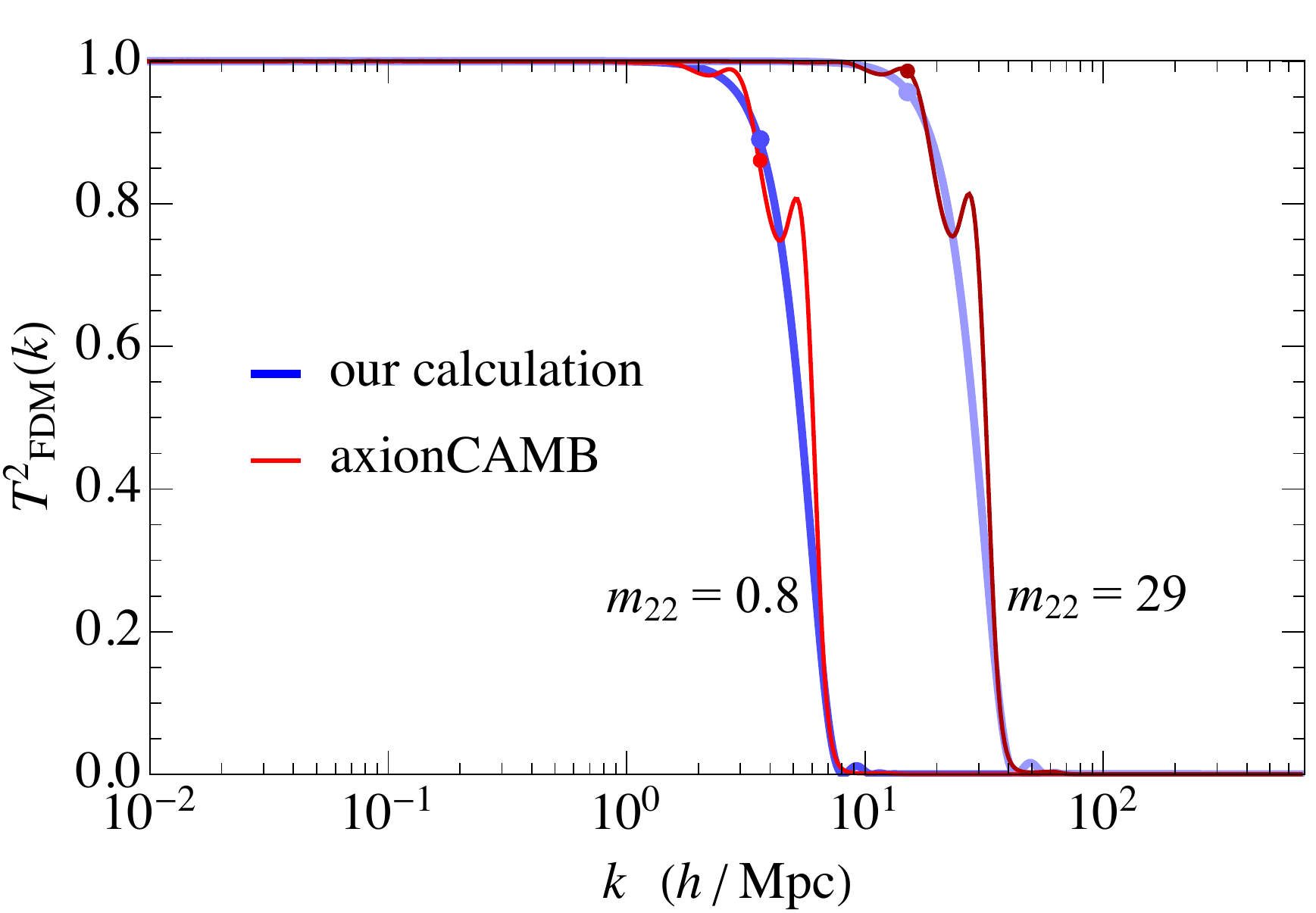}
    \caption{\textbf{Linear transfer functions.} \textbf{(a) (top panel)} Normalized transfer functions $T_\text{SFDM}^2(k,a)$ for SFDM-TF 
    vs. comoving wavenumber 
    $k$, for $R_\text{TF} = \{1000, 100, 10, 1\}$ pc (left to right). For comparison, we plot FDM transfer functions $T_\text{FDM}^2(k,a)$
    for $m_{22}= \{0.8, 29\}$, generated by \textsc{axioncamb}.
    Dots locate values of $k_\text{cut}$ derived in \S\ref{sec:cutoff}
    to estimate the cut-off wavenumber above which high-frequency modes are suppressed relative to CDM. 
    \textbf{(b) (middle panel)} SFDM-TF transfer functions in (a), 
    but with $k$-axis rescaled by dependence of $k_\text{cut}$ on $R_\text{TF}$, 
    so that locations on the horizontal axis 
    at which $k = k_\text{cut}$ are the same
    for all $R_\text{TF}$. 
    SFDM-TF curves for larger $R_\text{TF}$ start to deviate from CDM more sharply than for smaller 
    $R_\text{TF}$, so $T^2_\text{SFDM}(k_\text{cut})$ is closer to 1 for smaller $R_\text{TF}$.
    \textbf{(c) (bottom panel)} 
    FDM transfer functions   
    in (a) (light and dark red lines),
    generated by \textsc{axioncamb}, are 
    compared to our solutions of subhorizon
    linear perturbation growth for
    FDM (dark and light blue), 
    i.e. same procedure we used for
    SFDM-TF, but using the FDM linear perturbation equation (\ref{eq:linpertFDM}), instead.}
    \label{fig:transfer}
\end{figure}

\subsubsection{Normalized transfer functions}
\label{sec:transferfunctionresults}

In Figs. \ref{fig:transfer}(a) and (b), 
our 
solutions for linear perturbation growth for different $k$-modes are used to plot the 
normalized transfer functions 
$T_\text{SFDM}^2(k,a)$ for SFDM-TF vs. $k$, for several illustrative cases, with $R_\text{TF} = 1000, 100, 10, \text{and}~1$ pc (left to right). 
For comparison, 
we also plot FDM transfer functions
$T_\text{FDM}^2(k,a)$, for the same two masses, $m_{22}=0.8$ and $29$, as in Fig. \ref{fig:delta_ex}
above.   In Fig. \ref{fig:transfer}(a),
transfer functions for FDM were generated by \textsc{axioncamb}, while in 
Fig. \ref{fig:transfer}(c), we compare those
with our solutions
for FDM, to show their close agreement.
This agreement 
serves to validate our
solutions for both FDM and SFDM-TF, 
since the former were derived
by the same procedure as for our SFDM-TF 
solutions.  Dots locate values of $k_\text{cut}$ derived in \S\ref{sec:cutoff} to estimate the cut-off wavenumber above which high-frequency modes are suppressed relative to CDM.

In Fig. \ref{fig:transfer}(b) (the middle panel),
we replot the SFDM-TF transfer functions in 
Fig. \ref{fig:transfer}(a) (the top panel) 
with the $k$-axis rescaled by the dependence 
of $k_\text{cut}$ on $R_\text{TF}$, 
so that locations on the horizontal axis 
at which $k = k_\text{cut}$ are the same
for all $R_\text{TF}$. This shows that the smaller
the value of $R_\text{TF}$ is, the more gradual
is the decline relative to CDM at wavenumbers
where the SFDM-TF transfer function begins to
decline relative to CDM at high $k$.
In addition, since curves for larger $R_\text{TF}$ start 
to deviate from CDM more sharply than for smaller 
$R_\text{TF}$, $T^2_\text{SFDM}(k_\text{cut})$ 
is closer to 1 for smaller $R_\text{TF}$.
{We can also see from this figure that the amplitudes of the oscillations at high wavenumber are smaller for larger $R_\text{TF}$. 
This may be due, in part, to the fact that, for larger $R_\text{TF}$, modes with $k \gg k_\text{cut}$ have more of their sub-Jeans evolution occur during the matter-dominated era ($a > a_\text{eq}$).
Since CDM modes grow faster during this era (linearly with scale factor, as opposed to logarithmically for $a < a_\text{eq}$), the $k \gg k_\text{cut}$ SFDM-TF modes for larger $R_\text{TF}$ cases ``miss out'' on more of this comparatively rapid, linear growth phase than the corresponding modes of the smaller $R_\text{TF}$ cases, and so are more suppressed relative to CDM. }

According to our analytical estimates in \S\ref{sec:LinPertOverview}, it is possible
to choose values of $R_\text{TF}$ for SFDM-TF
and $m_{22}$ for FDM so as to facilitate their direct comparison,
by selecting values that make the
small-scale cut-offs in their respective
transfer functions occur at roughly the same 
wavenumber (see equation \ref{eq:RTFversusmFDM}).  In that case, for example, our solution for 
FDM with $m_{22} = 0.8$ can be compared with that
for SFDM-TF with $R_\text{TF} \simeq 10$ pc.
In this way, observational constraints on
FDM can also serve as a rough proxy for constraints on 
SFDM-TF, since both models have perturbation growth 
which is CDM-like at small $k$ but suppressed 
at high $k$, i.e. values of $k$ above 
the cut-off. 

This estimator for the $R_\text{TF}$-values that make
the transfer function cut-offs for SFDM-TF and 
FDM for a given $m_{22}$ ``align'' in wavenumber by
solving equation (\ref{eq:RTFversusmFDM})
should be used with caution, however. 
The \textit{shapes} of their transfer functions below the cut-off wavenumber differ somewhat. 
As a result, if different observables measure the
amount of structure on different scales, 
i.e. at different $k$, then the correspondence between 
the $m_{22}$- and $R_\text{TF}$-values that make
the two model predictions agree may depend upon which
observable is used to constrain them.  
As an example,
the two models can have very
different half-power wavenumbers, $k_{1/2}$
in their transfer functions,
even when their $k_\text{cut}$-values are identical.  

The two illustrative values of
FDM particle mass we chose for Figs. \ref{fig:delta_ex} and \ref{fig:transfer}
will appear again in later figures, too,
in which we compare the SFDM-TF and FDM models 
further.  These values roughly
span the current range of constraints in the
FDM literature, placed by various observational probes,  namely, $m_{22}=0.8$ 
(\citealp[e.g. fiducial case of ][]{VBB19} and \citetalias{paper1}) and $m_{22}=29$
\citep[e.g. lower bound from][]{Nadler21}.
For these values of $m_{22}$, de~Broglie wavelengths 
inside a present-day dwarf galaxy of mass $10^{10}\text{ M}_\odot$ with
velocity dispersion $\sigma_\text{v} \approx 30$ km s$^{-1}$, for 
example, are $\approx 4$ kpc and $0.1$ kpc, respectively, so haloes of this mass that
formed in FDM would have solitonic cores of these sizes.
If we chose $R_\text{TF}$-values of this size, 
so as to make the SI-polytropic cores inside 
SFDM-TF haloes as large as this, then
the $k_\text{cut}$-values for SFDM-TF will be
much smaller than the $k_\text{cut}$-values for
these FDM models.  The transfer function for
SFDM-TF in that case would begin to cut-off at a much higher
mass scale than FDM, so the amount of structure formed  
on such small mass scales in SFDM-TF would be well below
that in the corresponding, core-size-matched FDM model. 
On the other hand, if we align the $k_\text{cut}$-values 
of the two models, instead, which Fig. \ref{fig:transfer} tells us requires $R_\text{TF} \lesssim 1$ pc, then much
\textit{more} small-scale power is retained below the
cut-off in SFDM-TF than in FDM, as will be even
clearer from the results for their power spectra
and HMFs in the sections below.  Hence,
lower limits like those above on the 
particle masses for FDM, derived to keep FDM from \textit{underproducing} structure on small scales, 
when used to infer upper limits on
$R_\text{TF}$ based upon $k_\text{cut}$-matching 
FDM, might then be overly restrictive in limiting $R_\text{TF}$.

\subsubsection{Power spectra and RMS mass fluctuations}
\label{sec:powerspecsigmaresults}

\begin{figure}
    \centering
    \includegraphics[width=\columnwidth]{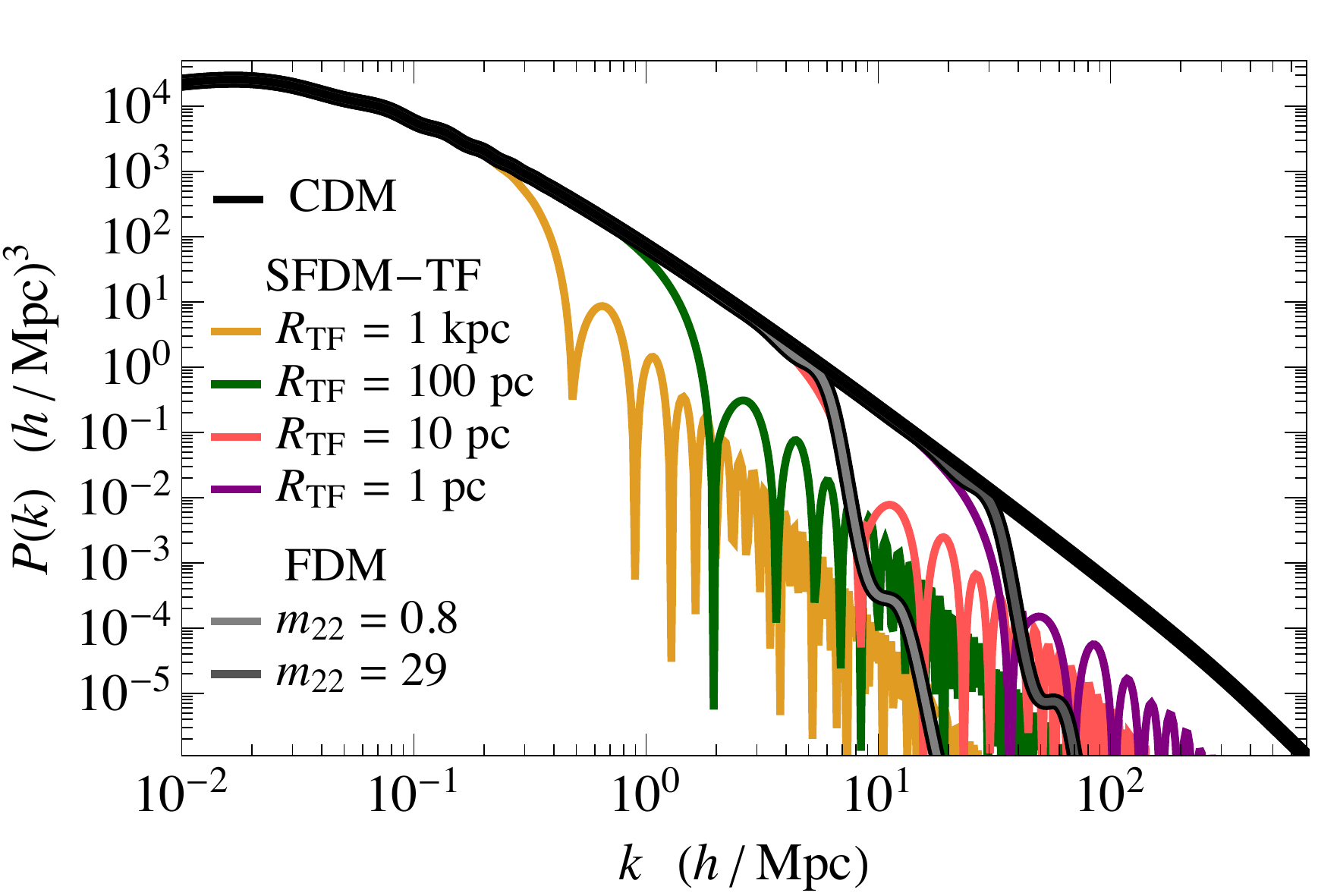}
    \includegraphics[width=\columnwidth]{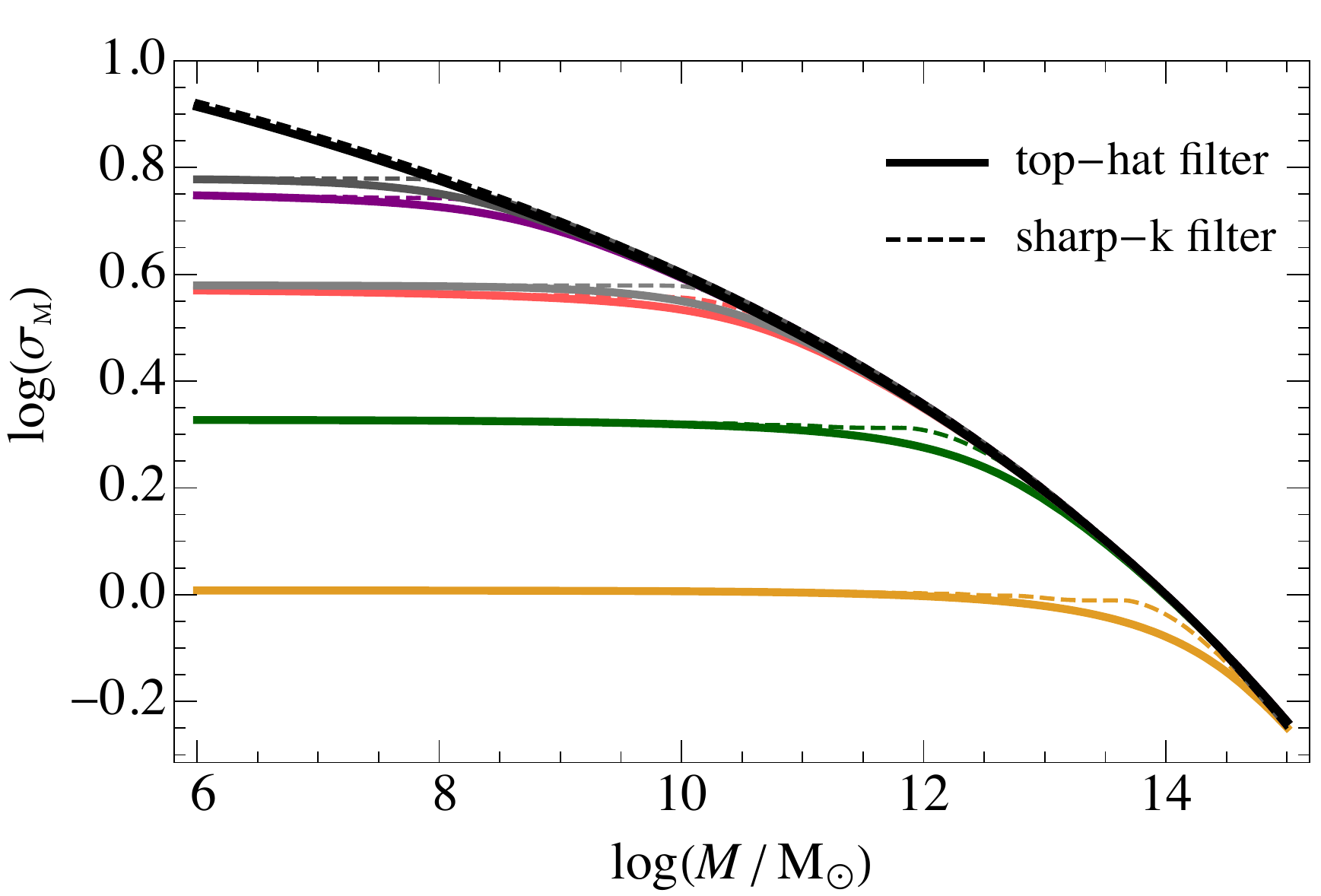}
    \includegraphics[width=\columnwidth]{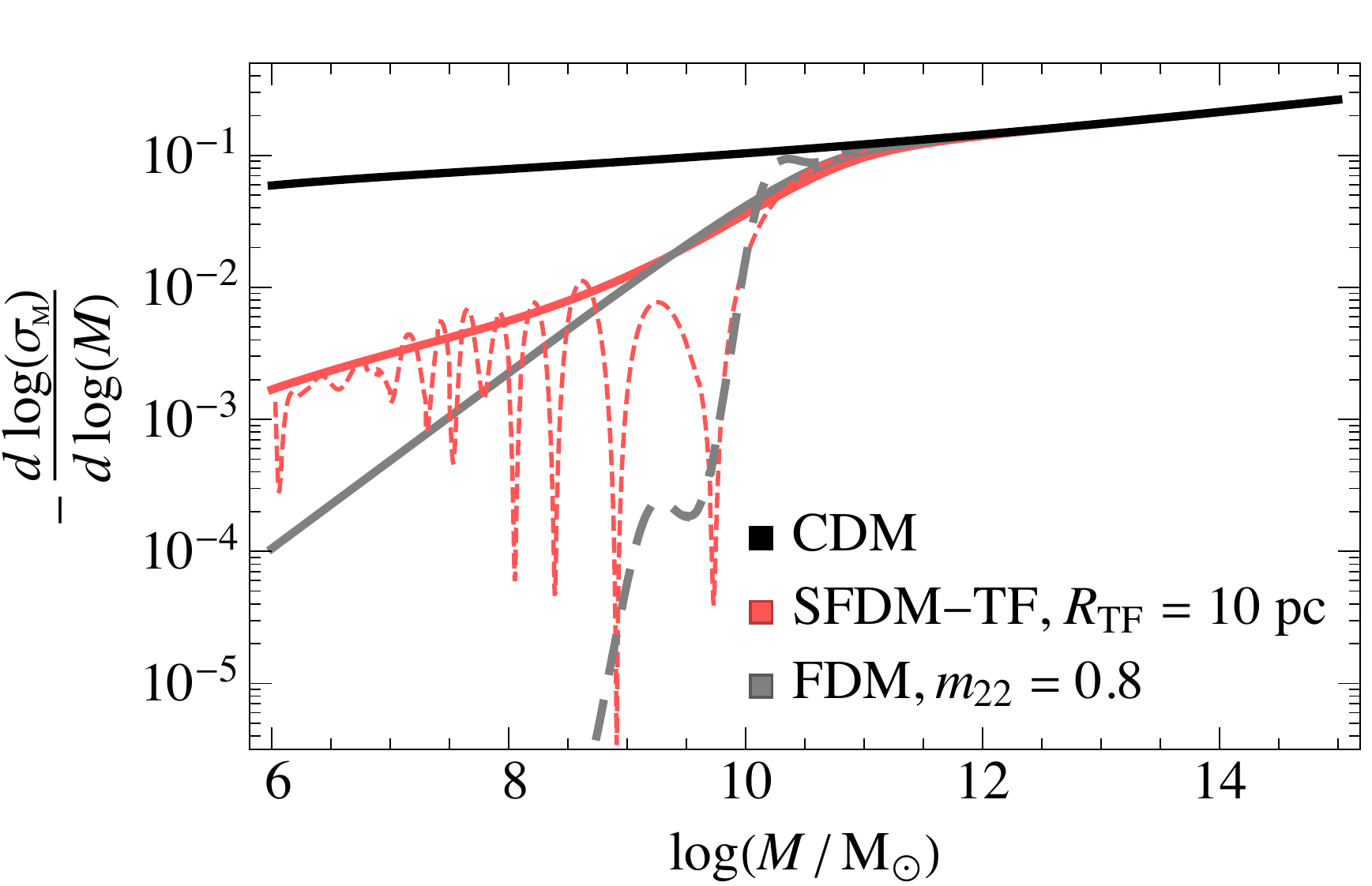}
    \caption{\textbf{Power spectrum and RMS mass fluctuation at $\bm{z=0}$}. \textbf{(a) (top panel)} Present-day matter power spectra $P(k)$, \textbf{(b) (middle panel)} RMS mass fluctuation $\sigma_\textsc{m}(M)$, and \textbf{(c) (bottom panel)} the logarithmic derivative of $\sigma_\textsc{m}$,  for CDM (black), FDM (grey), and SFDM-TF (colored) models, as labeled. The CDM and FDM curves are generated using \textsc{axioncamb}. The SFDM-TF curves are generated by multiplying the transfer functions of Fig.~\ref{fig:transfer} (obtained through our linear perturbation calculation) by the CDM power spectrum from \textsc{axioncamb}. In the middle and bottom panels, solid lines represent top-hat filtering, while dashed lines represent sharp-$k$ filtering. The logarithmic derivative is much steeper for sharp-$k$ filtering than for top-hat filtering for FDM, while for SFDM-TF, top-hat filtering captures the envelope of the rapidly-oscillating sharp-$k$-filtered result,
    so they are essentially the same.}
    \label{fig:PandSigma}
\end{figure}

In \S\ref{sec:transferfunctionresults} above,
we presented the normalized transfer function
results for SFDM-TF and FDM, which are,
by definition, expressed in units of the 
transfer function for CDM.  In Fig.
\ref{fig:PandSigma}(a), we also present the power
spectra $P(k)$ vs. $k$ for each model, separately,
by multiplying those normalized transfer functions
by the CDM transfer function generated by the
publicly-available \textsc{camb} code.
We present these power spectra, for several
illustrative values of $R_\text{TF}$ for
SFDM-TF and $m_{22}$ for FDM, respectively,
as labeled, linearly-extrapolated to $z=0$.
This takes proper account of the effect of
the smooth transition of our observed
background Universe at late times, from MD
to $\Lambda$-dominated (``$\Lambda$D''), 
since all modes of interest to us have,
by then, long-since
transitioned to pressure-free CDM-like 
perturbation growth according to linear theory.

With these power spectra, we can also compute
the RMS mass fluctuations $\sigma_\textsc{m}(M)$
and their logarithmic derivatives, 
as plotted vs. $M$ 
in Figs.~\ref{fig:PandSigma}(b) and (c).
The curves of $\sigma_\textsc{m}(M)$ for SFDM-TF
and FDM follow that for CDM at high mass, 
increasing towards smaller $M$, 
but they flatten to a 
contant value in the mass range
associated with wavenumbers of modes which are
suppressed by Jeans-filtering relative to the
growth of CDM perturbations of the same
wavenumber.  From the plots of $\sigma_\textsc{m}(M)$, alone, it is difficult
to appreciate the differences between the SFDM-TF
and FDM results, 
even when overplotting them so as to compare them 
directly, by choosing model parameters
for SFDM-TF and FDM which give them similar
values of $k_\text{cut}$.
In particular, it is difficult to see from the
curves in Fig.~\ref{fig:PandSigma}(b), the
important distinction between the behaviors of
the two models when switching between 
top-hat and sharp-$k$ filtering.  The difference
is more evident, however, when viewed by
plotting the logarithmic derivatives of $\sigma_\textsc{m}(M)$, instead, as
shown in Fig. \ref{fig:PandSigma}(c).  
For FDM, this logarithmic derivative 
at values of $M$ associated with 
wavenumbers above the cut-off in
the transfer function
approaches zero (from below) much more rapidly as $M$ decreases, for sharp-$k$ filtering than for top-hat filtering. We will see just how
significant this difference is below, 
when we apply these results to calculate the
HMF for FDM.  For SFDM-TF, however, the logarithmic
derivative shown in Fig. \ref{fig:PandSigma}(c) for top-hat filtering captures the envelope of the rapidly-oscillating behavior when a sharp-$k$-filter is adopted, instead, so, when averaged over these oscillations, the results for the
two filter functions are essentially the same.

\subsubsection{Halo mass functions}
\label{sec:HMFresults}

\begin{figure}
    \centering
    \includegraphics[width=\columnwidth]{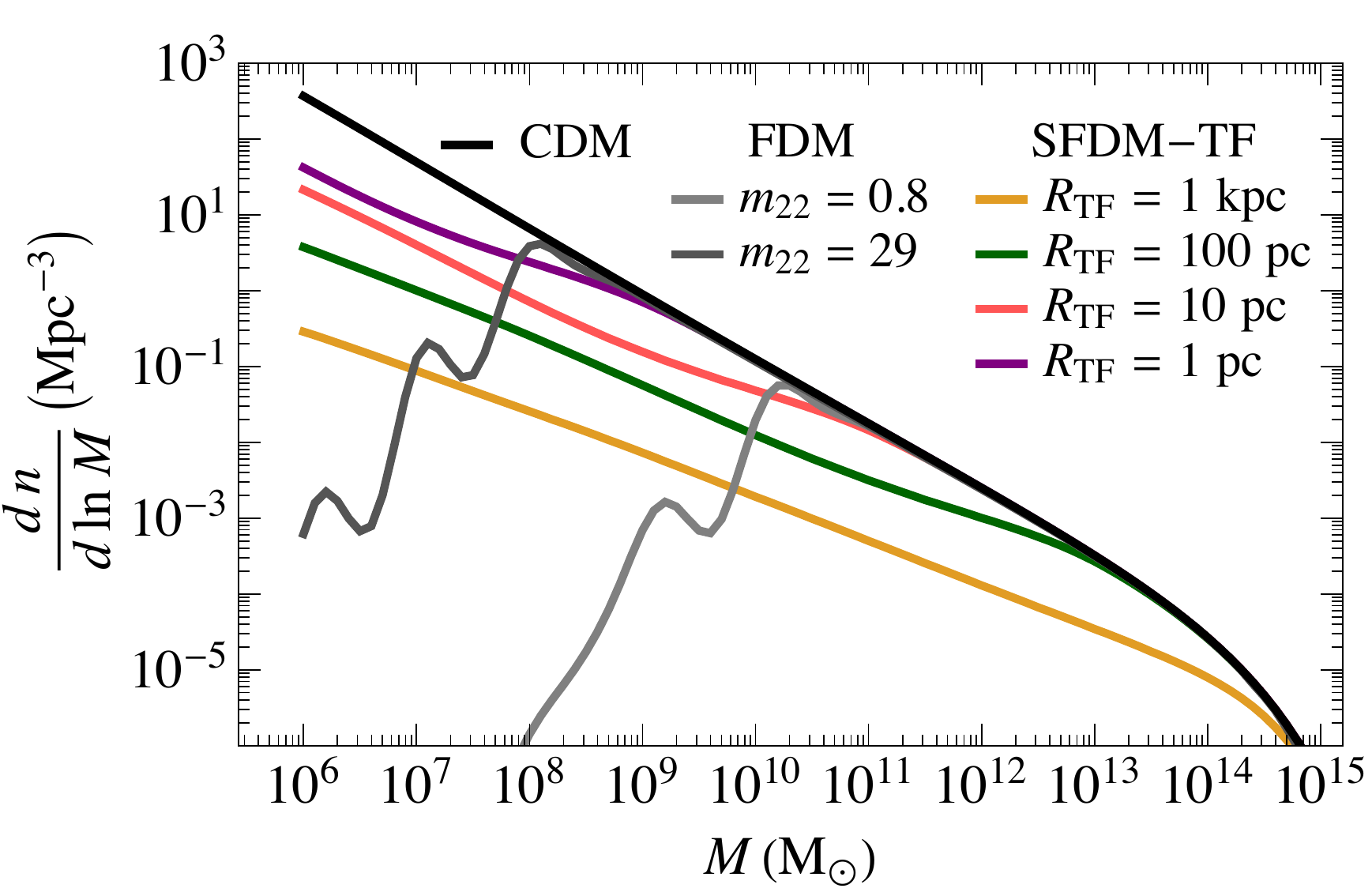}
    \includegraphics[width=\columnwidth]{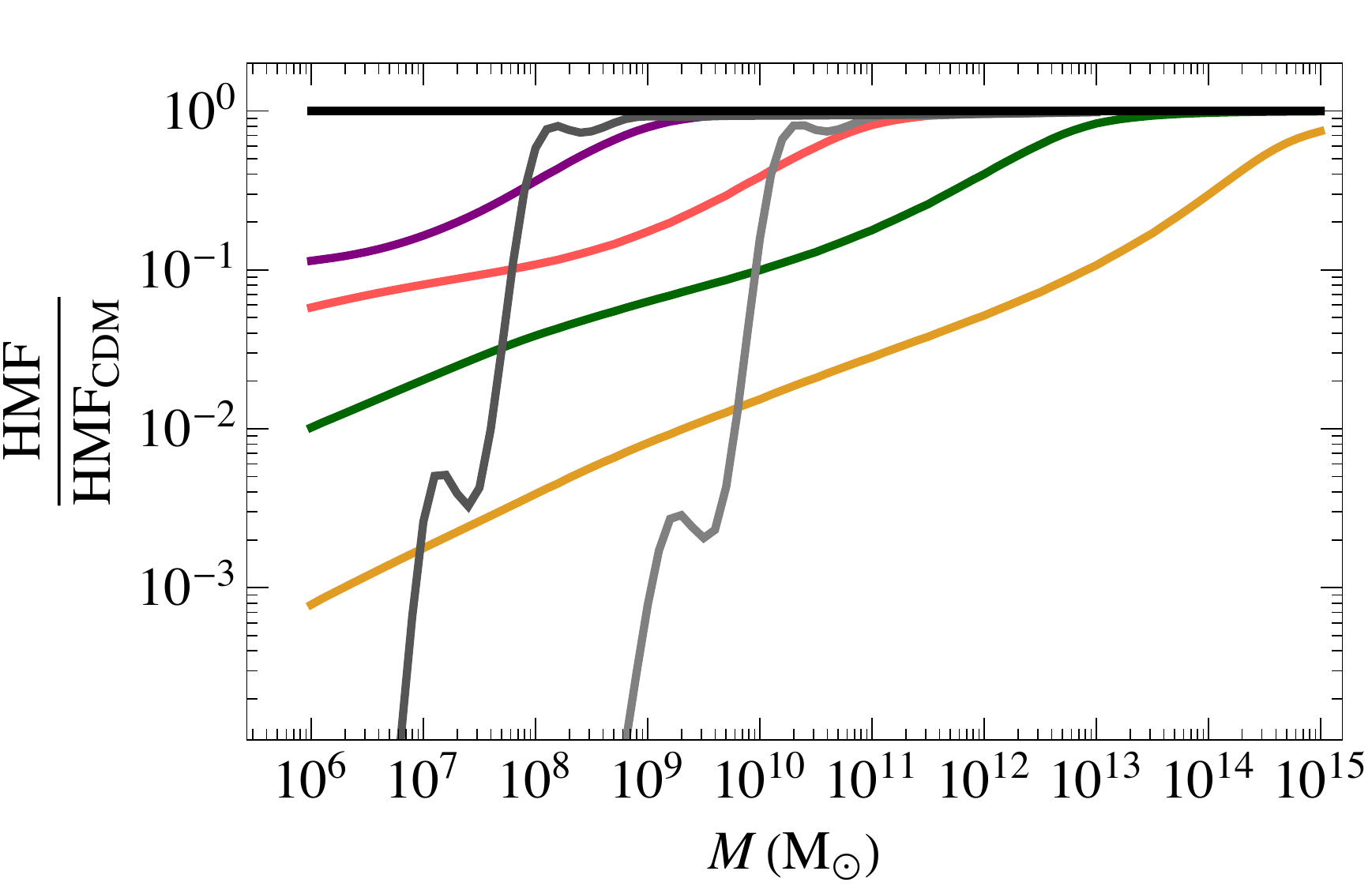}
    \caption{\textbf{Halo mass function at $\bm{z=0}$.}
    \textbf{(a) (top panel)} HMFs, 
    $dn/d\ln M$, are 
    plotted vs. halo mass $M$ for SFDM-TF, CDM
    and FDM. SFDM-TF curves are for different $R_\text{TF}$-values,  $R_\text{TF} = \{1000, 100, 10, 1\}$ pc (colored curves from lowest to highest). CDM HMF (black curve) is computed from the transfer function generated by \textsc{camb}. FDM HMFs (light and dark grey curves) are for different particle masses  $m_{22}=\{0.8, 29\}$,
    respectively, as computed from the transfer functions generated by \textsc{axioncamb}. \textbf{(b) (bottom panel)} Same as (a), but normalized by the CDM HMF.}
    \label{fig:HMF}
\end{figure}

\begin{figure}
    \centering
    \includegraphics[width=\columnwidth]{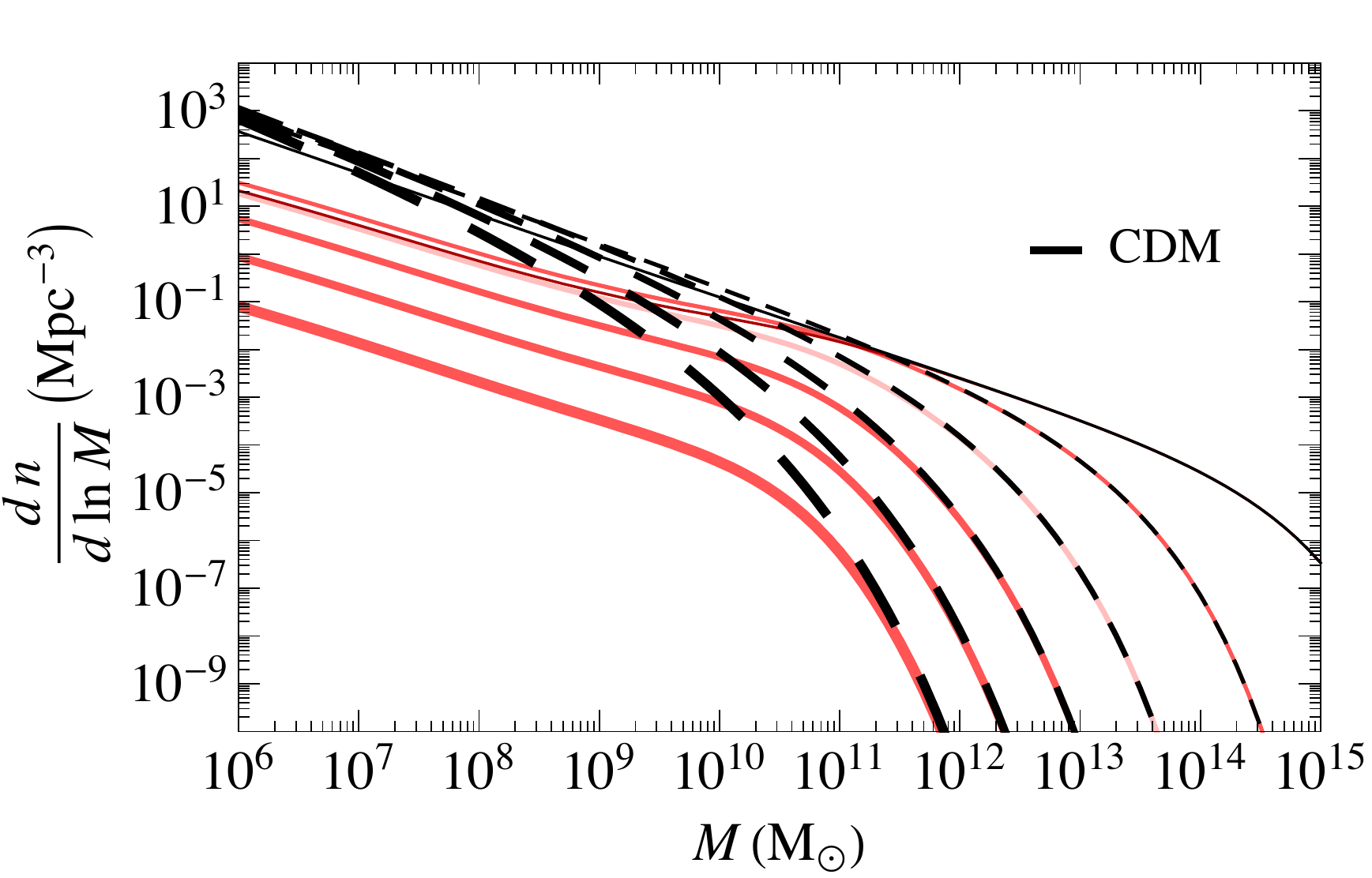}
    \includegraphics[width=\columnwidth]{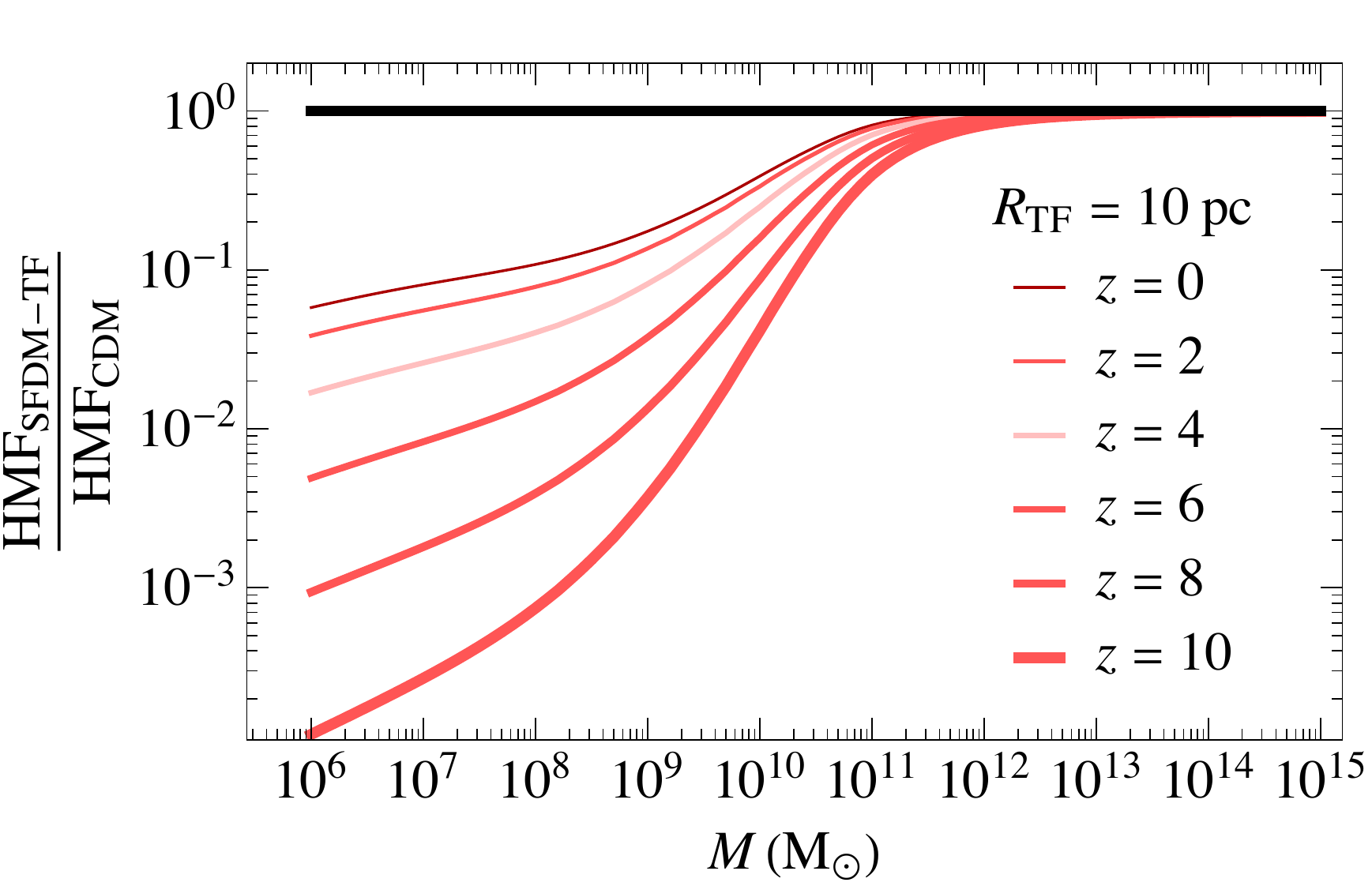}
    \includegraphics[width=\columnwidth]{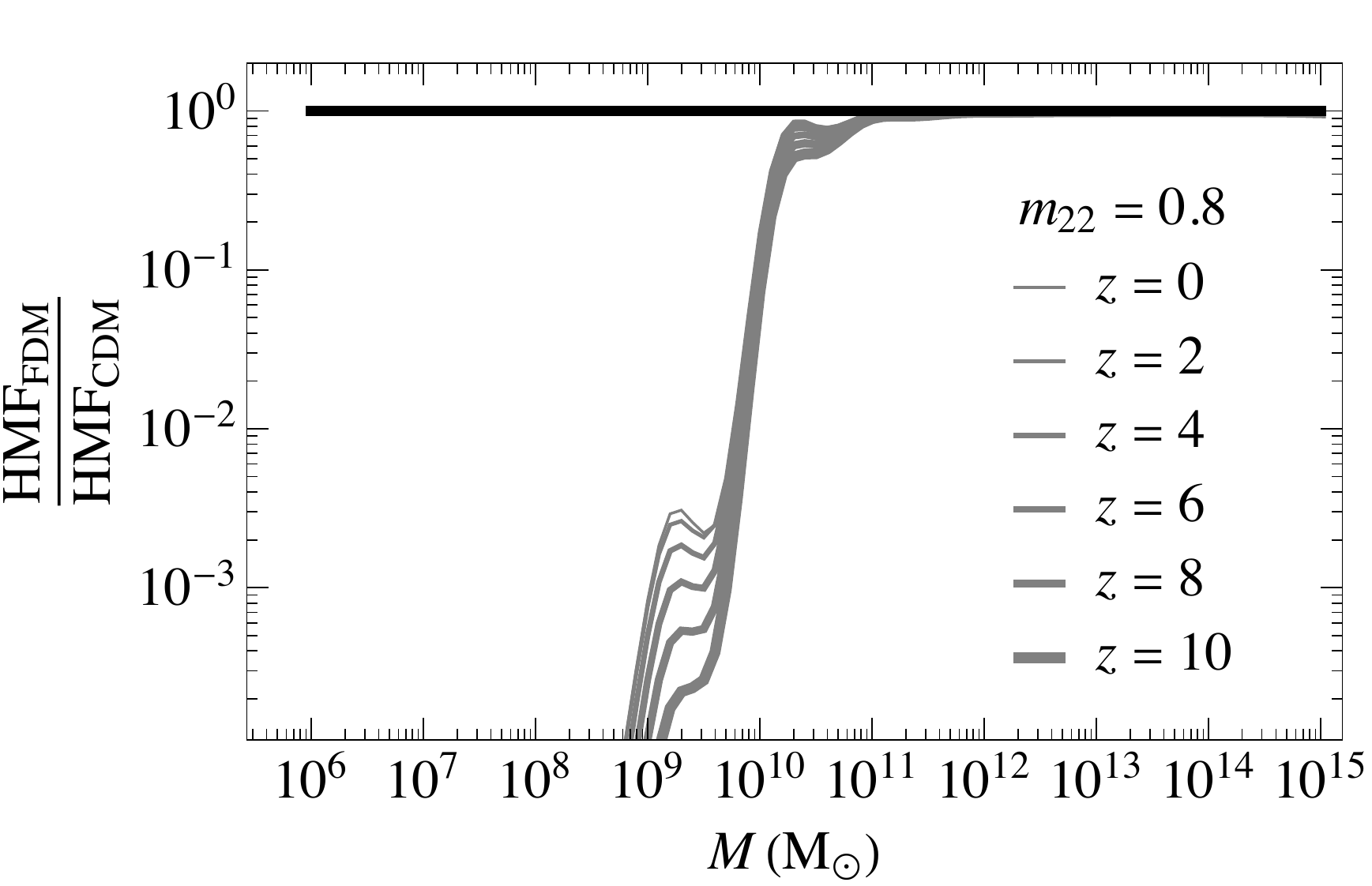}
    \caption{\textbf{Evolution of the halo mass function.} Same as Fig.~\ref{fig:HMF}, except for different redshifts as labeled, and the SFDM-TF curves are all for $R_\text{TF} = 10$~pc, while the FDM curves are all for $m_{22}=0.8$. 
    \textbf{(a) (top panel)} Same as Fig.~\ref{fig:HMF}(a) (but note the expanded $y$-axis scale), except for different redshifts and a single SFDM-TF case with $R_\text{TF} = 10$~pc.
    CDM curves are dashed for readability. \textbf{(b) (middle panel)} Same as
    Fig.~\ref{fig:HMF}(b), except for different redshifts and a single SFDM-TF case with $R_\text{TF} = 10$~pc.
    \textbf{(c) (bottom panel)} Same as (b), except for a FDM case with $m_{22} = 0.8$.}
    \label{fig:HMFevo}
\end{figure}

In Fig. \ref{fig:HMF}, we plot HMFs
at $z=0$, computed from the RMS mass fluctuations and their logarithmic
derivatives plotted in Fig. \ref{fig:PandSigma},
for SFDM-TF (for $R_\text{TF} = 1000, 100, 10$ and $1$ pc), FDM (for $m_{22}=0.8$ and 29), 
and CDM.  As expected, all three models produce
the same HMF at high mass, well above the scales
affected by Jeans-filtering. Since 
Jeans-filtering of small-scale structure produces
cut-offs at high wavenumber in the power spectra
and transfer functions of both FDM and SFDM-TF, 
cut-offs appear in both their HMFs at small halo
mass $M$, as well.  
However, as Fig. \ref{fig:HMF} makes
clear, the shape and depth of this HMF cut-off
for SFDM-TF are dramatically different
from that for FDM.  In particular, the cut-off
for SFDM-TF is much weaker and more gradual
than that for FDM.   This will have strong
consequences for the comparison between any
observables for the two models that reflect
a turn-down in their abundance of small-mass
galactic haloes relative to that in the 
standard CDM model.

The origin of this difference can be understood
from the curves for $\sigma_\textsc{m}(M)$ and
its logarithmic derivative for the two models,
in Fig. \ref{fig:PandSigma}(b) and (c). As we
noted above, the sharp-$k$-filtered result
for FDM is much more sharply cut-off at small
mass than is the top-hat-filtered result.
This is why sharp-$k$-filtered results are 
strongly preferred for this model \citep{KO20}.
This is generally true for other models with
a similar cut-off of power on
small scales, e.g. warm dark matter (``WDM''),
as well.  For this reason, we used
the sharp-$k$-filtered results to compute the
HMF for FDM in Fig.~\ref{fig:HMF}.
For SFDM-TF, however, the choice of 
filter does not make a difference in how
sharply $\sigma_\textsc{m}(M)$ cuts off.
Using the top-hat filter for SFDM-TF merely
smooths over high-frequency 
oscillations in the logarithmic derivative
of the sharp-$k$-filtered
$\sigma_\textsc{m}(M)$, but still
traces their envelope very well, so the 
sharpness of the cut-off is the same for the two filters.  \textit{Regardless} of which filter is
used, that cut-off is much more gradual
and less severe for the SFDM-TF model than for FDM.
We chose to plot the SFDM-TF HMF using the
top-hat filter, rather than the sharp-$k$-filter,
to avoid giving false prominence to the
corresponding oscillations that appear
in the HMF in this case, if the latter were used,
instead. Their prominence and phases are 
artificially related to our thin-horizon
and thin-EOS-transition approximations.  
However, our conclusion that the overall
cut-off for the SFDM-TF case is much shallower
and more gradual than for FDM is robust, 
as it is the same for both filters.   

In view of the importance of this difference in
the $z=0$ HMFs for SFDM-TF and FDM, 
we have explored the
time-dependence of this difference in Fig. \ref{fig:HMFevo}.  By comparing
the HMFs for the two models at
redshifts ranging from
$z=0$ to $z=10$, it is apparent
that the small-mass cut-off
for FDM, relative to that for
CDM, is much sharper and much 
deeper than that for SFDM-TF, 
at \textit{all} redshifts in this range.   
Moreover, there is a strong trend in the
HMF for SFDM-TF at masses below the cut-off,
for it to recover over time, gradually rising
toward that for CDM.  There is no comparable
recovery for the more
sharply-reduced HMF in the FDM case, 
sufficient to fill in
the missing halo abundance at small mass
relative to that of CDM.

\begin{table*}
    \centering
    \begin{tabular}{l|c c c c c}
        \quad\quad\quad\quad\quad\quad\quad\quad & \multicolumn{2}{c}{$10^6 \le M/\text{M}_\odot \le 10^{11}$} & \quad\quad & \multicolumn{2}{c}{$10^8 \le M/\text{M}_\odot \le 10^{11}$} \\
        \hline
        \hline
         &  $n_\textsc{lmh}$ \big(Mpc$^{-3}$\big) & $\rho_\textsc{lmh}$ \big($10^9$ M$_\odot$ Mpc$^{-3}$\big) & & $n_\textsc{lmh}$ \big(Mpc$^{-3}$\big) & $\rho_\textsc{lmh}$ \big($10^9$ M$_\odot$ Mpc$^{-3}$\big)\\
         \hline
        $R_\text{TF} = 10$ pc & 29    & 3.2 & & 1.1   & 3.0\\
        $m_{22} = 0.8$        & 0.086 & 2.6 & & 0.086 & 2.6\\ \\
        $R_\text{TF} = 1$ pc  & 61    & 7.1 & & 4.2   & 6.7\\
        $m_{22} = 29$         & 7.6   & 7.0 & & 6.0   & 6.9\\
        \hline
        & $n_\textsc{lmh}/n_{\textsc{lmh},\textsc{cdm}}$ & $\rho_\textsc{lmh}/\rho_{\textsc{lmh},\textsc{cdm}}$ & & $n_\textsc{lmh}/n_{\textsc{lmh},\textsc{cdm}}$ & $\rho_\textsc{lmh}/\rho_{\textsc{lmh},\textsc{cdm}}$\\
        \hline
        $R_\text{TF} = 10$ pc & 0.07   & 0.32 & & 0.14  & 0.39\\
        $m_{22} = 0.8$        & 0.0002 & 0.26 & & 0.011 & 0.33\\ \\
        $R_\text{TF} = 1$ pc  & 0.14   & 0.71 & & 0.55  & 0.87\\
        $m_{22} = 29$         & 0.017  & 0.70 & & 0.79  & 0.90\\
        \hline
    \end{tabular}
    \caption{\textbf{Integrated halo number density and mass density of low-mass haloes} (``LMH'')  in the range $10^6 \le M/\text{M}_\odot \le 10^{11}$ (left 2 columns) and $10^8 \le M/\text{M}_\odot \le 10^{11}$ (right 2 columns) at $z=0$ for SFDM-TF and FDM, including their ratios to the CDM quantities, for
    illustrative values of $R_\text{TF}$ and $m_{22}$.}
    \label{tab:halocounts}
\end{table*}

For some purposes, when comparing predictions of
``cut-off models'' like these to CDM and to
observations, the integrated HMFs over some 
low-mass halo (``LMH'') mass 
range affected by the cut-off is 
also of interest.  In
Table \ref{tab:halocounts}, we show
the integrated halo number and
mass densities, given by
\begin{equation}
 n_\textsc{lmh} = \int \frac{dn}{d \ln M} d\ln M = \int \frac{1}{M}\frac{dn}{d \ln M} dM
 \label{eq:nlmh}
\end{equation}
and 
\begin{equation}
 \rho_\textsc{lmh} = \int \frac{dn}{d \ln M} d M
 \label{eq:rholmh}
\end{equation}
evaluated at $z=0$, for halo masses in the ranges $10^6 \le M/\text{M}_\odot \le 10^{11}$ and $10^8 \le M/\text{M}_\odot \le 10^{11}$,
for two illustrative sets of
pairings of model parameters
for SFDM-TF vs. FDM, $R_\text{TF}=10$ pc vs.
$m_{22}=0.8$, and 
$R_\text{TF}=1$ pc vs.
$m_{22}=29$, respectively,  
plotted in
previous figures, including their results relative
to CDM. When the minimum halo mass
is $10^8\text{ M}_\odot$, 
SFDM-TF with $R_\text{TF}=1$ pc
has a halo number comparable to
that of FDM with $m_{22}=29$, and within a factor of 2 of that for CDM, while the masses in all three
models are very close. For $R_\text{TF}=10$ pc, however, SFDM-TF has many more haloes
than does FDM with $m_{22}=0.8$,
by an order of magnitude, and 
$14\%$ of the CDM number, while
the masses in both
SFDM-TF and FDM are comparable
and $\approx{1/3}$ of that for CDM. If the minimum halo mass is lowered to $10^6\text{ M}_\odot$, 
however, SFDM-TF retains
many more haloes than FDM for
\textit{both} example cases --
by an order of magnitude for
$R_\text{TF}=1$ pc vs. $m_{22}=29$,
and by more than a factor of 300 for
$R_\text{TF}=10$ pc vs. 
$m_{22}=0.8$ --
and somewhat more halo mass 
in this low-mass range than the corresponding FDM models, despite the fact that 
the SFDM-TF cases have slightly smaller cut-off mass scales ($M_\text{cut}$) than the corresponding FDM cases. 
In fact, the $R_\text{TF} = 10$ pc case even retains more haloes (by number) than 
the $m_{22} = 29$ case. 
These examples make it clear that one must be careful
when treating lower limits on $m_{22}$ for FDM as
a proxy for providing upper limits on $R_\text{TF}$
for SFDM-TF.  If the former is 
derived to ensure that FDM has a high enough 
HMF at low mass, it may be overly restrictive if
used to place an upper limit on the SFDM-TF $R_\text{TF}$
simply by solving for the
value that makes their cut-off wavenumbers the same.

This gradual cut-off of the HMF is a defining feature of SFDM-TF and distinguishes it from any other DM alternative with a cut-off in its power spectrum that reduces small-scale power relative to CDM as sharply as does FDM.
Its origin is rooted in the ``incredible shrinking Jeans mass'' of SFDM-TF, which leads to a ``recovery'' of small-scale perturbations that
were initially 
retarded below the cut-off when the Jeans scale was large, then
free to grow like CDM after exceeding the Jeans scale, with plenty of time left to close the gap with CDM, at least partially. As a
result, even haloes of mass 
well below $M_\text{cut}$ in the transfer function can still be present in significant numbers
at late times.

In principle, this suggests that SFDM-TF might have an advantage over FDM, when it comes to resolving the problem CDM has of overpredicting dwarf satellites compared with observations of the Milky Way
and Local Group.  
In particular, if FDM uses its
cut-off to suppress the abundance
of dwarf satellites so as to
reconcile with observations
at the higher-mass end (i.e.
at the scale of the so-called
``too big to fail'' dwarfs -- henceforth, ``TBTF''), it 
tends to eliminate lower-mass
haloes (e.g. the ultra-faint dwarfs)
too drastically.  Since the
more gradual cut-off in the HMF of SFDM-TF might avoid this difficulty, we consider this further, as follows. 

Predicted dwarf satellite abundances
can be expressed in terms either of
the number of haloes per interval
of halo mass or their number per
interval of halo circular velocity
(as measured at some radius), which decline with increasing halo mass
and velocity, respectively.
The observational problem for CDM
is usually expressed in terms of the latter, more observable property, involving
velocity, since it can be determined from a rotation curve, observed
only part-way out in total mass.  There are, in principle, two ways
to modify the predicted velocity distribution function
in the TBTF mass range
so as to reduce the overabundance
in CDM N-body simulations to 
make them agree with the observed
distribution: 
\begin{enumerate}[labelwidth=2em, leftmargin=2em, listparindent=\parindent, label=(\arabic*)]
    \item reduce the halo number per unit mass without 
    altering their internal densities relative to CDM haloes, 
    or 
    \item reduce the internal densities of the haloes 
    without changing their number,
    so haloes of a given mass appear at
    lower velocity but match the
    lower abundance of CDM haloes
    of higher velocity.
\end{enumerate}
(TBTF haloes are so named because
they are too massive to be made
undetectable by suppressing their ability to form stars, which would otherwise be a third option for avoiding their apparent overabundance in CDM.)  

FDM has
trouble with both options, 
as follows.  If $m_{22}\approx1$,
TBTF haloes (e.g. with $M\approx{10^{10}\text{ M}_\odot}$) have their internal densities lowered by the presence of solitonic cores as large as their de~Broglie wavelengths of a few kpc or more.  However, while
this helps to solve the TBTF problem by option (2),
it also reduces the 
abundance in the HMF, including a catastrophic reduction at lower-mass where ultra-faint dwarfs are detected. 
If, on the other hand, FDM tries
to avoid the latter problem by choosing a large enough $m_{22}$,
$m_{22} \geq 29$ \citep{Nadler21},
then neither option (1) nor (2)
are available to 
solve the TBTF problem. 
While the value of $m_{22}=29$ 
is just large enough to make sure
the HMF in the range of ultrafaint dwarfs is not over-reduced, the
cut-off is so sharp that
the number of haloes is then
not significantly reduced in the
TBTF mass range, either, so
option (1) fails.
And, unfortunately, the 
solitonic cores of the TBTF
mass range are smaller 
in that case, since their de~Broglie wavelengths are then 
$\lesssim0.1$ kpc,
which is too small to flatten the central density profiles enough to lower the rotation-curve velocities to make option (2) possible. 
Such small cores are also too small for FDM to solve the cusp-core
problem of CDM.  For that problem,
it turns out, even the smaller value of $m_{22}\approx1$ is not a good
solution, since the solitonic
cores of galaxies even \textit{more} massive than TBTF galaxies 
(e.g. $M\gtrsim{10^{11}\text{ M}_\odot}$) are then
too dense, even denser than in CDM \citep{VBB19}.

For SFDM-TF, on the other
hand, option (2) requires $R_\text{TF}\gtrsim1$~kpc, in
order to make large enough
polytropic cores to lower the densities of TBTF haloes sufficiently, which would also
help to resolve the cusp-core problem \citepalias{paper1}.
However, for halo formation
from cosmological
initial conditions, 
our results 
in Fig. \ref{fig:HMF} show that,
if $R_\text{TF}$ is this large,
the cut-off mass in the HMF is
so large that halo abundances at
all galactic halo masses
are substantially reduced
relative to CDM. Hence,
even though it is
a much more gradual cut-off 
than FDM's, it still over-reduces 
the dwarf satellite abundance
relative to CDM.  So, option (2)
is not available to SFDM-TF, either.
According to the results 
in Table \ref{tab:halocounts}, in fact, if $m_{22}\geq29$ is required for FDM to avoid
under-predicting the observed
abundance of ultrafaint dwarfs, then SFDM-TF probably requires $R_\text{TF} \lesssim 1$ pc.
According to
the HMF in Fig. \ref{fig:HMF} for
$R_\text{TF}=1$ pc, however, the halo abundance at the TBTF mass scale is then just the same
as for CDM, so option (1) is
unavailable just as it was
for FDM with $m_{22}=29$.  This
is true despite the fact that
the FDM HMF cut-off just below the assumed ultrafaint dwarf mass
range of $\gtrsim{10^{8}\text{ M}_\odot}$ is much sharper 
than that for SFDM-TF 
with $R_\text{TF}=1$ pc.
That property does
give SFDM-TF one advantage
over FDM, however, since it
enables SFDM-TF to contribute a population of even smaller-mass subhalos which may help to account for perturbers of stellar streams observed in the Milky Way, which
FDM may not \citep[e.g.][]{Banik21}.

While this may suggest that SFDM-TF 
is not much better at resolving the TBTF problem than FDM, in the face
of the recently tightened constraint on the latter imposed by  \citet{Nadler21} based on the abundance of ultrafaint dwarfs in the Milky Way and Local Group, there are caveats to this conclusion.
Firstly, our discussion is so far based
only upon the comparison of
the global HMFs of SFDM-TF, FDM
and CDM, while
dwarf satellites are subhalos
within these larger halos.
A conditional HMF which
predicts the abundances of subhalos
of a given mass within a halo of
a larger mass at a given epoch
might make the comparison between
FDM and SFDM-TF different from
that based upon comparing
their global HMFs. 
Secondly, if the masses
of ultrafaint dwarfs, which are
uncertain, are determined to be below $10^{8}\text{ M}_\odot$, then
the more gradual
cut-off of SFDM-TF gives it a much
better chance of explaining TBTF than FDM,
by allowing a 
larger value of $R_\text{TF}$
(e.g. 
$R_\text{TF}\gtrsim{10}$ pc) without then fatally
underproducing the ultrafaint dwarfs. This is apparent 
if we compare the results for the two 
halo mass ranges in Table \ref{tab:halocounts}:
in the range $10^6 \leq M/\text{M}_\odot \leq 10^{11}$, the $R_\text{TF} = 10$ pc SFDM-TF model
contains \textit{more} haloes than the $m_{22} = 29$ FDM limit, while this is not true in the range 
$10^8 \leq M/\text{M}_\odot \leq 10^{11}$.
Thirdly, comparisons of the
observations with these predicted
HMFs also says nothing about the impact
of dynamical processes inside the parent
halo which can alter the subhalo HMFs
there over time, e.g. by tidal
disruption, which may also be
different for SFDM-TF and FDM.
Ultimately, 3D-cosmological simulations with high resolution
and dynamic range, from initial
conditions that led to the formation
of the Local Group and its
surroundings, may be required to
settle the matter. 

For values
of $R_\text{TF}\ll1$ pc or $m_{22}\gg29$, the HMFs and
internal halo structure of both models approach that of CDM,
for most observables.  In that
case, the same solutions
to the small-scale structure 
problems of CDM that depend upon
modifications of halo structure
and abundance at the low-mass
end by baryonic effects, are
also available to solve the 
same problems for SFDM-TF or FDM.  However, it may still be possible to
distinguish the three models at
even smaller halo mass, e.g. in the 
minihalo mass range, below $\approx{10^{8}\text{ M}_\odot}$.
According to the results for $z=0$ in Table
\ref{tab:halocounts}, the more
gradual cut-off of the HMF for SFDM-TF, relative to FDM, means
that, for $R_\text{TF}=1$ pc, there is still a significant population of minihalos for SFDM-TF, albeit several times reduced from that for CDM, but
an order of magnitude more than
for FDM with $m_{22} = 29$.

\section{Summary and Conclusions}
\label{sec:Conclusion}

In this paper and its companion \citetalias{paper1}, we considered the gravitational dynamics
and quantum hydrodynamics of an alternative form of cosmic dark matter to standard CDM -- 
SFDM-TF -- a scalar field comprised
of ultralight bosons with a 
repulsive self-interaction (SI)
strong enough to make the
length scale, which characterizes
that SI, larger than the de~Broglie
wavelength (i.e. the Thomas-Fermi (TF)
regime). 
Here, in Paper II, we studied structure formation in a universe with this kind of dark matter in a cosmological context, including the dynamics of both nonlinear collapse leading to halo formation and
the linear perturbations which
form the initial conditions for
structure formation in this model.
For the former, we applied the tools developed in \citetalias{paper1}
to simulate cosmological infall and collapse to form individual virialized haloes.  As the first
step in understanding how cosmological initial and boundary
conditions affect the outcome,
we replaced the non-cosmological initial conditions of \citetalias{paper1} with CDM-like
linear initial conditions.
The latter were derived in order to make spherical infall occur at the
rate which matches the empirical mass-accretion history (MAH)
of haloes derived from
N-body simulations of CDM.  Our previous application of this method
for fixing the initial perturbation for CDM simulations showed that
CDM haloes formed in this way 
have NFW-like mass and phase-space density profiles, including the time-dependence of the NFW 
parameters of an individual halo as it grows, in close agreement with
CDM N-body simulations
\citep{AAS03,Shapiro04,Shapiro06}.  So, this provided an
excellent test-bed for determining the impact on this problem of replacing CDM with SFDM-TF in the cosmological context of the expanding universe.

The SFDM-TF haloes that resulted
(see \S\ref{sec:simulationresults})
shared some properties of those in \citetalias{paper1}, namely, polytropic cores within
radius $R_\text{TF}$ and outer envelopes that matched those 
of a simulation with CDM dynamics from the same initial conditions,
although in this case, the
outer-envelope profiles were NFW-like, as expected.
In \citetalias{paper1}, the envelopes were power-law
profiles, $\rho \propto r^{-12/7}$,
everywhere outside the cores,
just as for CDM simulations
from the same \textit{non}-cosmological (i.e. static) initial conditions.
In the cosmological case here, 
however, between the core
and the NFW-like outer envelope
was an intermediate zone with the same power-law profile, $\rho \propto r^{-12/7}$, as 
\citetalias{paper1} found
for its entire envelope outside
the core. The result
of \citetalias{paper1} that haloes
form with polytropic cores surrounded by CDM-like envelopes is still an accurate description of the cosmological case, too, because the infall and collapse that
leads to accretion shocks and virialized regions inside
the shock are essentially CDM-like
dynamics outside the polytropic cores in both cases.  In other words, the term \textit{CDM-like} characterizes the structure outside the cores in both cases, just from different initial conditions.  The reason the
cosmological SFDM-TF 
outcome exhibits
\textit{both} kinds of CDM-like
envelope structure in the same
halo -- a power-law like the CDM simulation from non-cosmological initial conditions, surrounded 
by the NFW-like profile of CDM
simulated in the cosmological case -- is explained by the early
smoothing effect that SI has on
those  
initial conditions
while they are still linear. 
An SI-pressure-driven sound wave was found to smooth-out the mass within a proper radius $R_\text{TF}$
of the center, in
a time short compared to the time for that mass to expand  cosmologically, turn around and recollapse.  Once smoothed,
that mass continues expanding, but thereafter evolves like a top-hat perturbation, at the smoothed-out overdensity.
When this mass collapses, the
outcome for it is just like
that in \citetalias{paper1}, while
the mass outside it evolves just
as it would for those CDM-like initial conditions with no smoothing effect.

This has profound consequences for
halo formation from cosmological perturbations in SFDM-TF.
If SI is too strong, i.e. $R_\text{TF}$ is too large, the
SI-smoothing sound wave can
smooth-out a mass as large as that
of the halo destined to form at some epoch in CDM from the same initial conditions,
preventing its formation,
altogether, or at least delaying
it significantly.  In \citetalias{paper1}, we showed
how SFDM-TF haloes with the right value of $R_\text{TF}$ could resolve
the cusp-core and TBTF problems 
of CDM, without suffering
from the problem identified for FDM
by \citet{VBB19} in which the
solitonic cores of larger-mass FDM haloes were too dense to be consistent with the rotation curves
of more massive galaxies, if FDM
particle mass were tuned to 
$m_{22}\simeq1$ to solve
the other problems.  We found
that $R_\text{TF}\gtrsim1$ kpc is
required.  It was left to Paper II,
here, to determine if such a value
of $R_\text{TF}$ was consistent
with the cosmological formation of
those haloes.  Our cosmological simulations here indicate
that the SI-smoothed mass 
at early times is large 
enough to suppress
the formation of such dwarf galaxies if $R_\text{TF}\gtrsim1$ kpc, if
we start from CDM initial conditions.  At the other extreme,
if $R_\text{TF}\lesssim1$ pc, then
the internal structure of haloes
that form from such CDM-like 
initial conditions would closely resemble those of CDM, over most
of the observable range of radii.

This still left open 
the question of what
realistic cosmological initial conditions are for SFDM-TF halo formation, as distinct from those
for CDM.  To answer that requires
linear perturbation theory, 
from which the statistical likelihood of finding haloes of different masses at different epochs, can be determined. That
was the subject of \S\ref{sec:LinPertsTransferFn}.
There, we derived the transfer function for SFDM-TF, normalized
by that for CDM, and, for comparison, following the same
procedure, did the same for
FDM, to contrast the three models. With the transfer function, we were able to calculate statistical measures of structure formation for the SFDM-TF model, for the first time, including its halo mass function (HMF).  Just as
the FDM model is fully parameterized
by its particle mass $m$, so
SFDM-TF is parameterized entirely
by its SI strength parameter,
$g/m^2$, which is directly related
to the radius $R_\text{TF}$ of the
($n=1$)-polytropic sphere in hydrostatic equilibrium in which 
gravity is balanced by SI
pressure. Like FDM, SFDM-TF is
a model with a cut-off of power
on small scales, reflected as a
cut-off at high wavenumber in
the transfer function.  As a
result, we used these findings
to make a first cut at estimating
the range of $R_\text{TF}$-values
which are consistent with 
astronomical observations, by
using constraints on the FDM
model derived elsewhere, 
expressed as constraints on its $m_{22}$, as a proxy.

According to our linear perturbation theory presented in \S\ref{sec:LinPertsTransferFn},
the Fourier mode at which SFDM-TF cuts off relative to CDM has a comoving mass associated with it that equalled the time-varying Jeans mass at the moment that mode entered the horizon (i.e. filled the Hubble volume).  With a Jeans length equal to $R_\text{TF}$, fixed in proper coordinates, the comoving
Jeans mass during this phase decreased rapidly with scale factor, as $M_\text{J,0}\propto a^{-3}$. 
Modes of smaller wavenumber entered the horizon later, with larger associated masses, \textit{above} the Jeans mass, 
so they always grew like CDM. 
Modes of higher wavenumber entered earlier, however, when their associated masses were \textit{below} the Jeans mass and, so, were initially retarded relative to CDM, but, as soon as that rapidly shrinking Jeans mass reached their mass scale, they, too, were free, thereafter, to grow like CDM.
The cosmological story of structure formation in SFDM-TF is, therefore,
very much a tale of \textit{``the incredible shrinking Jeans mass''}, which is the key to understanding how it differs
from other cut-off models 
like FDM \big(for which the Jeans
mass only drops weakly, as
$M_\text{J,F}\propto a^{-3/4}$\big)
which do not share this unique property.
Another factor that distinguishes
SFDM-TF from FDM, in particular,
is that, even when SFDM-TF
modes are \textit{sub-Jeans}, 
they still grow, albeit modestly, as $a^{1/4}$ (when the field is
matter-like), while for FDM, there is \textit{no} sub-Jeans growth. 

This explains why, when we 
applied the same treatment to FDM to identify its cut-off wavenumber (as a function of $m_{22}$), and solved for the $R_\text{TF}$-values required to align the cut-off wavenumbers of the two models, SFDM-TF haloes had SI-polytropic halo cores
(of radius $R_\text{TF}$) much
smaller than the solitonic cores of
FDM haloes of the same mass.  
For example,
FDM with $m_{22}\simeq 1$, for 
which a $10^{10}\text{ M}_\odot$ halo has a solitonic core size
(of order the de~Broglie wavelength) of $\simeq 4$ kpc, cuts off at close
to the same wavenumber as SFDM-TF
with $R_\text{TF}\simeq3$ pc, 
which gives a halo of that mass a polytropic core  $\approx$\,1000 times smaller.  This is
consistent with our simulations in \S\ref{sec:simulationresults},
which showed that halo formation
in SFDM-TF from a CDM-like perturbation experiences early SI-smoothing that, in effect, ``Jeans-filters'' a mass closer to the \textit{initial} value
of the Jeans mass than to its 
smaller value after the collapse is
finished.  As a result, although
SFDM-TF haloes with
flattened cores as large
as $R_\text{TF}\!\gtrsim\!1$ kpc,
formed by Jeans 
instability and collapse from 
a static initial condition,
were shown in \citetalias{paper1}
to be capable of resolving the 
cusp-core problem of CDM while \textit{simultaneously} solving its
TBTF problem, as well,
this unified explanation 
is disfavored, 
once formation is placed in the context of
cosmological perturbation growth.

Observational constraints on FDM
can be used as a proxy for those
on SFDM-TF by solving for the matching $m_{22}$ and $R_\text{TF}$ values that make their respective transfer functions cut off at the same wavenumber, e.g. as estimated analytically by equation (\ref{eq:RTFversusmFDM}).  Care
should be taken to 
avoid FDM constraints that are based upon phenomenology that might distinguish it from SFDM-TF because of their different dynamics.  
Since lowering 
$m_{22}$ increases the de~Broglie
wavelength and extends the mass range suppressed by FDM to larger scales, many of its observational constraints are expressed as lower limits on $m_{22}$ required to avoid over-suppressing structure.   
There is
a long list of attempts to
estimate or limit $m_{22}$, with results roughly spanning
the range from 1 to 30 
(\citealp[as summarized, e.g., in][]{AWP}). Some are based
upon the comparison of the abundance of haloes or subhaloes with the
predicted HMF for FDM.  \citet{Nadler21}, for example,
compare the predicted
HMF of subhaloes to the observed
dwarf galaxies in the Local
Group and find that $m_{22}\gtrsim{29}$ is required,
corresponding to $\lambda_\text{deB} \lesssim 0.5$~kpc. Others
are based upon more complicated,
nonlinear dynamical phenomena. 
For example,
the requirements of not dynamically over-heating the Milky Way disk or its stellar streams were
used to place bounds of $m_{22} \gtrsim 0.6-1.5$ \citep{AL18, Church19}. Still
others are based upon structure on larger scales, at
epochs when it is only
quasi-linear, but where one must model effects on the baryonic component, too, such as
fluctuations in the matter-density
field inferred from the
power spectrum of measured fluctuations in the Lyman-$\alpha$
forest. Bounds inferred from the Lyman-$\alpha$ forest were the 
first to pose the ``catch-22 problem'' for FDM, in fact, 
finding $m_{22} \gtrsim 20-30$ \citep{sc22,sc23}, now
consistent with the recent
lower limits
quoted above from the abundance
of dwarf satellites in the Local
Group. 

If the observational 
constraints on FDM we use as our proxy for those on SFDM-TF
are taken to be 
$1 \lesssim m_{22} \lesssim 30$,
then the range of $R_\text{TF}$-values for which
the two models have the same cut-off wavenumber corresponds
roughly to 
$3 \text{ pc} \gtrsim R_\text{TF} \gtrsim 0.2$ pc.  Of all the methods for
constraining the FDM model we 
might use, as a more \textit{specific} proxy than this
to constrain SFDM-TF, those comparing
their HMFs are,
to first approximation, less affected by additional differences in their dynamics
than others.  For this reason,
we focused most of our additional
comparison here on the results of
their HMF predictions, as discussed
in \S\ref{sec:HMFresults}.
We showed there that, for the same
cut-off wavenumbers, the SFDM-TF 
HMF is much more gradually cut off
than that of FDM, so it retains
many more haloes at the low-mass
end.  As a result, if we
were, instead, to solve for the
$R_\text{TF}$-value that yields the
same halo number as FDM
for some value of $m_{22}$, the
corresponding $R_\text{TF}$-value
is generally larger than the
one that makes the cut-off wavenumbers match, described
above.  For example, if we
integrate over their HMFs
to count haloes above 
$10^8\text{ M}_\odot$, 
SFDM-TF with $R_\text{TF}=1$ pc
has a halo number comparable to
that of FDM with $m_{22}=29$, and within a factor of 2 of that for CDM, while the integrated collapsed
fractions in this halo mass range
in all three models are very close.
In that case, we should
multiply the $R_\text{TF}$-value
above that matches the cut-off wavenumber for this $m_{22}$ by
a factor of 5, from 0.2 to 1 pc,
to use the HMF FDM constraint as a proxy for SFDM-TF. Moreover, 
if $R_\text{TF}=10$ pc, 
SFDM-TF has many more haloes
than does FDM with $m_{22}=0.8$,
by an order of magnitude, and
still has $14\%$ of the CDM number.
In that case, we should refine
the first proxy range 
estimated above, upward,
at least to $10 \text{ pc} \gtrsim R_\text{TF} \gtrsim 1$ pc.  If we compare
the integrated halo counts 
in the two models down to even smaller halo mass, the corresponding $R_\text{TF}$-values shift upward
even further. If the minimum halo mass is lowered to $10^6\text{ M}_\odot$, for example,
then SFDM-TF with $R_\text{TF}=1$ pc has an order of magnitude more haloes than does FDM with $m_{22}=29$, while 
if $R_\text{TF}=10$ pc, it has 
more by a factor of 300 than FDM
with $m_{22}=0.8$. 
This indicates that one must be
careful not to take the simple
alignment of the cut-off wavenumbers of SFDM-TF and FDM as a basis
for using the observational
constraints on $m_{22}$ to solve
for the corresponding constraints
on $R_\text{TF}$ -- that will generally be overly restrictive,
with upper limits on $R_\text{TF}$
which are too low. It further
suggests that SFDM-TF may
have some advantages 
over FDM since, even when
their model parameters are tuned
so as to \textit{match} HMFs 
at larger
mass scales, SFDM-TF may still have
more smaller-mass haloes.  For
example, when FDM and SFDM-TF
are both limited so as to avoid
over-suppressing the
dwarfs in the Local Group, 
SFDM-TF may still have enough even-smaller-mass subhalos to account for the perturbers of stellar streams observed in the Milky Way, even if FDM does not \citep[e.g.][]{Banik21}. 

To test SFDM-TF against observations more directly in future work, 
it will be necessary to revisit
the full list of phenomenological
constraints that have been applied
to FDM and other models that
suppress small-scale structure
as a built-in feature (e.g. warm dark matter),
with SFDM-TF initial conditions and dynamics, instead.  This will require us to apply the cosmological initial conditions derived here for SFDM-TF in fully 3D simulations of galaxy and large-scale structure formation, which are under development. This includes the coupling of dark matter and baryonic components, as well as pure dark matter simulation.  
For ``cut-off models'' like these,
the nature of structure formation
on scales at or below the cut-off
is different for different types
of dark matter, even when they
start from transfer functions that
have similar cut-off scales,
as our results here demonstrate
by comparing the HMFs for SFDM-TF and FDM.   Additional effects,
like fragmentation of 
pancakes and filaments, for
example, can also introduce novel small-scale structures, unfamiliar from the hierarchical clustering paradigm of CDM \citep[e.g.][]{Mocz19,Valinia97}.  The small-$R_\text{TF}$-regime favored 
thus far by our results presented above, for which 
the polytropic core sizes are
subgalactic (i.e. sub-kpc), may
present other novel features
which are detectable, as well,
worth exploring further.   As discussed in \citet{Padilla20}, for example, such cores, 
if dense enough, may
be subject to general relativistic
instability leading to collapse to
form supermassive black holes or
their seeds.

{
In this paper, we considered the r\^ole of SI in structure formation in the SFDM model, in the TF regime.   We also compared the results in this regime with the other limiting case, of FDM, in which SI is absent.   As we showed in \S\ref{sec:LinPerts}, as long as the particle mass and interaction strength place the scalar field in this regime at late times (e.g. the present), it is in the TF regime at earlier times, as well. In the future, we will consider the possibility of a transitional regime between these two limits.
}

\section*{Acknowledgements}

This material is based upon work supported by the National Science Foundation Graduate Research Fellowship Program under Grant No. DGE-1610403. Any opinions, findings, and conclusions or recommendations expressed in this material are those of the authors and do not necessarily reflect the views of the National Science Foundation. 
T.R.-D. is supported by the Austrian Science Fund FWF through an Elise Richter fellowship, Grant No. V 656-N28. 
Simulations presented here were conducted on the Texas Advanced Computing Center's Stampede2 supercomputer under accounts asoz-630 and NSF XSEDE account TG-AST090005.  
The authors gratefully acknowledge  
\textit{The Incredible Shrinking Man }(1957)
~[Film], directed by Jack Arnold, Universal-International Pictures Co., Inc., as the inspiration for our subtitle.

\section*{Data Availability Statement}

The data underlying this article are available in the article and in its online supplementary material.



\bibliographystyle{mnras}
\bibliography{ref} 



\appendix

\section{Overdensity profile for non-self-similar spherical infall: NFW-producing initial conditions}
\label{sec:deli}

In order to obtain an exact expression for equations~(\ref{eq:NFWPert}) and (\ref{eq:deli}), we calculate the initial mean overdensity of Lagrangian mass shells that accrete onto the halo (i.e. reach close to $R_{200}$) at times that are consistent with the MAH of equation~(\ref{eq:MAH}). Prior to the scale factor at which each shell has accreted, $a_{200}(M)$, the pressure forces are negligible, so the shell's trajectory is well-described by pressure-free, spherical top-hat collapse, governed by Newtonian gravitational motion:
\begin{align}
    \frac{d^2r}{dt^2} &= -\frac{GM}{r^2} \\
    &= -\frac{2}{9}\frac{r}{t^2}\big(1+\Delta\big) \label{eq:tophat}
\end{align}
where the last expression comes from the definition of the mean overdensity of a shell embedded in an EdS background universe for which $\rho_\text{crit} = (6\pi Gt^2)^{-1}$,
\begin{equation}
    1+\Delta = \frac{M}{\frac{4\pi}{3}r^3\rho_\text{crit}} = \frac{9}{2}\frac{GMt^2}{r^3}
\end{equation}
Equation~(\ref{eq:tophat}) yields parametric solutions
for each $M$ \citep[see, e.g.,][\S5.1.1]{MVW10}:
\begin{align}
    r(M,a) &= \frac{1}{2}\frac{r(M,a_i)}{5\Delta_\textsc{l}(M,a_i)/3}(1 - \cos{\theta}) \\
    t(a) &= \frac{3}{4}\frac{t(a_i)}{[5\Delta_\textsc{l}(M,a_i)/3]^{3/2}}(\theta - \sin{\theta}) \label{eq:time_param}\\
    1+\Delta(M,a) &= \frac{9}{2}\frac{(\theta-\sin{\theta})^2}{(1-\cos{\theta})^3}
\end{align}
where $\theta = \theta(M,a)$.
Replacing cosmic time for scale factor in equation~(\ref{eq:time_param}) ($a\propto t^{2/3}$ for EdS), we can express the initial overdensity in terms of $a_{200}(M)$:
\begin{equation}
    \Delta_\textsc{l}(M,a_i) = \frac{3}{5}\frac{a_i}{a_{200}(M)}\bigg(\frac{3}{4}(\theta_{200}-\sin{\theta_{200})}\bigg)^{2/3}
    \label{eq:laststep}
\end{equation}
where $\theta_{200} \simeq 4.8$ is the $\theta$ parameter for which $1+\Delta = 200$ (i.e. the parameter at which the shell reaches $r = R_{200}$, and at which $a=a_{200}$). For a given shell at this point in its trajectory, its enclosed mass will now be identical to the total mass of the halo, so by equation~(\ref{eq:MAH})
\begin{align}
    &M = M_{200} = M_\infty e^{-s a_f/a_{200}(M)} \\
    &a_{200}(M) = \frac{s a_f}{\ln{(M_\infty/M)}}
\end{align}
Finally, substituting this into equation~(\ref{eq:laststep}), we obtain the exact form of equation~(\ref{eq:NFWPert}):
\begin{align}
    \Delta_\textsc{l}(M,a_i) &= \frac{3}{5}\frac{a_i}{s a_f}\bigg(\frac{3}{4}(\theta_{200}-\sin{\theta_{200})}\bigg)^{2/3}\ln{(M_\infty/M)} \\
    &\simeq 0.8 \frac{a_i}{a_f}\ln{(M_\infty/M)}
\end{align}

\section{Setting the horizon-entry scale factor}
\label{sec:aH}

\begin{figure}
    \centering
    \includegraphics[width=\columnwidth]{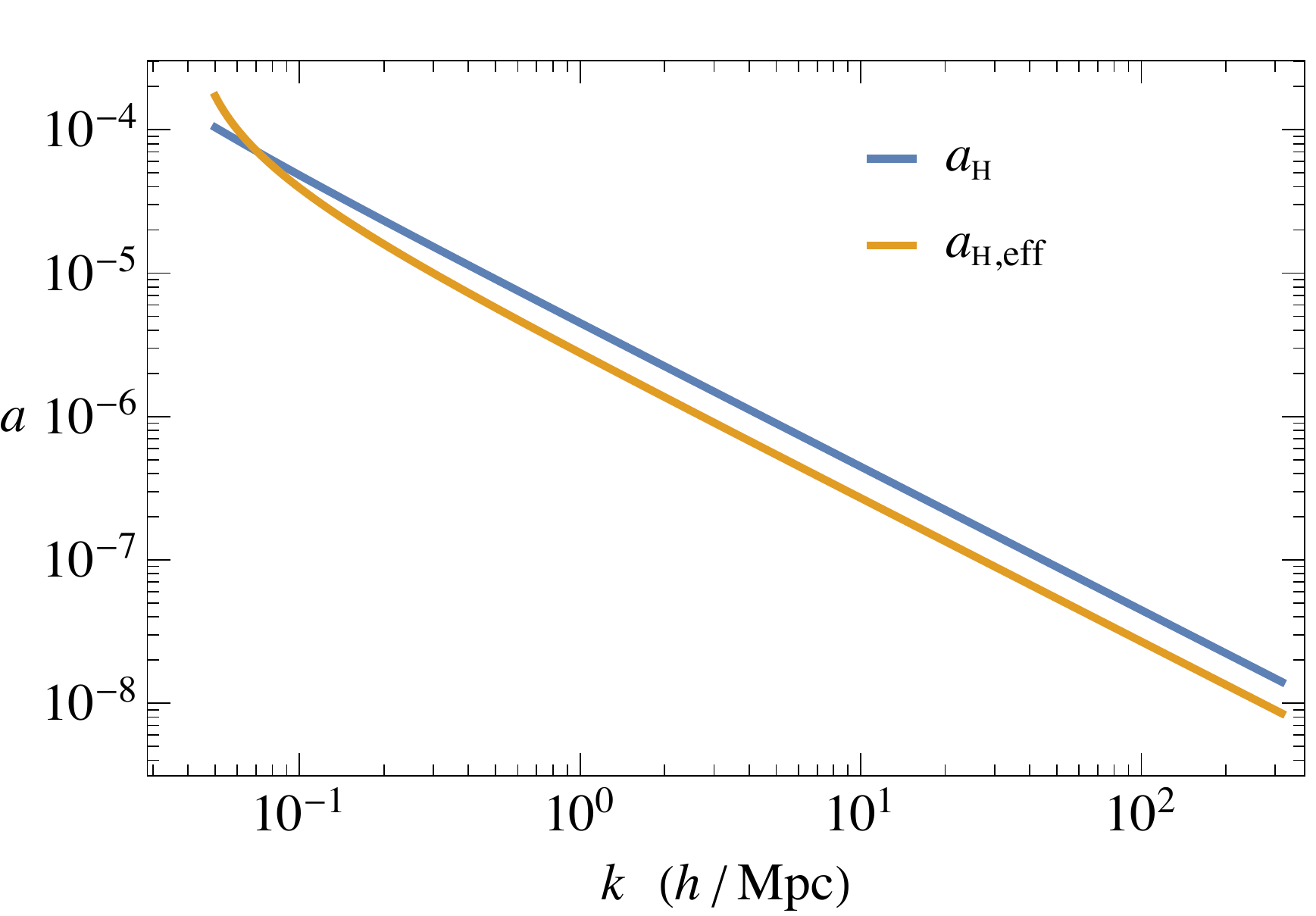}
    \caption{Comparison of the horizon-entry scale factors, $a_\textsc{h}$ and $a_{\textsc{h},\text{eff}}$, as defined by equations~(\ref{eq:aH}) and (\ref{eq:aHeff}), respectively.}
    \label{fig:aHeff}
\end{figure}

In our treatment of linear perturbations in the SFDM model, we use the thin-horizon approximation and define a ``horizon-entry'' scale factor that specifies when a mode transitions from superhorizon to subhorizon. 
A physically-motivated definition of horizon entry is roughly the scale factor at which the wavelength of a mode (with wavenumber $k$) equals the diameter of the Hubble sphere, so that:
\begin{equation}
    k = \frac{\pi}{c} a_\textsc{h} H(a_\textsc{h}) \label{eq:aH}
\end{equation}
In the radiation-dominated era, this can be approximated by
\begin{equation}
    a_\textsc{h} \approx \pi \frac{H_0 \sqrt{\Omega_\text{rad}}}{ck}
\end{equation}
However, in order to ensure that the perturbation amplitude is continuous across the horizon-entry transition, we must require that the horizon-entry scale factor be such that the equation describing the subhorizon evolution of the perturbation be equal to its superhorizon amplitude at the horizon-entry scale factor.
For CDM perturbations, the subhorizon evolution is given by equation~(\ref{eq:deltaCDM}), so we define the effective horizon-entry scale factor, $a_{\textsc{h},\text{eff}}$, as that which makes the right side of this equation equal to 1.
In the radiation-dominated era, this can be approximated by
\begin{equation}
    a_{\textsc{h},\text{eff}} \approx e^{2/3-\gamma_\textsc{e}} \sqrt{3} \frac{H_0 \sqrt{\Omega_\text{rad}}}{ck} \simeq 2 \frac{H_0 \sqrt{\Omega_\text{rad}}}{ck}
\end{equation}
which is not far from the physically-motivated definition of $a_\textsc{h}$.
We compare both of these definitions as a function of $k$ in Fig.~\ref{fig:aHeff}.

\bsp	
\label{lastpage}
\end{document}